\documentclass
[preprint,pre,tightenlines,showkeys,titlepage,nobibnotes,twocolumn,10pt]{revtex4}%
\usepackage{amsfonts}
\usepackage{amsmath}
\usepackage{amssymb}
\usepackage{graphicx}%
\newtheorem{theorem}{Theorem}

\newtheorem{conjecture}[theorem]{Conjecture}

\newtheorem{remark}[theorem]{Remark}

\begin{document}
\preprint{UATP/07-03}
\title{Exact Statistical Mechanical Investigation of a Finite Model Protein in its
environment:\ A Small System Paradigm}
\author{P.D. Gujrati,$^{1,2,\email{pdg@arjun.physics.uakron.edu}}$Bradley P. Lambeth, Jr.,$^{1}$ Andrea Corsi,$^{1,2,\altaffiliation{A. Corsi is currently at Molecular Stamping, Via Einstein, 26900 Lodi, Italy.}}$, and
Evan Askanazi,$^{2}$}
\affiliation{$^{1}$The Department of Polymer Science, $^{2}$The Department of Physics, The
University of Akron, Akron, OH 44325}

\begin{abstract}
We consider a general incompressible finite model protein of size $M$ in its
environment, which we represent by a semiflexible copolymer consisting of
amino acid residues classified into only two species (H and P, see text)
following Lau and Dill. We allowing various interactions between chemically
unbonded residues in a given sequence $\chi$ and the solvent (water), and
exactly enumerate the number of conformations $W(E)$ as a function of the
energy $E$ on an infinite lattice under two different conditions: (i) we allow
conformations that are restricted to be compact (known as Hamilton walk
conformations), and (ii) we allow unrestricted conformations that can also be
non-compact. It is easily demonstrated using plausible arguments that our
model does not possess any energy gap even though it is supposed to exhibit a
sharp folding transition in the thermodynamic limit. The enumeration allows us
to investigate exactly the effects of energetics on the native state(s), and
the effect of small size on protein thermodynamics and, in particular, on the
differences between the microcanonical and canonical ensembles. We find that
the canonical entropy is much larger than the microcanonical entropy for
finite systems. We investigate the property of self-averaging and conclude
that small proteins do not self-average. We also present results that (i)
provide some understanding of the energy landscape, and (ii) shed light on the
free energy landscape at different temperatures.

\end{abstract}
\date{\today}
\maketitle

\section{ Introduction}

\subsection{Proteins as Semiflexible Heteropolymers}

Proteins are organic compounds made of amino acids, also known as residues,
bound in a chain-like structure by peptide bonds. Self-assembling small
proteins can fold into their native states (of minimum free energy) without
any chaperones, and have been extensively investigated recently using lattice
models by thermodynamic principles \cite{Anfinsen}. They differ from flexible
polymers, which collapse to a compact disordered state; they are similar to
\emph{semiflexible} polymers in which semiflexibility forces an ordered
(crystalline) compact structure at low temperatures \cite{note00}.

Let $N_{\text{R}}$ denote the total number of residues in $N$ proteins in a
volume $V;$ the residue concentration is
\[
c\equiv N_{\text{R}}/V.
\]
To ensure that the boundary of the volume $V$ does not affect the behavior of
the system, we need to take the limit $V\rightarrow\infty$. This limit will be
usually implicit in the following, unless mentioned otherwise. In many cases,
we deal with a dilute solution so that the concentration of proteins is
exceedingly small. Accordingly, the proteins are far apart with no appreciable
inter-protein interactions. It is then safe to consider a single protein by
itself in its environment, i.e. in the presence of water. The presence of
inter-protein interactions in a solution, which is not dilute, and in a bulk
means that these systems (both of which we will not consider in this work)
containing many proteins should be distinguished from that containing a single
protein, as their thermodynamics will be very different.

\subsubsection{Protein as a Small System}

Our focus in this work is on a single protein ($N=1$) containing $M$ residues
so that $N_{\text{R}}=M$. As proteins are usually small in size, we need to
recognize that the behavior of a single protein is governed by the
thermodynamics of a \emph{small} system (defined as a system in which
$N_{\text{R}}$ does not grow with the volume $V$ as $V\rightarrow\infty$)\ and
not of a macroscopic system, such as formed by a bulk (in which $N_{\text{R}%
}\equiv NM$ grows with the volume $V$); the latter will be governed by the
thermodynamics of a macroscopic system \cite{note03}. It is well known that
predictions of different ensembles describing a macroscopic system are the
same, except at some singular points such as where phase transitions occur.
Therefore, it is important to understand the ways in which different
statistical ensembles differ from each other for small systems. This is one of
the important issues motivating this investigation: how to distinguish small
system thermodynamics from a macroscopic system thermodynamics in various
ensembles. For this purpose, it is sufficient to consider only two ensembles:
the microcanonical (ME) and the canonical (CE) ensembles.

\subsubsection{Structures and the Standard Model}

The \emph{residue sequence} (known as the primary structure) in a protein is
defined by a gene and is encoded in the corresponding genetic code.
Understanding the relationship between the sequence and protein functionality
is an unsolved problem though major progress has been made \cite{Finkelstein}.
A first-principle study of primary, secondary (regularly repeating local
structures, such as helices and $\beta$-sheets) and tertiary (the overall
shape or \emph{conformations }of a single protein) structures requires short
(local) and long (nonlocal) ranged model energetics that, while remaining
independent of protein conformations, temperature and pressure, determines the
native state(s), and has to be judiciously chosen to give a unique and correct
native state \cite{Scheraga0}.

The simplest model that can be used is the \emph{standard model }of Lau and
Dill \cite{Dill}, which classifies the 20 different amino acid groups or
residues into two subsets, H (\emph{hydrophobic} residues) and P
(\emph{hydrophilic/polar} residues), and allows only nearest-neighbor
attractive HH interaction (whose strength is set equal to 1 in some
predetermined unit) to provide good hydrophobic cores; however, consideration
of local energetics of the 20 residues \cite{Miyazawa} is also common. It is
also found that the introduction of multi-body interaction enhances
cooperativity \cite{Kolinski}, and should not be neglected.

The protein in the standard model is an example of a copolymer of a prescribed
sequence. It is this simplified copolymer model and its variants proposed in
this work that will be the subject of investigation here, even though the work
can be extended to a more general case.

\subsection{Energetics and Energy Distribution $W(E)$ of Conformations}

\subsubsection{Microscopic Interaction Energies}

The\emph{ microscopic energies} that appear in the model energetics, while
determining the thermodynamics, must themselves be independent of the
thermodynamic state, i.e., of protein conformations, temperature, pressure,
concentration, etc. to be truly microscopic. In addition, a proper model
should satisfy certain principles \cite{Scheraga1}, one of which is the
requirement of \emph{cooperativity} needed for the existence of a first-order
transition (a latent heat) at the folding transition to the native state. The
residue sequence plays an important role in determining the native state
\cite{Shakhnovich} and, therefore, the thermodynamics. Thus, we are driven to
treat proteins as semiflexible heteropolymers with certain specific sequences
\cite{Venkat}. However, there is no consensus for general energetics to
describe all proteins, and there remains a certain amount of freedom in the
choice for a theoretical investigation. It is widely recognized that secondary
structures are also important in the folding process \cite{Dill}, yet they are
not always incorporated in determining the energetics.

In view of the above discussion, it is important, therefore, to investigate
the effects of energetics on the behavior of \emph{small} proteins, an issue
that, to the best of our knowledge, has not been studied fully.

Protein stability and function are the results of extensive evolutionary
changes. In other words, the natural evolution has over a long period
eventually found the most optimal energetics for an individual protein with a
given sequence to fold fast into its native state. The energetics must be
tuned to the particular sequence in addition to the protein \emph{structure};
the latter is defined as a particular conformation of the protein alone
without any regard to the surrounding environment or the sequence. Thus, the
study of the structure without accounting for the environment such as water
inside the cell or the sequence will not provide a complete understanding of
protein thermodynamics. This is because the true interactions of a real
protein determine the equilibrium structure for a given sequence. For the
energetics to be truly microscopic, it must also be independent of the
sequence. This means that not all sequences will form natural proteins.

It has also been argued that conflicts among interactions also play a
significant role in folding \cite{Clementi}. The interplay of intra-protein
molecular interactions, the interaction with the surrounding, and the residue
sequence to give rise to the folded native state is quite intricate and far
from being understood. A complete understanding will enhance not only our
ability to find cures, but also to design proteins with a desired behavior.
For this, we need a true appreciation of the underlying molecular interactions
and the resulting thermodynamics, not specific to a particular folding. This
is a key ingredient in obtaining a detailed understanding of folding, as the
energetics determines the \emph{energy landscape} that presumably dictates the
path to folding.

As the knowledge of the general energetics that controls folding in all
proteins is an unsolved problem, progress can only be made by constructing a
model or models with a goal to explain some desired or important features of
the folding process as is common with any complex physical system. In general,
the model should contain various interactions relevant not only for various
secondary substructures like helix formation in the native state, but also for
proteins as semi-flexible heteropolymers.

For the standard model and its variants that we consider here, the proteins
are treated as semiflexible \emph{copolymers}. The model should also contain
solvation effects, as all protein activity occurs in the presence of water or
solvent. The compressibility also plays an important role. However, as we will
discuss later, this makes the problem very complicated. Therefore, in this
work we only consider an \emph{incompressible} model, and propose such a model
and investigate its behavior in different limits, one of which is the standard
model described above. However, the central focus of the work remains to be
the investigation of small system thermodynamics, since proteins form small
systems( $M<\infty$). We will demonstrate that the thermodynamics of small
proteins differs from that of its macroscopic analog in some unexpected but
substantial ways.

\subsubsection{Energy Distribution $W(E)$ and Small System Thermodynamics}

Pairwise residue contact energies or potentials are commonly used in
theoretical studies of protein folding as an important simplification because
of the complexity of the problem. These potentials are derived from the
knowledge of conformations in the crystal structures of proteins in the
protein data bank, but the procedure comes with serious limitations
\cite{Scheraga0}. One such limitation is the small number of conformations
that describe the ordered state of the protein. A better way would be to use
all the conformations $W\geq1$ of the protein \cite{note01}. This requires the
determination of the distribution $W(E)\geq1,$ the number of the conformations
of a given energy $E$ \cite{note}.

Once $W(E)$ is known, the complete thermodynamics is determined. This is
certainly believed to be true for macroscopic systems, systems in which the
volume of the system becomes macroscopically large to suppress boundary
effects, while keeping the density of participating particles such as $c$
fixed in the limit; in mathematical terms, the volume must diverge to infinity
(thermodynamic limit) \cite{note03}.

\begin{conjecture}
\label{Conj_1}We will take the viewpoint that $W(E)$ also provides the
complete thermodynamics for small systems \cite{note03}.
\end{conjecture}

We will demonstrate, however, that care must be exercised as not all that is
valid for a macroscopic system remains valid for small systems. It should be
stressed \ that $W(E)$ depends on the particular sequence\ $\chi$ of the
residues, even if $W$ does not \cite{note01}. An interesting question arises
about the property of self-averaging in heteropolymers
\cite{Rensburg,Kardar,Bryngelson}; see Sect. \ref{Sect_SelfAve} for details.
For small proteins, there is evidence that certain properties of interest
depend on the sequence $\chi$ in important ways \cite{Bryngelson}.

\subsection{Exact Approach for\ small Proteins}

Usually, one attempts to determine the distribution $W(E)$ by carrying out
several simulations. Because of the limitations inherent in the simulation, an
alternative approach is to determine $W(E)$ by \emph{exact enumeration }on a
lattice. Such enumerations allow us to do exact calculations; no approximation
has to be made. This has the added benefit that we can verify various
conjectures about the form of entropy, self-averaging, landscape, etc. The
enumeration is, however, feasible only for short proteins. The smallest known
natural protein (at least to us) is Trp-Cage derived from the saliva of Gila
monsters. It has only 20 residues. Our approach is to consider the protein to
be a small thermodynamic system containing $M\,<\infty$ residues or ammino
acids \cite{note03}, even if the lattice on which it is embedded is infinite.
(As discussed later, we cut down the number $W(E)$ by \emph{rooting} the
protein by fixing one of its end at the origin of the lattice and exploiting
some symmetry properties.)\ This approach also allows us to investigate how
the thermodynamics of small proteins differ from that of macroscopic polymers,
with some unexpected results. In particular, we need to recognize that small
proteins cannot undergo a sharp (i.e., discontinuous or first-order) folding
transition. Thus, there will, in principle, be no latent heat. One can only
look for some unambiguous signature of a latent heat (i.e., of cooperativity),
which can justify a sharp transition in the thermodynamic limit of a
macroscopic protein. We must also consider the effects of residue sequences on
the degeneracy of the lowest energy state and the nature of any possible
transition in the thermodynamic limit.

\subsection{Layout}

The layout of the paper is as follows. In the next section, we provide a
discussion of the required thermodynamic background to appreciate what may
happen differently for small systems compared to a macroscopic system. In
Sect. III, we discuss a very general incompressible lattice model of a protein
of a given sequence. The incompressibility brings about certain
simplifications as we will discuss later. We will only consider a small
protein. We introduce three models that include the standard model and two
variants due to weak and strong perturbations. We consider random, ordered and
fixed sequences. We consider compact conformations or all conformations
(compact and non-compact) separately, and label them as restricted or
unrestricted to distinguish them. In the following section, we discuss the
issue of self-averaging and test it for small proteins. In Sect. V, we study
the effects of energetics on native conformations. In Sect. VI, we introduce
small system entropies in the microcanonical and canonical ensembles, and
discuss various thermodynamic laws that remain valid for small systems. In the
following section, we compare the entropies in the two ensembles. In Sect.
VIII, we study various densities and the specific heat. We introduce the
notion of a distance in Sect. IX and use this to project the multi-dimensional
configuration space onto a two-dimensional space from which we draw some
conclusions about the configuration space and the landscape. We construct the
free energy landscape from our numerical results in Sect. X. The last section
contains a brief summary and discussion of our results.

\subsection{Results}

\begin{enumerate}
\item We show that the conformations associated with native states of a given
fixed energy depend on the residue sequence.

\item Under very mild assumptions, we show that there is \emph{no energy gap}
in our model of a macroscopic protein; see Sect. \ref{Sect_Absc_Energy_Gap}.

\item The self-averaging does not seem to occur in small proteins, at least
for the native state energy, so that the sequence $\chi$ plays an important
role; see Sect. \ref{Sect_SelfAve}.

\item Different energetics can give the same native state (Sect.
\ref{Sect_Energetics_NativeConf}).

\item For small proteins, the entropy and energy densities are not only
discrete but also depend on $M$ strongly; see Sect.
\ref{Sect_Small_System_Equilibrium}. In addition, the entropy density $s(e)$
is higher for larger $M$ over a wide range of energies; see Sect.
\ref{Sect_Small_System_Equilibrium}.

\item Justification for using the Boltzmann entropy and the Gibbsian entropy
and the partition function formalism for small system is given in Sect.
\ref{Sect_Justification_Small_System}. We follow this approach in this investigation.

\item For small systems, we prove that $\overline{S}(\overline{E})\geq
S(\overline{E})$ where $S(T)=\overline{S}(\overline{E})$ is the canonical or
the Gibbsian entropy at $T$, while $S(\overline{E})$ is the Boltzmann entropy
at the average energy $\overline{E};$ see Sect. \ref{Sect_Both_S_E_Inequality}%
. For \ a macroscopically large system, the two entropies are the same. We
also prove that $\overline{S}(E)$ is a concave function, but $S(E)$ is not.

\item One cannot trust the Gaussian form of the ME entropy following the
random energy model, as it predicts a vanishing entropy at an energy above the
native state, thereby suggesting an energy gap and a frozen native state, both
of which are not correct for a finite protein; see Sect.
\ref{Sect_Gaussian_Fit}.

\item The net effect of the perturbations is to make the native state more
robust to perturbations: Stronger the perturbation is, more robust the native
state is to the perturbation, i.e., it has less excitations. See Sect.
\ref{Sect_Excitations_NativeState}.

\item The behavior of the specific heat suggests a discontinuous folding
transition; see Sect. \ref{Sect_SpHeat}.

\item The two-dimensional projection of the energy landscape $%
\mathbb{C}
_{2\text{S}}$ is more symmetric than $%
\mathbb{C}
_{2\text{0}}$; see Sect. \ref{Sect_Distance}.

\item The energy landscape for the standard model has energy barriers in the
radial direction for only low-lying microstates; see Sect.
\ref{Sect_Distance_Standard}.

\item The energy landscape may not be relevant for folding in small proteins;
see Sect. \ref{Sect_Distance_Convexity}.

\item The thermodynamic relation $\partial S(E)/\partial E=1/T$ for the
microcanonical entropy $S(E)$ is not valid for small proteins; see Sect.
\ref{Sect_T_Relation}.
\end{enumerate}

\section{Thermodynamic Background}

\subsection{Configurational Approach on a Lattice}

\subsubsection{Configurational Partition Function}

In classical statistical mechanics, the canonical partition function, the
partition function (PF) in the canonical ensemble, factors into two
independent factors: one factor depends only on the kinetic energy, and the
second factor depends only on the interaction energy, provided the
interactions do not depend on particle momenta as happens with magnetic
interactions; see \cite{GujFedor} for a recent discussion of this issue. The
same is true of other ensembles; however, we are only going to consider the
microcanonical and canonical ensembles in this work We will assume here that
factorization occurs. This factorization establishes a very important aspect
of classical statistical mechanics: the free energies corresponding to the two
factors are \emph{additive}. Thus, one can study them separately. Furthermore,
since the contribution from the kinetic energy is independent of the
interactions, it has no bearing on studying energetics. Because of this, one
needs to focus only on the second factor, commonly known as the
\emph{configurational partition function}, and totally disregard the kinetic
energy of the system. This allows us to consider a lattice model where the
focus is on the configurational partition function, since there is no kinetic
energy in a lattice model. On a lattice, therefore, the entropy refers to the
\emph{configurational entropy}. In the context of a single protein
investigation, it is commonly known as the \emph{conformational} entropy. The
volume $V$ of the system is then determined by the number of lattice
$N_{\text{L}}$ sites on the lattice. We will set the \emph{lattice spacing}
$a=1$ in some predetermined unit of the length so that $V=N_{\text{L}}%
a^{3}=N_{\text{L}},$ where $a^{3}$ is the lattice cell volume$.$For general
dimension $d$, we have $V=N_{\text{L}}a^{d}=N_{\text{L}}.$

The absence of kinetic energy does not mean that dynamics cannot be studied on
a lattice. All one needs to do is to introduce some configurational moves to
change one configuration into another. This is quite common in a lattice
investigation of any physical model. However, we are not interested in
studying dynamics in this work.

\subsubsection{Most Probable and Average Energies May Not be Same}

The total number of conformations $W$ of a rooted protein with a given number
$M$ of residues depends only on the lattice geometry, the boundary conditions
imposed on the lattice, and $M$ \cite{note01}. For a small protein, $W$ is
most certainly finite. It also does not depend on the sequence of the residues
\cite{note01}, regardless of the size of the protein, even though $W(E)$ does
depend on the sequence strongly. This is an important observation, as its
implications are not well appreciated. At sufficiently high temperatures, a
protein will explore almost all the conformations, regardless of the model
energetics. It is only at lower temperatures that the energetics allow the
protein to explore only a selected set of conformations $W(\overline{E})$ of a
given \emph{average energy} $\overline{E}$\ that itself depends on the
temperature. It is a well-known fact that the average energy is the energy of
the most probable conformations, and that the average energy is also the
\emph{most probable energy}. If the energetics strongly favors the native
state, such as in the G\={o} model \cite{Go}, then the majority of the
conformations are going to resemble the native conformation(s). Thus, the
number of probed configurations is expected to be smaller in such models,
which will then provide a very efficient way to approach the native state by
reducing the configurational search \cite{Skolnick1}.

\subsubsection{Twists due to the small size}

However, there are two twists. The above reasoning is justified from a
thermodynamic point of view only if the system is macroscopically large as we
have recently pointed out \cite{Guj0412548,GujLambeth}. This is not true of a
protein, which constitutes a small system due to its small size. This point
will be discussed further below. The other twist has to do with the existence
of cooperativity or a first-order folding transition in such models. Not all
energetics and/or sequences will give rise to such a folding transition to an
ordered state.

\subsection{Small System Discreteness and the Thermodynamic Limit}

\subsubsection{Configurational Space discretization}

It should be stressed that the evaluation of the number $W,$ an integer
quantity, requires some sort of \emph{discretization} of the configurational
space. In the absence of any discretization, the entropy in classical
statistical mechanics will always be infinite due to the continuum nature of
the space. It is only when we use quantum statistical mechanics that the
entropy can be properly calculated. However, at present, there is no hope of
studying a single protein using quantum statistical mechanics, and we are
forced to confine ourselves to the classical statistical mechanics. Thus, a
lattice formulation allows us to calculate the entropy, and not only just the
change in the entropy \cite{GujFedor}.

For a lattice model, the configurational energy $E$ is going to be discrete in
that the difference $\Delta E$ between two neighboring energies is going to be
a finite, but non-zero quantity. In addition, for a small protein, $\Delta
e\equiv\Delta E/N_{\text{R}}$ per residue will also remain non-zero; recall
that for a single protein, $N_{\text{R}}=M$. Therefore, the energy spectrum
will be discrete, whether we consider the energy $E$ or the energy
\[
e(N_{\text{R}})\equiv E/N_{\text{R}}%
\]
per residue. It is only in the limit of an infinitely large macroscopic system
($N_{\text{R}}\rightarrow\infty,$ with the understanding that $N_{\text{L}%
}\geq N_{\text{R}}$ so that the proteins can be accommodated on the lattice)
that the energy per residue will give rise to a continuum spectrum
\cite{note0}. In addition, it is in this limit that $e$ also becomes
independent of $N_{\text{R}}$ \cite{note01}; see Fig. \ref{F18} later for
direct evidence for a single protein case.\ As long as we are dealing with a
small protein, we are forced to consider a discrete spectrum of $e(N_{\text{R}%
})$ or $E.$ Consequently, $W(E)$ is a \emph{discrete} function of $E,$ and as
said above, $e(N_{\text{R}})$ continues to depend on $N_{\text{R}}$
\cite{note01} for finite $N_{\text{R}}$.

\subsubsection{Thermodynamic Limit}

To obtain a proper thermodynamic description which is insensitive to the
boundary (i.e., surface) effects, we need to consider a macroscopically large
volume ($N_{\text{L}}\rightarrow\infty$). This limit by itself does not
automatically require the limit $N_{\text{R}}\rightarrow\infty$, as long as
$N_{\text{L}}\geq N_{\text{R}}$. The proper thermodynamics is obtained
formally by taking the \emph{thermodynamic limit, }which requires considering
a macroscopically large volume ($V\rightarrow\infty$), such that the residue
density $c$ (per unit volume) and the energy density $e$ (per residue) are
either \emph{fixed} or reach their respective \emph{limits }that are
independent of $N_{\text{R}}$. At this point, we need to emphasize that a
clear distinction between a single protein (finite $M$) and its bulk
counterpart (which we do not consider in this work) containing many proteins
should be made, as their thermodynamics would be very different. The
thermodynamic limit for the bulk containing a large number of fixed size
proteins, each in a given sequence $\chi,$ requires\ the number of proteins to
increase with the volume to keep the residue density $c$ fixed. In the
simultaneous limit $N_{\text{R}}\rightarrow\infty,V\rightarrow\infty,~$such
that the limiting densities $c\geq0,$ and $e,$ both of which are continuous,
are kept \emph{fixed}, $E$ becomes infinitely large, and one cannot use it or
other extensive quantities (which are also infinitely large) to study
thermodynamics \cite{note0} in this limit; one must consider corresponding
densities, which remain bounded. The standard approach is to consider a
sequence of systems of increasing volume $V_{k}$ constructed so that the
resulting sequence of densities\ $\left\{  c_{k}\right\}  ,\left\{
e_{k}\right\}  $\ converge to their respective limiting densities%
\begin{align*}
\left\{  c_{k}\right\}   &  \rightarrow c,\\
\left\{  e_{k}\right\}   &  \rightarrow e
\end{align*}
in the thermodynamic limit. This approach is equivalent to the following
alternative description commonly employed in thermodynamics. In this approach,
one considers finite extensive quantities such as the configurational energy
$E$ by considering a large but finite size system containing $N_{\text{R}}$
residues in a finite volume $V.$ The configurational energy $E$ of the system
is almost identical to%
\begin{equation}
E=N_{\text{R}}e,\ \ N_{\text{R}}<\infty. \label{Macro_E}%
\end{equation}
Here\ $e$ is the energy per residue in the thermodynamic limit $N_{\text{R}%
}\rightarrow\infty$ as shown above$.$ The accuracy of (\ref{Macro_E})
increases as $N_{\text{R}}$ increases, and ensures that $E$ is in general
bounded ($N_{\text{R}}<\infty$) and can be approximately treated as a
continuous variable since $\Delta E=M\Delta e=0$ \cite{note0}, which follows
from the fact that $e$ is continuous. This is the case, for example, for the
random energy model to be discussed below. However, even in this approach, one
formally needs to take the limit as $N_{\text{R}}\rightarrow\infty$ to
properly treat $e$ as a continuous variable, but is never done in practice as
the system under consideration is finite though large. Since $E$ and other
extensive quantities are now approximately treated as continuous variables,
though they are finite in magnitude, one can carry out thermodynamic
investigation which requires taking derivatives of various (continuous) functions.

\subsubsection{Single Protein as a Small System}

The limit, however, causes a very serious problem when we wish to consider a
single protein, which is characterized by $N_{\text{R}}=M$ and $\chi$. To
maintain a fixed non-zero density $c$, we need to consider the protein size
$M$ to also increase with the volume. Thus, the thermodynamic limit will
require $M$ to diverge simultaneously with the volume of the system. This also
means that the sequence $\chi$\ will also change. If it happens that the
sequence is relevant in determining thermodynamics, then we are dealing with
different proteins as $M$ increases. For example, the energy is usually
determined not only by $M$ but also by the sequence $\chi.$ The sequence
$\chi$ associated with a protein of size $M$ will be different for different
$M$ and also from that of a protein of an infinite size. The way to avoid this
problem is to fix both $M$ and $\chi$ and let the volume diverge \cite{note03}
so that the boundary effects become irrelevant. In this case, $c\rightarrow0$
in the limit, but $E$ remains bounded and discrete. Therefore, in the
following, we will consider our system to consist of a small protein of size
$M$ in a given sequence $\chi$. However, we let $V\rightarrow\infty,$ so that
our system forms a small system in which $E$ remains bounded. \ The same holds
for all other extensive quantities \cite{note3} in the following for our small
system. In the rest of the work, all extensive quantities must be interpreted
in the above sense, even though the volume or the size of the lattice may be
infinite large. Thus, $M\rightarrow\infty$ is never going to be implied in the
following whenever we talk about a small system. This should cause no
confusion. As we will see below, the incompressibility condition allows us to
take the volume infinitely large for any $M$.

From now on, we will only consider a single protein system, unless specified otherwise.

\subsection{Energy Landscape, Conformation Space and "Distance" between
Conformations}

The number $W(E)$ (or $W(E)dE$ for continuous energy spectra) also
characterizes the potential energy landscape for the protein
\cite{Miller,Wales,Sali}, which has become very useful for describing the
equilibrium properties. Each conformation of the protein of energy $E$ is
represented by a point of energy $E$ on the energy landscape. The number of
such points is precisely $W(E)$ (or $W(E)dE$ for the continuum case) and
represents the element of the "hypersurface area" of energy $E.$ The entire
"hypersurface area" of the landscape directly determines the number of
conformations $W$ \cite{Guj0412548}$.$ The native state(s) represents the
global minimum (minima) of the landscape. The projection of the energy
landscape in the direction orthogonal to the energy axis represents the
\emph{conformation space} $%
\mathbb{C}
\ $of the protein. Each point in the conformation space represents a
conformation of the protein, and its energy is given by the height of the
point on the energy landscape directly above it in the direction of the energy
axis. As discussed above, the energy is a discrete variable on a lattice, so
that $W(E)$, and therefore the entropy are also discrete functions
\cite{note0}. For a macroscopic system, one can usually treat both as
continuous. But this is not possible for a small system. Thus, the concept of
the potential energy landscape must be modified in important ways. In
particular, the investigation of the landscape requires knowing the "distance"
between conformations in the conformation space $%
\mathbb{C}
$. While this distance is trivial to define for monomeric systems, this is not
so for a polymeric system due to its connectivity. Thus, one of our tasks
would be to introduce the concept of a "distance" between different
conformations of a protein. In particular, we need to define a "distance" for
all conformations from the native state or from various native states. The
notion of a "distance" allows us to partially understand why a protein in a
given conformation may not fold into its native state when its energetics or
its sequence has been altered due to a disease or some other reasons.

\subsection{Pathways}

To understand the dynamics of protein folding, we follow Anfinsen
\cite{Anfinsen}. According to Anfinsen, proteins get into their native state
following a time-ordered sequence of conformations, now called a "pathway".
The pathway may have a fractal nature \cite{Lidar} and is supposed to dictate
the kinetics of protein folding. Two consecutive conformations $\Gamma$ at
time $t$ and $\Gamma^{\prime}$ at the next time $t+\Delta t$ in the pathway
must differ by some local movements, provided $\Delta t$ is chosen
sufficiently small to allow only for some local movements of the protein.
Thus, the concept of \ a "distance" between two conformations must be such
that a small distance between two conformations is consistent with allowing a
conformation to turn into a "nearby" conformation using only a few local
movements. It is easy to be convinced that because of the connectivity of the
protein, such local movements can most often occur near the ends of the
protein, but not so often in its interior. The movement at an interior point
(away from the ends) would most often require a large portion of the protein
from the interior point to the end to participate in a cooperative movement.
This must require a much longer time duration than the smallest time interval
$\Delta t$\ chosen above. However, some local internal movements such as a
reflection along a diagonal of the square cell, is possible between nearby conformations.

Usually, in the folding problem, one is interested in following the pathway to
the native conformation from a nonnative conformation of much higher energy.
Thus, the entire pathways would correspond to an eventual lowering of the
energy. However, there is no guarantee that $\Gamma^{\prime}$ will always be
of a lower energy than $\Gamma$. There is also no guarantee that
$\Gamma^{\prime}$ will be closer (in distance) to the native state than
$\Gamma$. The only constraint is that $\Gamma$ and $\Gamma^{\prime}$ are close
in distance. It is possible that two conformations are closer in energy but
have much different distances from the native state. Thus, the "distance" and
energy are going to be independent. The pathway most certainly will include
non-native contacts, which disappear as the protein gets into its native
state. It will also depend crucially on various energies in the model, since
the energetics uniquely govern the partitioning of $W$ into a distribution
$W(E)$ of the number of conformations of energy $E$ on the energy landscape$:$%
\begin{equation}
W=%
{\textstyle\sum\limits_{E}}
W(E)\geq1. \label{Partition}%
\end{equation}
Different energetics will usually lead to different pathways. Thus, it is
possible to extract information about energetics from a knowledge of pathways.

A pathway will contain conformations that are not all going to be compact, so
the aqueous interactions will also play an important role in determining the
pathway along with other bonded and non-bonded interactions. As the relative
strengths of various interactions change, so do the partitioning of $W$ in the
distribution $W(E)$: wifferent models will assign different energies to
various conformations with the result that different conformations contribute
to $W(E)$.

\subsection{Random Energy Model of a Macroscopic System and Concavity of its
Entropy\label{REM}}

\subsubsection{Random Energy model}

A common distribution is the Gaussian distribution of the \emph{random energy
model} \cite{Derrida}, which has been extensively employed for proteins (see
\cite{Sali} for example), and which will be discussed later in the work. In
this model, $W(E)$ is given by the following continuous function of the
continuous variable $E$%
\begin{equation}
W(E)=A\exp\left[  -a(E-\widetilde{E})^{2}\right]  , \label{Gaussian_W(E)}%
\end{equation}
where $A$, $a,$ and $\widetilde{E}$ are constants \cite{note02}. In general,
$A$\ depends exponentially and $a$ inversely on the size $M$ of the protein:%
\begin{equation}
\ln A\propto M,\ a\propto1/M. \label{Gaussian_Aa}%
\end{equation}
This ensures that $W(E)$ grows exponentially with $M$. It is easy to envision
situations, however, in which one can obtain non-Gaussian distributions with
unusual properties, not commonly associated with such a distribution. In
particular, some distributions would be completely irrelevant for proteins.
Hopefully, some energetics will allow the model protein to behave like a real
protein. The current investigation is a first step towards identifying such
realistic energetics.

\subsubsection{Entropy Concavity}

The configurational entropy in the random energy model, following the
Boltzmann relation%

\begin{equation}
S(E)\equiv\ln W(E), \label{B_S}%
\end{equation}
is given by
\begin{equation}
S(E)=\ln A-a(E-\widetilde{E})^{2}; \label{Gaussian_S}%
\end{equation}
see (\ref{Gaussian_W(E)}); both terms in (\ref{Gaussian_S}) are extensive. The
form of this entropy is an inverted parabola so that it is \emph{concave}
\cite{note1}. Mathematically, this requires%
\begin{equation}
\partial^{2}S/\partial E^{2}\leq0 \label{Concave_S}%
\end{equation}
for a macroscopic system to ensure its thermodynamic stability. Observe that
$\widetilde{E}$\ is where the entropy has its maximum. It should be noted that
the number of states $W(E)$ in (\ref{Gaussian_W(E)}) vanishes at the extremes
of the allowed energies \cite{note02}. In these neighborhoods, $S(E)$ becomes
negative. To avoid a negative $S(E),$ one uses the above form over the range
\[
(\widetilde{E}-\alpha,\widetilde{E}+\alpha),\alpha\equiv\sqrt{\ln A/a},
\]
where $\alpha$ is extensive so that $S(E)$ is non-negative over this range,
and supplements it by $S(E)=0$ outside this range. In the following, we will
only focus on the low energy range.

\subsubsection{Energy Gap}

The supplementary function $S(E)=0$ requires making the \emph{assumption} that
the lowest allowed energy $E_{0}$ in the energy spectrum is below the lower
end of the above range:%
\[
E_{0}<E_{\text{G}}\equiv\widetilde{E}-\alpha.
\]
This assumption gives rise to an \emph{energy gap} between $E_{0}$ and
$E_{\text{G}}$, the width of the gap itself being \emph{extensive}. The
presence of the energy gap makes the modified entropy function \emph{convex}
in the region about $E_{\text{G}}$. It is this modified form of the random
energy model that has been extensively used in studying protein folding; the
resulting concavity violation around $E_{\text{G}}$ is interpreted as a
folding transition, as we will show below. The folding temperature
$T_{\text{F}}$ is given by the inverse of the slope of the tangent drawn from
$E_{0}$ so that it touches the entropy function (\ref{Gaussian_S}); see
\cite{Sali} for example. The modified Gaussian model also shows that the
energy gap above the ground state is crucial for foldability. It should be
noted, however, that there are idealized physical models, such as the KDP
model, that freeze into the ground state through a first-order transition at a
finite non-zero temperature\ \cite{KDP,Nagle}, something similar to the
protein folding.

It is well known that the energy gap in the KDP model is extensive in size
just as in the random energy model. It is this extensive size of the gap that
makes the macroscopic entropy non-concave in the neighborhood of the gap in
the random energy model.

The temperature at which $S(E)$ vanishes represents the ideal glass transition
temperature $T_{\text{G}}.~$The ideal glass is a frozen state of zero entropy
and exists below this temperature and has a constant energy $E_{\text{G}%
}>E_{0}$ and zero specific heat.

\subsubsection{Equality of $S(\overline{E})$ and $S(T)$}

The Gaussian form (\ref{Gaussian_S}) of the entropy has been used to suggest
the following form of the average energy $\overline{E}$ \cite{Clementi1}:%
\begin{equation}
\overline{E}=\widetilde{E}-1/2aT \label{Gaussian_E}%
\end{equation}
above the folding temperature. As we will see later, this form of the energy
can be justified for a macroscopic system. This form cannot apply near
absolute zero where it becomes unbounded, and the problem is avoided by a
folding transition. As a sharp folding transition cannot occur in small
proteins, it is also desirable to understand the limitation of the above
energy form for small proteins.

The Helmholtz free energy $F(T)$ is obtained by evaluating $F(T)\equiv
\overline{E}-TS(\overline{E}),$ and is given by%
\begin{equation}
F(T)=\widetilde{E}-T\ln A-1/4aT, \label{Gaussian_F}%
\end{equation}
from which $S(T)$ can be obtained directly, see below (\ref{Can_S}):%
\begin{equation}
S(T)=\ln A-1/4aT^{2}, \label{Gaussian_S_T}%
\end{equation}
so that
\begin{equation}
S(\overline{E})=S(T), \label{S_EQ}%
\end{equation}
see (\ref{Gaussian_S}), as said above. The ideal glass temperature is given
by
\[
T_{\text{G}}=1/2a\alpha.
\]

This equality is only valid for a macroscopic system and, as shown recently
\cite{GujLambeth} and will also be discussed further in this work, does not
hold for small systems such as a finite protein that is of our interest here.
Their equality, however, is crucial as direct experimental approaches (such as
crystallography or NMR techniques) are used to provide information about the
typical conformations associated with the average or most probable energy.
Thus, it is also important to know if the two concepts of entropy are
equivalent for small proteins. If not true, the interpretation of experimental
data for the energetics would be incorrect. This will become a limitation of
any direct experimental technique in determining the energetics and its
association with conformations.

\subsubsection{Limitations of the Model}

The random energy model can be justified for a macroscopic system by appealing
to the central limit theorem and assuming that various energies are random
variables. Accordingly, this model is not applicable to small proteins.
Therefore, it is far from obvious how relevant the random energy model is for
small proteins. Moreover, there are other limitations of the model in addition
to those noted in \cite{note02}. One of the problems with the random energy
model becomes evident from its free energy (\ref{Gaussian_F}), which does not
reduce to $E_{0}$ at absolute zero as required by thermodynamics. Note that
the free energy continues to satisfy the condition of stability everywhere%
\[
\partial^{2}F/\partial T^{2}<0,
\]
which follows from the non-negativity of the specific heat. Therefore, the
above thermodynamic violation is not a consequence of any thermodynamic
instability. The violation has to do with its unphysical entropy in
(\ref{Gaussian_S_T}), which does not satisfy the thermodynamic requirement
$TS(T)\rightarrow0$ as $T\rightarrow0$ \cite{GujGlass}$.$ To avoid the above
violation, a first-order folding transition is invoked at $T=T_{\text{F}}$
given by
\[
F(T_{\text{F}})=E_{0}.
\]
Above $T_{\text{F}},$ one uses the free energy (\ref{Gaussian_F}), and below
$T_{\text{F}}$ one uses $F(T)=E_{0}.$ The folding transition is in reality a
freezing transition in that the low-temperature phase is a frozen state of
zero specific heat, similar to the ideal glass, except that the ideal glass
has a much higher energy $E_{\text{G}}$ due to the energy gap discussed above.
It should be clear that $E_{\text{F}}=\overline{E}(T_{\text{F}})>E_{\text{G}%
}.$ However, it should at the same time be stressed that the energy gap is not
present in the random energy model, but has been put in "by hand" to avoid a
negative $S(E).$ This energy gap then makes the entropy $S(E)$
\emph{non-concave}, which is then responsible for the first-order folding
transition. If there were no energy gap, i.e. if $E_{0}\geq E_{\text{G}},$
then there would be no loss of concavity. In that case, there would be no
folding transition. However, the condition $E_{0}\geq E_{\text{G}}$ would make
the model quite unphysical as no equilibrium state would exist in the model
below a non-zero temperature at which $\overline{E}=E_{0},$ but the entropy is
not zero.

It should be noted that the\ random energy model itself does not specify the
value of $E_{0}.$ Indeed, (\ref{Gaussian_W(E)}) is valid for all $E\geq
-\infty.$ This suggests that $E_{0}\rightarrow-\infty$ . If this is accepted,
then the tangent construction to locate the folding temperature will give
$T_{\text{F}}\rightarrow\infty.$ This is not meaningful. For a meaningful
discussion, we need the following conjecture.

\begin{conjecture}
We need to treat $E_{0}$ as finite.
\end{conjecture}

This should not come as a surprise. Indeed, it follows from our earlier
discussion of the energy in (\ref{Macro_E}). We need to apply the random
energy model to a finite but large system so that $E_{0}$ can be treated as finite.

At the same time, a physical requirement for $W(E)$ is that for allowed
energies, $W(E)\geq1.$ If this is taken literally \cite{note02}, then
(\ref{Gaussian_W(E)}) must be restricted to the energies in the range
$(\widetilde{E}-\alpha,\widetilde{E}+\alpha),$ so that the lowest allowed
energy is $E_{0}=E_{\text{G}}.$ In this case, there will not be any energy
gap, and no loss of concavity. This is usually not the interpretation adopted
in the literature. Invariably, one adopts the conventional choice
$E_{0}<E_{\text{G}},$ the actual value of $E_{0}$\ itself being irrelevant, as
long as it is taken to be finite. But this is merely a convention, which then
justifies the folding transition in the model.

It should also be noted that an energy gap is not the only mechanism by which
a first-order transition and an ideal glass transition can occur. Both can
occur without an energy gap as we will discuss below. Here it is sufficient to
note that all one needs is a lack of concavity in the entropy for a folding transition.

\subsection{Small System Microcanonical and Canonical Entropies}

\subsubsection{Microcanonical Entropy and Energy Landscape}

The \emph{microcanonical} entropy is given by the Boltzmann relation
(\ref{B_S}), and has played a very important role in our attempts to
understand the way folding occurs into compact native states along a very
large number of microscopic pathways that connect a native state to myriad
unfolded conformations. This entropy definition is useful when the system
(such as a protein) is\ forms an \emph{isolated }system so that its energy
remains fixed, along with $N_{\text{R}},$ and $V$. The system occupies each of
the various conformations $\Gamma\in\boldsymbol{\Gamma}(E)$, all of energy
$E$, with equal probability
\begin{equation}
p(\Gamma)\equiv1/W(E). \label{ME_probability}%
\end{equation}
Here, $\boldsymbol{\Gamma}(E)$ represents the set of conformations, each of
energy $E$ (for given $N_{\text{R}},$ and $V,$ which we do not show below for
simplicity$),$ and contains $W(E)$ distinct conformations. The corresponding
ensemble containing these conformations is called the \emph{microcanonical
ensemble} (ME).

\begin{conjecture}
\label{Conj_3}The ME entropy via (\ref{B_S}) can most certainly be defined
even for a small system such as a protein.
\end{conjecture}

This makes the Boltzmann entropy (\ref{B_S}) a very useful quantity to study
for proteins. There is an additional significance of this entropy or of the
number $W(E),$ as noted earlier. The number $W(E)$ also characterizes the
potential energy landscape for the protein \cite{Miller,Wales,Sali}.

It is a well-established tenant of macroscopic thermodynamics that in the
physically relevant range of the energy $W(E)$\ decreases with falling energy
$E$ so that
\begin{equation}
\partial S/\partial E\geq0; \label{S_E_Relation}%
\end{equation}
consequently, the energy landscape for a macroscopic system in the physically
relevant range of the energy is expected to possess a structure that narrows
down with falling energy. An example of such a landscape could be a funnel
such as the surface of an inverted hyper-cone (a cone in a high-dimension
space). The hypersurface area of such a cone at height $E-E_{0}$ in a $p$
dimensional space\ is proportional to $\left(  E-E_{0}\right)  ^{p-2}$, which
satisfies the property (\ref{S_E_Relation}). Whether this property is also a
characteristic of a landscape associated with a small system remains to be
investigated. This is one of the aims of this work. It should be noted that
the "energy landscape" for a lattice model will be discrete and not a
continuous hypersurface \cite{note0}.

\begin{remark}
Property (\ref{S_E_Relation}) should be interpreted not as a differential
property, but merely implying that $S(E)$ decreases with $E$ for the discrete case.
\end{remark}

In the following, all differential relations will have such an interpretation
for the discrete case, if applicable.

It is known that the entire thermodynamics is contained in $S(E)$, which is
supposed to \ be \emph{concave} \cite{note1} for a macroscopic system. Its
violation is a signature of a phase transition in the model. Whether this
\emph{concavity} is also a characteristic of a small system ME\ entropy
remains to be investigated.

In view of the above discussion, it is important, therefore, to investigate
the form of $S(E)$ and the effects of energetics on it for small proteins,
which to the best of our knowledge has not been studied fully.

\subsubsection{Canonical Entropy}

The direct experimental approaches (primarily, crystallography) used to
determine energetics in proteins at a given temperature $T$ provide
information about the conformations associated with the average energy
$\overline{E}$ at $T$. In this work, $T$ is always going to represent the
temperature in the units of the Boltzmann constant. The protein is no longer
isolated, but interacts with its environment at a given temperature $T$ so
that the energy can be exchanged but $N_{\text{R}},$ and $V$ still remain
fixed$.$\ The system now requires the canonical ensemble (CE) for its
thermodynamic description. Thus, one needs to know the dependence of the
\emph{canonical} entropy $S(T)$ on the average energy $\overline{E}$ at a
given temperature $T$. This entropy is given by the Gibbsian relation
\begin{equation}
S(T)=-%
{\displaystyle\sum}
p(\Gamma)\ln p(\Gamma), \label{G_S}%
\end{equation}
where $p(\Gamma)$\ is the \emph{probability} to be in the conformation
$\Gamma$ and is controlled by the energetics and the temperature $T$ of the
system$;$ we have suppressed the latter dependence in $p(\Gamma)$ for
notational simplicity$.$ It is also equivalent to the conventional entropy in
the canonical ensemble given by
\begin{equation}
S(T)\equiv-\partial F(T)/\partial T, \label{Can_S}%
\end{equation}
as we will show later; here $F(T)$ is the Helmholtz free energy, the
thermodynamic potential in the canonical ensemble.

It is important to appreciate the significance of the form of the Gibbsian
definition (\ref{G_S}). It can also be applied to the equilibrium
microcanonical ensemble. In this case, $p(\Gamma)$ is independent of $T$, and
is given by (\ref{ME_probability}). It is easily seen that the Gibbsian
entropy, applied to ME, is exactly the same as the Boltzmann entropy
(\ref{B_S}). This is true regardless of the size of the system. Thus, we will
take the Gibbsian definition (\ref{G_S}) to be applicable for systems of any size.

For a macroscopic system, $S(T)$ given by the Gibbs formulation is
\emph{identical} to the Boltzmann entropy $S(\overline{E})$ at the average or
the most probable energy $\overline{E}$ at the temperature $T$; see
(\ref{S_EQ}) in the random energy model for an example. The general equality
(\ref{S_EQ}) allows us to relate the energetics with configurational
properties: the canonical entropy at $T$ provides information about the
conformations of average energy $\overline{E}.$

\begin{itemize}
\item \textbf{Warning}: There should be no confusion in distinguishing $S(T)$
and $S(E)$, as their arguments will always be exhibited. This is important to
note as we will show that the two quantities are very different for small systems.
\end{itemize}

\section{Model}%

\begin{figure}
[ptb]
\begin{center}
\includegraphics[
trim=1.713398in 2.741251in 2.402668in 3.436141in,
height=3.3313in,
width=3.0113in
]%
{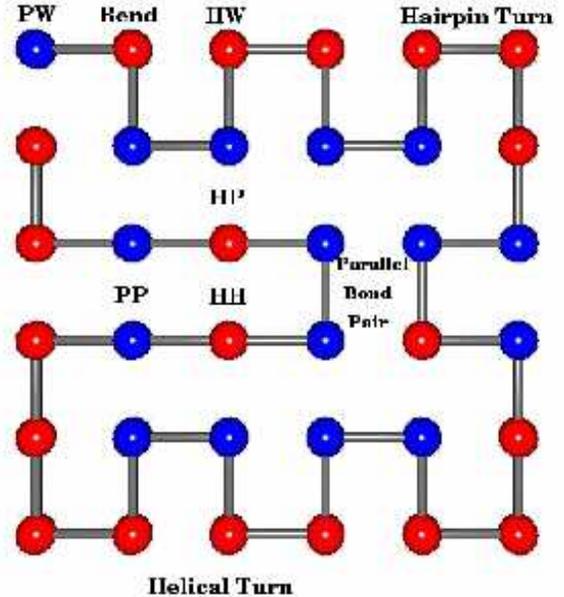}%
\caption{A 2-d model of a finite protein on a square lattice. The red spheres
represent hydrophobic sites and the blue spheres represent hydrophilic sites.}%
\label{F1}%
\end{center}
\end{figure}

\subsection{Rooted or Anchored Protein}

A proper model for protein folding will require using semiflexibility of the
protein, for which we will use a recent model developed in our group
\cite{GujSemiflex}. It is the semiflexibility which gives rise to a
crystalline phase; the latter represents the ordered native state of the
protein at low temperatures. Therefore, we will treat a protein as a
semiflexible \emph{self-avoiding copolymer chain} on a lattice to study its
folding by properly extending the above model \cite{GujSemiflex}. The lattice
is taken to be infinitely large ($N_{\text{L}}\rightarrow\infty$) so that the
protein will never feel the effects of its boundary. Each amino acid residue
(including any side group) is represented by a tiny sphere, which must lie on
a lattice site; see Fig. \ref{F1}. Each solvent also occupies a lattice site.
We will consider an \emph{incompressible model} so that no voids are allowed.
A site is either occupied by a residue or by a solvent. The self-avoidance
condition means that a lattice site \emph{cannot} be occupied by more than one
residue or a solvent. We consider a two-state model \cite{Dill,Dill1999} in
which each amino acid is classified either as a \emph{hydrophobic} site (red
spheres in Fig. \ref{F1} and denoted by H) or a \emph{hydrophilic/polar }site
(blue spheres in Fig. \ref{F1} and denoted by P). Due to the chemical
structure of an amino acid, a protein is directional. One end of the protein
has a free carboxyl group and is known as the C-terminus or carboxyl terminus.
The other end of the protein has a free amino group and is known as the
N-terminus or amino terminus. Proteins are always biosynthesized from the
N-terminus to the C-terminus. On the other hand, most chemically synthesized
proteins grow from the C-terminus to the N-terminus. Thus, a proper model
should account for this directionality. Accordingly, in this work, we will
incorporate the directionality of the protein, and treat both ends as
dissimilar. This condition can always be relaxed without much complication.
Treating both ends dissimilar basically doubles the number of distinct
conformations of the protein, without any useful implication for the way the
entropy behaves.

\subsubsection{Compact and Unrestricted Protein Conformations}

In our enumeration, we only consider a square lattice in this work. We will
consider a protein to have either no restriction on its allowed conformations,
or restrict it to only take a compact form, which we take to be rectangular.
In the former case, the protein will be allowed to take all shapes including
compact shapes by having it probe all allowed sites on an infinite lattice. In
the second case, the protein will be restricted to have only compact shapes so
that there are no solvent molecules in its interior; the surrounding of a
compact region will be occupied by the solvent, i.e., water. The compact
conformations are also present in the former unrestricted case. We will say
that the conformations are unrestricted in the former case and compact in the
latter case. In both cases, the end of the protein is \emph{rooted} and is not
allowed to move. There is a simple reason for\emph{ rooting} or
\emph{anchoring} the protein. The process of folding in vivo often begins
co-translationally, so that the N-terminus of the protein begins to fold while
the C-terminal portion of the protein is still being synthesized by the
ribosome. Thus, it is the C-terminus that we root or anchor at the origin, and
allow the N-terminus to be free to begin folding.

To generate compact rectangular shapes, we allow all possible rectangular
shapes that could accommodate a given protein of size $M$. We give an example
to clarify this point. Consider $M=24.$ For this case, we consider the
following rectangular shapes in two dimensions:\ $1\times24,2\times
12,3\times8,$ and $4\times6$. We do not need to separately consider
$24\times1,12\times2,8\times3$, and $6\times4$ because of the rotational symmetry.

The anchoring has three important consequences for our computation. In the
first place, this reduces the number of conformations that need to be counted.
On an infinite lattice, an unanchored protein can start from any of the
infinite lattice sites, making $W$ infinitely large. This trivial infinity due
to nonanchoring has no bearing on thermodynamics. In the second place,
anchoring allows us to uniquely define the \emph{distance} between two
conformations as we will discuss below. From now on, we will always root our
protein at one of its ends on the lattice. In addition, we will also restrict
the protein conformations so that its first bond from the root is along a
fixed direction, which we take to be to the right, to limit the number of
conformations. In order to further reduce the number of distinct
conformations, we also restrict the first bend, as we start from the root, to
be in the down direction of the square lattice. It is easily seen that any
other conformation of the protein is related to one of the generated
conformations by some trivial rotation. The last consequence of rooting is the
following. There will be no doubling of conformations due to directionality
that was discussed above.%

\begin{figure}
[ptb]
\begin{center}
\includegraphics[
trim=0.535284in 3.001967in 1.339432in 3.007288in,
height=2.6212in,
width=3.5397in
]%
{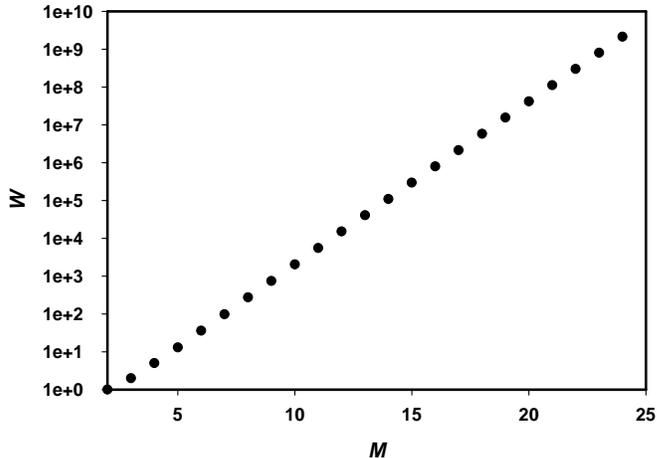}%
\caption{The rapid growth of $W$ (shown in the common log scale) with $M$ for an unrestricted protein on an
infinite lattice. The allowed conformations are grown as described in the
text.}%
\label{F3}%
\end{center}
\end{figure}

The number of conformations $W$ for rooted proteins increases rapidly with the
protein size, as is seen in Fig. \ref{F3}. The number of conformations $W$ for
rooted proteins increases rapidly with the protein size, as is seen in
Fig.\ref{F3} below. The growth of $W$ for the rooted protein with its first
bond in a specified direction on an infinite lattice can be fitted by
\[
W=0.102272\exp(0.990933M),
\]
with $R^{2}=0.999876$ \cite{note2}. Correspondingly, the time required to
generate all the conformations $W$ (but no other computation such as their
energies, distances, etc.) also increases rapidly with the size $M$ as the
following Table \ref{Table_1} shows. The time reported here is on a PC. The
time obviously increases if other computations are also carried out.\ 

\begin{center}%
\begin{align*}
&  \text{Table \ref{Table_1}}-\text{Size and Computation Time on a PC}\\
&  \ \ \ \ \ \ \ \ \ \ \ \ \ \ \ \ \ \ \ \ \ \ \ \ \ \
\begin{tabular}
[c]{|c|c|c|}\hline
$M$ & Finite & Infinite\\\hline
16 & 1 s & 10 s\\\hline
18 & 1 s & 2 min\\\hline
20 & 1 s & 1 hour\\\hline
24 & 1 s & 3 days\\\hline
26 & 1 s & 5 days\\\hline
36 & 10 s & -\\\hline
49 & 45 min & -\\\hline
64 & 5 weeks & -\\\hline
\end{tabular}
\end{align*}
\label{Table_1} \hspace{0in}
\end{center}

\subsection{Microscopic Interaction Energies}

To account for the presence of water surrounding the protein, water molecules
(to be denoted by W) are also allowed in the model. Each water molecule
occupies a site of the lattice. To incorporate compressibility, voids can also
be incorporated in the model. In that case, each void will be allowed to
occupy a site of the lattice. We now turn to the complications induced by the compressibility.

\subsubsection{Simplification Resulting from Incompressibility}

Each conformation of the protein on the lattice results in certain sites of
the lattice being occupied by the protein. In the incompressible model, rest
of the sites will be occupied by the solvent. Thus, each conformation of the
protein is associated with only one possible distribution of the solvent
molecules on the lattice. Accordingly, there exists one and only one
\emph{microstate} of the system (the lattice containing the rooted protein)
for each conformation of the protein. In other words, the number of possible
microstates of the entire system is the total number of conformations $W$ of
the rooted protein. It should be stressed that for sufficiently large volume
$V$ or $N_{\text{L}}$ compared to $M$, the number of conformations $W$ will
depend only on $M$ but not on $V$ or $N_{\text{L}}.$ This is a major
simplification. The Gibbsian definition (\ref{G_S}) of the entropy of the
system refers to the sum over the microstates of the system. This means that
the sum in (\ref{G_S}) for the system is nothing but the sum over the
conformations belonging to $W$.

This simplification is lost if we consider a compressible model containing
voids. Then, there will be many more possible distributions of the solvent for
each conformation of the protein. Let $k$ denote one of the microstates of the
system, and $k(\Gamma)$\ the set of microstates that are associated with a
conformation $\Gamma$\ of the protein. The set $k(\Gamma)$\ depends not only
on $M$ as above, but now it also depends on $N_{\text{L}}$ and $N_{0},$the
number of voids, even if $N_{\text{L}}$ is sufficiently large. This is very
different from the situation above for the incompressible limit. The entropy
of the system is now given by the Gibbsian definition
\begin{equation}
S(T)=-%
{\displaystyle\sum}
p_{k}\ln p_{k}, \label{G_S_gen}%
\end{equation}
where $p_{k}$ is the probability of the $k$th microstate. This entropy can be
reexpressed as follows:%
\[
S(T)=-%
{\displaystyle\sum_{\Gamma}}
\sum_{k\in k(\Gamma)}p_{k}\ln p_{k}.
\]
\ The number of microstates of the system which determine the sum in the
Gibbsian definition (\ref{G_S_gen}) will far exceed the sum $W$ of protein
conformations. This will make the computation much more extensive, depending
on the amount of free volume (i.e. of the voids): larger the free volume, more
extensive the computation. Because of this complication, we only deal with the
incompressible model in this work.

\subsubsection{Equal Size Approximation for Residues and Solvent}

We do not allow voids in the present work, and take the solvent (water)
molecule and the residue each to occupy a lattice site. This is an
approximation as the water molecule and the residue do not have the same size.
In a more realistic model, the water molecule and a residue may be allowed to
occupy more than one lattice sites, depending on their relative size. While we
can incorporate size difference in our lattice model, it makes the calculation
harder. To avoid this, we adopt the simplification of equal size in this work.

\subsubsection{Interaction Energies}

The \emph{excluded-volume effects} are accounted by enforcing that a lattice
site cannot be occupied by more than one residue or water molecule. The
interaction energies are restricted between chemically unbonded particles
(residues H and P, and water molecules W) that are nearest neighbors of each
other. Long range interactions are neglected, but can be incorporated later if
so desired. We will not do that here. There are three species of particles
(H,P, and W) in our model. As shown elsewhere \cite{Guj2003}, we need to only
consider three independent energies of interaction between three chemically
unbonded pairs of species. We have decided to use the following three van der
Waals energies $e_{\text{HH}},$ $e_{\text{HW}},$ and $e_{\text{PH}}$ between
the three unbonded pairs HH, HW, and PH. In the standard model due to Lau and
Dill, only the first one in non-zero, as shown in Table \ref{Table_2}. To
account for the semiflexibility of the protein, we use the model recently
developed by us to study crystallization and glass transition in polymers
\cite{GujSemiflex}, but extend it to include preference of helical formation.
The original model has a penalty $e_{\text{b}}>0$ for making a bend, an
attractive energy $e_{\text{P}}<0$ between two parallel protein bonds, an
attractive energy $e_{\text{hp}}<0$ for a hairpin turn (on top of the penalty
for two consecutive bends in the same circulation direction), and an
attractive energy $e_{\text{hl}}<0$ for a helical turn (on top of the energy
for four bends and two hairpin turns).

We consider a protein with $M$ residues in a given sequence $\chi$ of H and P
associated with the residues on a square lattice, with one of its end fixed at
the origin so that the total number of conformations $W$ for a small protein
remains finite even on an infinite lattice. We only consider the case in which
the number of H and P are equal. This can be considered as the condition of
charge neutrality. We generalize a recent model \cite{GujSemiflex}, in which
the number of bends $N_{\text{b}},$ pairs of parallel bonds $N_{\text{p}},$
and hairpin turns $N_{\text{hp}}$\ characterize the semiflexibility; see
Fig.\ref{F1}, where we show a protein in its compact form so that all the
solvent molecules (W) such as water are expelled from the inside and surround
the protein. The dark spheres denote hydrophobic residues (H) and light
spheres denote hydrophilic (i.e., polar) residues (P). The nearest-neighbor
distinct pairs PP, HH, HP, PW and HW between the residues and the water are
also shown, but not the contact WW. Only three out of these six contacts are
independent on the lattice \cite{Guj2003}, which we take to be HH, HW, and HP
pairs. A bend is where the protein deviates from its collinear path. Each
hairpin turn requires two consecutive bends in the same direction (clockwise
or counterclockwise); see Fig. \ref{F1}. Two parallel bonds form a pair when
they are one lattice spacing apart. We also use the number of helical turns
$N_{\text{hl}}.$ On a square lattice, a "helical turn" is interpreted as two
consecutive hairpin turns in opposite directions as shown in Fig. \ref{F1}.
The corresponding energies are $e_{\text{b}},$ $e_{\text{P}},$ $e_{\text{hp}}%
$, and $e_{\text{hl}},$ respectively$.$ The interaction energies are
$e_{\text{HH}}=-1,$\ $e_{\text{HW}},$ and $e_{\text{HP}},$\ corresponding to
the HH, HW, and HP, respectively. The number of these pairs are $N_{\text{HH}%
},$ $N_{\text{HW}},$ and $N_{\text{HP}},$ respectively. We let $\mathbf{e}%
^{\prime}$denote the set containing all $\left\{  e_{i}\right\}  ,$ except
$e_{\text{HH}}=-1,$ and $\mathbf{e}$ the entire set $\{e_{i}\},$where $i$
stands for b,p,hp,hl,HH,HW, and HP$.$ Thus, $\mathbf{e,e}^{\prime}$ represent
the sets
\begin{align*}
\mathbf{e}  &  \equiv\left\{  e_{\text{b}},e_{\text{P}},e_{\text{hp}%
},e_{\text{hl}},e_{\text{HH}},e_{\text{HW}},e_{\text{PH}}\right\}  ,\\
\mathbf{e}^{\prime}  &  \equiv\left\{  e_{\text{b}},e_{\text{P}},e_{\text{hp}%
},e_{\text{hl}},e_{\text{HW}},e_{\text{PH}}\right\}  .
\end{align*}
Similarly, $\mathbf{N}\equiv\mathbf{N}(\Gamma)\equiv\{N_{i}(\Gamma)\}$ denotes
the set
\[
\mathbf{N}\equiv\left\{  N_{\text{b}},N_{\text{P}},N_{\text{hp}},N_{\text{hl}%
},N_{\text{HH}},N_{\text{HW}},N_{\text{PH}}\right\}  ,
\]
and $\mathbf{N}^{\prime}$ denotes all $\left\{  N_{i}\right\}  ,$ except
$N_{\text{HH}}:$%
\[
\mathbf{N}^{\prime}\equiv\left\{  N_{\text{b}},N_{\text{P}},N_{\text{hp}%
},N_{\text{hl}},N_{\text{HW}},N_{\text{PH}}\right\}  .
\]
Let $W(\mathbf{N})$ denote the number of protein configurations on a lattice
of size $N_{\text{L}}\geq M$. The energy of the configuration $\Gamma$
corresponding to the set $\mathbf{N}$ is given by%
\begin{equation}
E(\mathbf{N})=\mathbf{e\cdot N=}\sum_{i}e_{i}N_{i}. \label{Energy0}%
\end{equation}
The energy varies from configuration to configuration as it depends on
$\mathbf{N}$. But it does not depend on thermodynamic state parameters such as
the temperature, pressure, etc.

The dimensionless entropy function corresponding to configurations with a
given $\mathbf{N}$ is defined as%
\begin{equation}
S(\mathbf{N})\equiv\ln W(\mathbf{N}). \label{EntropyCof}%
\end{equation}
(This definition amounts to setting the Boltzmann constant equal to 1.) There
will in general be many sets $\mathbf{N}$\ that will result in the same energy
$E.$ We denote the collection of these sets by $\mathbf{N}(E).$\ Thus, the
number of configurations $W(E)$ for a given $E$ is obtained by summing
$W(\mathbf{N})$ over this collection $\mathbf{N}(E):\label{Table_2}$%
\begin{equation}
W(E)=\sum_{\mathbf{N}\in\mathbf{N}(E)}W(\mathbf{N}). \label{StateNumb_E}%
\end{equation}
The corresponding entropy function for a given $E$ is given, as usual, by
(\ref{B_S}). The total number of all protein configurations, regardless of the
energy $E$, is given by (\ref{Partition}).

\subsection{Various Model Energetics Choices}

The three choices we have most often made for energies are described below in
the form of three different models, the parameters for which are shown in
Table \ref{Table_2}.

\begin{center}%
\begin{align*}
&  \text{Table \ref{Table_2}}-\text{Possible Models and their parameters}\\
\ \ \  &
\begin{tabular}
[c]{|l|l|l|l|}\hline
& \emph{Standard }(A) & \emph{Weakly\ }(B$_{1}$) & \emph{Strongly \ }(C$_{1}%
$)\\\hline
Bend & $\ \ \ \ \ \ \ \ 0$ & $\ \ \ \ \ \ 1/50$ & $\ \ \ \ \ \ \ 1/3$\\\hline
Parallel & $\ \ \ \ \ \ \ \ 0$ & $\ \ \ \ -1/50$ & $\ \ \ \ \ -1/3$\\\hline
Hairpin & $\ \ \ \ \ \ \ \ 0$ & $\ \ \ \ -2/50$ & $\ \ \ \ \ -1/3$\\\hline
Helix & $\ \ \ \ \ \ \ \ 0$ & $\ \ \ \ -2/50$ & $\ \ \ \ \ -1/3$\\\hline
HH & $\ \ \ \ \ \ -1$ & $\ \ \ -50/50$ & $\ \ \ \ \ -3/3$\\\hline
HW & $\ \ \ \ \ \ \ \ 0$ & $\ \ \ \ \ 20/50$ & $\ \ \ \ \ \ \ 2/3$\\\hline
PH & $\ \ \ \ \ \ \ \ 0$ & $\ \ \ \ \ \ 5/50$ & $\ \ \ \ \ \ \ 1/3$\\\hline
\end{tabular}
\end{align*}

\end{center}

\subsubsection{Model (A)}

In the standard model, the set $\mathbf{N}$ contains only one quantity, the
HH\ contact number $N_{\text{HH}}.$ Thus, $\mathbf{e}^{\prime}=0,$ and the
adimensional energy in this model is simply given by $E=N_{\text{HH}}.$ As
$N_{\text{HH}}$ is going to be an integer, the corresponding density
\[
n_{\text{HH}}\equiv N_{\text{HH}}/M
\]
is going to be a discrete quantity, so will be the adimensional energy density
$e\equiv E/M=n_{\text{HH}}$. The number of conformations $W(N_{\text{HH}})$ of
a given $N_{\text{HH}}$ is%
\begin{equation}
W(N_{\text{HH}})\equiv%
{\textstyle\sum}
W(N_{\text{HH}},\mathbf{N}^{\prime}). \label{W_HH}%
\end{equation}
In the standard model, $E=N_{\text{HH}}$. It is clear from (\ref{W_HH}) that
the entropy $S(N_{\text{HH}})=\ln W(N_{\text{HH}})$ for a given $N_{\text{HH}%
},$ regardless of $\mathbf{N}^{\prime}$,\ is maximum in the standard model
\cite{Gujrati1,Guj0412548}. This feature of the standard model entropy is a
possible justification of the observation made in \cite{Kolinski}. As a
consequence, the protein with a given $N_{\text{HH}}$ will probe many more
states in the standard model than in any other model, which then slows down
its approach to the native state. Thus, it is important to have non-zero
$\mathbf{e}^{\prime}$ to step up the approach to the native state. (It is
highly likely that the native states in different models are different, but
this does not affect the above conclusion, provided the native states are
unique.) There is another important consequences of having the remaining
$\varepsilon_{i}=0.$ The fluctuations in the corresponding $N_{i}$ are maximum
as there is no penalty no matter what $\mathbf{N}^{\prime}$ is. Hence, the
protein will spend a lot of time probing a large number of conformations so as
to maximize fluctuations in $\mathbf{N}^{\prime}.$\ This also suggests that we
need to go beyond the standard model to describe proteins that fold fast.
Correspondingly, the entropy per residue is also discrete, with two successive
values differing in the argument by $1/M.$ In other words, for small proteins,
the entropy per residue $s(e)$ is not a continuous function, but a set of
discrete values, as shown in Figs.\ref{F18} and \ref{F19}. It is clear from
the figure that one can easily draw a concave envelop for the discrete values
of $s(e).$ However, one can also draw a variety of other envelop functions
that would not necessarily be concave such as those shown by the lines joining
these points in the figures.

\subsubsection{Weakly Perturbed Model (B$_{1}$,B$_{2}$)}

In this model, we allow for other energies to be non-zero, but still small
compared in strength. The model with the parameters in the above table will be
called B$_{1}$ in the following. Another common choice we have made is
$\mathbf{e}^{\prime}=(3/56,-1/56,-3/56,-3/56,21/56,5/56),$ and the
corresponding model will be called B$_{2}$ in the following. The two models
collectively will be simply denoted by B. The numerator of various energies
are integers and are used to determine the energy $E$ as an integer, which
makes it easy to classify energy levels in groups of a given energy. The
energy is divided by the denominator at the end to ensure that $e_{\text{HH}%
}=-1.$ The energy corresponding to a HW-contact is the only energy close to
$\left\vert e_{\text{HH}}\right\vert ;$ this is to account for the strong
repulsion between H and W. Otherwise, all other energies are extremely small
compared to $\left\vert e_{\text{HH}}\right\vert .$ Consequently, this model
will be identified as a model with weak perturbation on the standard model.

The model B$_{2}$\ can also be treated as a model with small perturbations on
the model B$_{1}$ (or vice versa) in which each residue is allowed to move
about within the small cell surrounding the lattice site on which it is
located. Such a disturbance will usually cause a small perturbation of B$_{1}$
(or vice versa) and can be described by the model B.

\subsubsection{Strongly Perturbed Model (C$_{1}$,C$_{2}$)}

In this model, we allow for other energies to be not only non-zero, but also
comparable in strength to $e=1$. The most common choice we have made is the
one shown in the Table \ref{Table_2}: $\mathbf{e}^{\prime}%
=(\ b,-b,-b,-b,2b,b),b=1/3(\simeq1).$ We will call this the model C$_{1}$.
Again, the numerators for various energies are integers for the reason
explained above. Another model called C$_{2}$ has only one non-zero element
$e_{\text{b}}=1$ in $\mathbf{e}^{\prime}.$ Both models will be collectively
denoted simply by C.

The model A is the standard model. In the model B, we have most other
interactions much weaker than $\left\vert e_{\text{HH}}\right\vert $, while
they are comparable to $\left\vert e_{\text{HH}}\right\vert $ in the model C.
Thus, the model B is closer to the model A than to the model C is. Despite
this, we will see that the models B and C behave very different from A. It
should be noted that $W$ does not depend on the model; it is its partition
into $W(E)$ that depends on the model. Thus, the shape of the energy landscape
changes from model to model, but not its total "area" which is given by $W$
\cite{Guj0412548}.

\subsection{Absence of Energy Gap\label{Sect_Absc_Energy_Gap}}

\subsubsection{Semiflexible Homopolymers and Absence of Energy Gap}

The semiflexibility of homopolymers has been exploited by Flory to explain
crystallinity by using a very simple model, which contained only the bending
penalty \cite{Flory}. The energy was simply given by
\[
E_{\text{Flory}}=e_{\text{b}}N_{\text{b}}.
\]
No other interaction such as with the solvent was considered. Thus, the lowest
energy is $E_{\text{Flory}}=0.$ At absolute zero, the polymer chains are going
to be all straight with no bends (provided the chains are finite in length).
Thus, it is anticipated that they would give rise to an ordered structure. One
possibility is that of an aligned configuration in which all chains are
parallel to each other, though this is by no means the only configuration as
one can envision many other configurations of the same energy $E_{\text{Flory}%
}=0.$ The aligned configuration was considered by Flory to represent the
crystalline state formed by linear polymers. Thus, it is expected that the
above simple model will give rise to a melting transition from a disordered
liquid state to a crystalline state at a melting temperature $T_{\text{M}}$.

To make connection with our protein model, we will henceforth consider the
limiting case of a single macroscopically large semiflexible homopolymer
chain. The original approximate solution due to Flory indeed shows such a
melting transition at a non-zero melting temperature $T_{\text{M}}$. The
approximation used by Flory gives rise to an energy gap, which is deduced by
the observation that the resulting entropy based on the approximation becomes
negative over the gap, similar to what happens in the random energy model
discussed earlier in Sect. \ref{REM}. Over the gap, the entropy is replaced by
$S(E)=0;$ we will use $E$ instead of $E_{\text{Flory}}$ in the following for
convenience. This gap then makes the entropy \emph{non-concave} and results in
a melting transition in the model.\ The transition turns out to be a
\emph{freezing} transition in that the entropy of the frozen state (the
crystal) remains zero below the melting temperature, just as was the case for
the random energy model.

It was later shown by Gujrati and coworkers \cite{GujGoldstein} that there was
no energy gap in the Flory model of semiflexible homopolymers. A macroscopic
chain with no solvent was considered. For the infinitely long polymer chain in
the absence of any solvent, the problem is also known as the \emph{Hamilton
walk problem}, the problem in which the walk visits all sites once and only
once. The demonstration of the absence of an energy gap was achieved by
demonstrating that the entropy was never negative over the entire energy range
in the model. The demonstration itself was done by obtaining a \emph{rigorous
lower bound} to the entropy $S(E)$. This required an explicit construction in
which local excitations, the \emph{Gujrati-Goldstein excitations} (GG
excitations) which are pairs of oppositely oriented hairpin turns, populate
the crystal. One such excitation is shown in Fig. \ref{F1} for the case of no
solvent in the interior. It is the local excitation represented by the two
hairpin turns where the parallel bond pair is shown in the figure: it is a
"bound" pair of oppositely oriented hairpin turns and represents a GG
excitation. These GG excitations should be distinguished from unpaired hairpin
turns. The unpaired hairpin turns either cannot be moved, or can be moved only
by changing the number of bends or of parallel bonds or by introducing voids;
see the hairpin turn in the second row (from the top) just above the shown HP
pair in Fig. \ref{F1}; it cannot be moved up or down without increasing the
number of bends or of parallel bonds or by introducing voids. In contrast to
these, the bound GG excitations are highly "mobile" in that they can be moved
about without changing the number of bends or of parallel bonds or by
introducing voids until they hit another defect or the wall; see the
excitation between the third and fourth row (from the top) in Fig. \ref{F1},
which can be freely moved to the left. This "agility" of the excitation
increases the entropy in the system without changing the energy in the model.
It should be noted, see Fig. \ref{F1}, that an isolated hairpin turn can be
turned into a GG excitation by increasing the number of bends by 4 and
parallel bonds by 2, after which the excitation becomes "agile" to move.

The distances over which the GG excitations can be moved can be easily
estimated in a crude fashion by the defect density. This is similar to the
interparticle distance between particles at a given concentration $c$, which
is given by $c^{-1/d}$, where $d$ is the dimension of the lattice. We can use
for $c$ the density $c_{\text{d}}$ of the defects (the bends, hairpin turns or
the GG excitations) in the crystal. Thus, the number of possible moves for a
single GG excitation is this distance and is on an average%
\begin{equation}
W_{\text{GG}}\sim c_{\text{d}}^{-1/d}/a=c_{\text{d}}^{-1/d}, \label{W_GG}%
\end{equation}
as we have set $a=1$. At $T=0,$ we surely have $c_{\text{d}}=0.$ The GG
excitations along with other defects like the bends, the hairpin turns, etc.
gradually populate and begin to destroy the perfect crystalline order by
increasing the entropy as soon the temperature rises above $T=0,$ and the
crystalline phase melts at the melting (or unfolding) temperature
$T_{\text{M}}$ into a disordered phase \cite{GujSemiflex}. The crystalline
state has been shown to occur via a sharp first-order transition if we have
either an infinitely long macroscopic polymer \cite{GujGoldstein} or a bulk
system containing a macroscopic number of finite length polymers
\cite{GujSemiflex} provided we allow \emph{other} energies besides that for
bending. As long as we have a single polymer, which is finite in length, the
folding transition is not going to be sharp, but diffuse.

\subsubsection{Semiflexible Copolymer and Absence of Energy Gap}

The constructive proof of no energy gap also works for the current protein
model, as we now discuss. The main difference is that while the calculation
discussed above for the homopolymer is done rigorously, we do not have a
rigorous calculation at present for the copolymer because of the complexity
produced by the sequence structure. Our results are based on plausibility
arguments, which we present below. As said earlier, the issue of an energy gap
in proteins requires studying macroscopic proteins. We, therefore, consider a
single macroscopic protein. We will also not consider any solvent, so that we
are dealing with a Hamilton walk problem. Accordingly, $M=N_{\text{L}},$ and
$c=1/a=1$. As we have just seen, the presence of the Gujrati-Goldstein
excitations in a homopolymer implies that there is no energy gap in our model
of melting for a homopolymer \cite{GujGoldstein,GujSemiflex}. We now extend
the constructive proof to the copolymer case (or to the heteropolymer case).
The complication arises from the presence of other interactions, such as the
HH interaction. Let us for the moment only consider the bending penalty and
the hairpin and parallel bond energies in addition to the contact interaction
energy due to the HH pair contacts. Thus, we consider the variant models B and
C and not the standard model in the following. We will return to the standard
model later.

Consider a macroscopically large copolymer of a given sequence $\chi$ on a
lattice. Let us consider the native state at $T=0.$ The attractive HH
interaction and a favorable (negative) energy for a hairpin turn compete with
the bending penalty in order to minimize the internal energy in the native
state. In contrast, one only need to maximize the HH contact number without
any regard to the number of bends in the standard model, and to only minimize
the number of bends in the Flory model without any regards to the HH contacts.
We will assume that there is only one unique native state (modulo any symmetry
operation). For example, for $M=24,$ we show the native state for the model
B$_{1}$ in (\ref{Native2}), which is related to the native state in
(\ref{Native3}) by a symmetry transformation (\ref{Conf_Trans}) as explained
later. This does not prove but strongly suggests a unique native state even
for larger $M$.

Because of the favorable nature of hairpin turns, the native state must have a
non-zero density of them. Thus, the defect density $c_{\text{d}}$ would be
non-zero at $T=0,$ which makes this problem inherently different from that of
the semiflexible homopolymer. Some of the hairpin turns must be in the bound
state\ in the form of the GG excitations. We assume that there is a non-zero
density $c_{\text{GG}}$ of these excitations in the native state at $T=0$. The
native state will usually have the maximum number of the HH contacts for most
of the sequences $\chi$ as $e_{\text{HH}}$ has the maximum strength. If we
move a GG excitation, this will require a rearrangement on the lattice of that
portion of the protein that is contained between the two hairpin turns of the
excitation under investigation. We can crudely estimate the number of residues
on this portion of the protein as
\[
n_{\text{R}}\sim M/c_{\text{d}}V=1/c_{\text{d}}a^{d}=1/c_{\text{d}}.
\]
Half of this number is the average number of H residues in this portion.

The positions on the lattice of the residues belonging to this portion of the
protein will change with the movement of the GG excitation. Even though this
movement does not change the number of bends and parallel bonds, it will
invariably reduce the number of HH contacts compared to that in the native
state. Thus, the energy of the deformed conformation due to the GG excitation
movement will be higher than that of the native state. Indeed, this is true of
any deformation of the native conformation (including that generated by the
movements of the GG excitations):\ Any deformation of the native state will
always raise the energy since by definition, the unique native state has the
lowest energy (at $T=0$). For the deformation due to the GG excitation
movement, this increase is due to breaking some of the HH contacts.

Not much can be said about how much the increase in the energy will happen in
displacing a GG excitation, as it depends strongly on the sequence $\chi$ and
on the topology of the native state. Furthermore, not all newly generated
conformations in $W_{\text{GG}}$ will have the same excess energy. We now pick
an extensively large number of GG excitations and move each of them, which
results in $W_{\text{GG}}$ new conformations. The new $W_{\text{GG}}$ is the
product of $W_{\text{GG}}$ \ in \ref{W_GG} over the set of selected
GG\ excitations in the construction. The resulting gain in the entropy density
will be
\[
\Delta s\sim(n_{\text{GG}}/d)\ln c_{\text{d}},
\]
where $n_{\text{GG}}$ is the density of GG excitations used in the
construction. We expect $n_{\text{GG}}$ to be proportional to the defect
density $c_{\text{d}},$ at least for small $c_{\text{d}},$ so that the above
entropy gain vanishes as $c_{\text{d}}\rightarrow0$.

Let $W_{\text{GG}}(E)$ denote the number of conformations in the above
construction to have the energy $E,$ where $E>E_{0},$ $E_{0\text{ }}$being the
energy of the native state. Obviously,
\[
W_{\text{GG}}\equiv\sum_{E}W_{\text{GG}}(E),
\]
where the sum is over possible energies that appear in the construction due to
the movement of the excitation. For a macroscopic system, the sum is going to
be dominated by some energy $E=\overline{E}>E_{0},$ so that
\[
W_{\text{GG}}\simeq W_{\text{GG}}(\overline{E}).
\]
But a little reflection will convince the reader that the excess energy
density $\overline{e}-e_{0}$ is also proportional to $n_{\text{GG}}.$\ Thus,
we will obtain a continuous energy density spectrum in our construction. As
the construction only generates some of the conformations of energy
$E=\overline{E},$ the actual entropy gain is at least as much as $\Delta s>0$
given above. Consequently, it does not seem possible to have an energy gap for
most of the sequences.

\section{Self-Averaging and Small Proteins\label{Sect_SelfAve}}

For a system with quenched randomness, which in our case is created by the
\emph{fixed }sequence of amino acids, an important question about
self-averaging has been probed. The idea is quite simple. Consider a protein
with $M$ amino acids in a given sequence $\chi.$ The sequence for a given
protein is fixed in Nature (or in the lab, where it is synthesized). However,
there are several possible sequences. For example, consider all possible
sequences for any given $M$ in which there are exactly $s$ H-type residues and
$(M-s)$ P-type residues. The number of possible distinct sequences is given by%
\[
C_{M,s}\equiv\frac{M!}{s!(M-s)!}.
\]
On the other hand, if we consider all possible sequences without any
restrictions on the number of H-residues, then the number of possible
sequences is $2^{M}$ corresponding to all possible values of $s.$ The most
probable value of $s$ is $s=\left[  M/2\right]  ,$ where $\left[  x\right]  $
is the integer part of $x,$ since $C_{M,\left[  M/2\right]  }$ is maximum. Let
us denote the set of corresponding sequences by $\widetilde{\boldsymbol{\chi}%
}$. Loosely speaking, we will call these sequences the \emph{most probable
sequences}, knowing well that it is the value of $s$ or the corresponding set
$\widetilde{\boldsymbol{\chi}}$\ that is most probable and not one of the
sequences.\emph{ }

Let $Q$ denote a certain thermodynamic property like the energy of the native
state, the free energy of the protein, the number of helices in the native
state, etc. This quantity will, in general, depend on the sequence $\chi,$ and
one can determine its \emph{quenched average}
\begin{equation}
<Q>_{\text{seq}}\equiv\frac{1}{\left\vert \chi\right\vert }\sum_{\chi}Q(\chi),
\label{QuenchedAv}%
\end{equation}
where $\left\vert \chi\right\vert $\ is the number of possible sequences over
which the averaging is done. The property $Q$ is said to be
\emph{self-averaging} if%
\begin{equation}
\lim_{M\rightarrow\infty}Q(\chi)=\lim_{M\rightarrow\infty}<Q>_{\text{seq}}
\label{SelfAv}%
\end{equation}
for \emph{almost} all $\chi$. As usually happens in the thermodynamic limit,
$\widetilde{\boldsymbol{\chi}}$ contains almost all the sequences. This is
evident from the behavior of \ $C_{M,s}$ for large $M$. The most probable
sequence contains $C_{M,\left[  M/2\right]  }\simeq2^{M}$ for $M>>1.$ This is
also the number of all sequences. Then, the above condition of self averaging
really refers to any sequence belonging to $\widetilde{\boldsymbol{\chi}}.$ It
is clear that the idea of self-averaging, which is not a trivial property,
requires considering a macroscopic copolymer. If the property is self
averaging, then the limit on the left in (\ref{SelfAv}) is independent of the
sequence $\chi$. This important property then gives rise to many
simplifications. For example, it allows one to use the replica trick
\cite{ReplicaTrick} to calculate the quenched averages of quantities such as
the free energy. The trick represents a major technical advantage that has
been extensively used quite successfully to study macroscopic random systems.
As shown in \cite{Kardar}, there are strong indications that self averaging is
valid for macroscopic proteins.

It is instructive now to see how well the equality (\ref{SelfAv}) (without the
limits on both sides) is obeyed for finite $M$. For this purpose, we consider
the native state energy $E_{0}$ to be the the thermodynamic property $Q,$ and
consider the quenched average of the native state energy density $e_{0}\equiv
E_{0}/M$:%
\[
<e_{0}>_{\operatorname{seq}}\equiv\,\frac{1}{M}<E_{0}>_{\operatorname{seq}%
}\equiv\frac{1}{M\left\vert \chi\right\vert }\sum_{\chi}E_{0}(\chi)
\]
over all sequences that belong to $\widetilde{\boldsymbol{\chi}},$ so that the
average is taken over all sequences with the restriction of equal H and P
(even $M$). Thus, not all sequences are allowed. This is done because of the
importance of the most probable sequence noted above and requires evaluating
$E_{0}(\chi)$ for each sequence in $\widetilde{\boldsymbol{\chi}}.$%

\begin{figure}
[ptb]
\begin{center}
\includegraphics[
trim=0.828591in 2.918964in 1.005389in 3.007288in,
height=2.6212in,
width=3.5008in
]%
{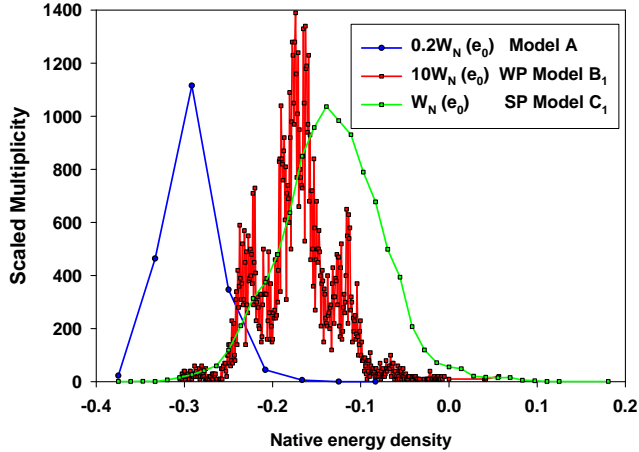}%
\caption{The scaled distribution of $\widetilde{W}(e_{0})$ as a function of
the native state energy $e_{0}$ for 10,000 different sequences for
unrestricted coformations of $M=24$. For the standard and the weakly perturbed
models, we show the scaled distribution $\widetilde{W}(e_{0})/5$ and
$10\widetilde{W}(e_{0})$ so that the scaled distributions can be shown on the
same scale. }%
\label{F12}%
\end{center}
\end{figure}

Let $W_{\text{N}}(e_{0})$ denote the number of times a given native energy
$e_{0}\equiv E_{0}/M$ appears among all sequences in $\widetilde
{\boldsymbol{\chi}}$. We then calculate the \emph{relative root mean square
}(\emph{rms})\emph{ fluctuation}%
\begin{equation}
\,\left\langle \delta e_{0}\right\rangle _{\text{seq}}\equiv\frac{\sqrt
{<e_{0}^{2}>_{\operatorname{seq}}-(<e_{0}>_{\operatorname{seq}})^{2}}%
}{\left\vert <e_{0}>_{\operatorname{seq}}\right\vert }, \label{Rel_FLuct0}%
\end{equation}
where
\[
<e_{0}^{2}>_{\operatorname{seq}}\equiv\frac{1}{M\left\vert \chi\right\vert
}\sum_{\chi}E_{0}^{2}(\chi).
\]
Standard arguments \cite{ReplicaTrick} show that the relative fluctuation
$\left\langle \delta e_{0}\right\rangle _{\text{seq}}$ should decrease as
$1/\sqrt{M}$ for large $M$:%
\begin{equation}
\left\langle \delta e_{0}\right\rangle _{\text{seq}}\propto1/\sqrt{M}.
\label{Rel_FLuct}%
\end{equation}
We have done the calculations for the three models for $M=16,$ and $M=24$ on
an infinite lattice$.$ For $M=16,$ we have considered \emph{all} the sequences
in $\widetilde{\boldsymbol{\chi}}$, each with equal number of H and P
residues. The total number of these restricted sequences is
\[
C_{16,8}=12,870.
\]
For $M=24$, we have only considered $10,000$ different sequences for the three
different classes of energetics, which is a small fraction of all allowed
sequences $C_{24,12}=2,704,156.$ We only show the distribution for $M=24$ in
Fig.(\ref{F12}). The results for various quenched averages and the relative
fluctuations are summarized in Table \ref{Table_3}.

\begin{center}%
\begin{align*}
&  \text{Table \ref{Table_3}}-\text{Quenched averages and relative
fluctuation}\\
&
\begin{tabular}
[c]{|c|c|c|c|c|}\hline
& Model & $<e_{0}>_{\operatorname{seq}}$ & $<e_{0}^{2}>_{\operatorname{seq}}$
& $<\delta e_{0}>_{\operatorname{seq}}$\\\hline
$M=16$ &
\begin{tabular}
[c]{c}%
Model A\\
Model B$_{2}$\\
Model C$_{1}$%
\end{tabular}
&
\begin{tabular}
[c]{c}%
$-0.3208$\\
$-0.1959$\\
$-0.1107$%
\end{tabular}
&
\begin{tabular}
[c]{c}%
$0.1058$\\
$0.0427$\\
$0.0174$%
\end{tabular}
&
\begin{tabular}
[c]{c}%
$0.1674$\\
$0.3344$\\
$0.6448$%
\end{tabular}
\\\hline
$M=24$ &
\begin{tabular}
[c]{l}%
Model A\\
Model B$_{1}$\\
Model C$_{1}$%
\end{tabular}
&
\begin{tabular}
[c]{l}%
$-0.2927$\\
$-0.1720$\\
$-0.1336$%
\end{tabular}
&
\begin{tabular}
[c]{l}%
$0.0867$\\
$0.0314$\\
$0.0212$%
\end{tabular}
&
\begin{tabular}
[c]{l}%
$0.1079$\\
$0.2440$\\
$0.4320$%
\end{tabular}
\\\hline
\end{tabular}
\end{align*}
\label{Table_3}
\end{center}

We see that the relative fluctuation increases as the strength of the
perturbation increases for both sizes. In addition, it appears that the
relative increase ($0.4320/0.1079=3.8519$ for $M=16$) or
($0.6448/0.1674=4.0037$ for $M=24$) does not appreciably change with the size.
This needs to be investigated further for other sizes. Moreover, the relative
fluctuation is not small, implying that the spread of the distribution
$W_{\text{N}}(e_{0})$ is not insignificant. If we calculate $\sqrt
{M}\left\langle \delta e_{0}\right\rangle _{\text{seq}}$ from Table
\ref{Table_3}, we observe that this product is much smaller for $M=24$ than
for $M=16,$ while according to (\ref{Rel_FLuct}), this product should not
change. There are two possibilities for this behavior. It is quite conceivable
that either $M=24$ is not large enough for (\ref{Rel_FLuct}) to be observed or
that the choice of only $10,000$ sequences for $M=24$ does not give a good
estimate of the relative fluctuation $\left\langle \delta e_{0}\right\rangle
_{\text{seq}}.$ Thus, our results may not be reliable enough to prove or
disprove self-averaging for a macroscopic protein. Nevertheless, the results
in Table \ref{Table_3} for the small proteins that we have considered in the
present work clearly show that the average native state energy $<e_{0}%
>_{\operatorname{seq}},$ though highly probable, does not represent the native
state energy of almost most of the random sequences in $\widetilde
{\boldsymbol{\chi}}.$ There is no reason to believe that other thermodynamic
quantities will have their sequence average equal the average of any randomly
selected sequence. Thus, small proteins are \emph{not} \emph{self-averaging.
}This is consistent with the accepted result in the literature, see for
example, \cite{Bryngelson}, that sequences play an important role in small proteins.

The situation in Fig.(\ref{F12}) does raise an interesting question. We see
the most probable native energy is far from the lowest native energy for each
of the three models. Does Nature prefer to design proteins whose native
energies are close to the most probable native energies or to the lowest
native energy? It should be remarked that all those sequences that have their
native energies close to the most probable native energy do not fold into one
unique native structure, though many sequences are found to have the same
structure (conformations without any regard to the sequence). The native
conformations, though compact, have varied structures.

For the standard model in Fig.(\ref{F12}), we observe that there are eight
different native energies for the $10^{4}$ random sequences for the standard
model. We find that $e_{0}\simeq-0.3$ is the most common native state energy;
all these native states differ only in their sequences. None of the models
ascribes a unique structure of the native conformation to a particular
sequence. However, the standard model does point to an interesting fact. The
number of sequences with the lowest native energy $e_{0}\simeq-0.38$ is an
extremely small fraction of the $10^{4}$ sequences considered here. It should
be remarked that the sequence $\chi_{0}$ described below in \ref{Chi0} gives a
much lower native state energy $e_{0}=-0.4167,$ and is not part of the
$10^{4}$ sequences whose results are shown in Fig.(\ref{F12}). For $M=16,$
there are seven different native energies between $e_{0}\simeq-0.44$ and
$e_{0}=0$ for the standard model. The most dominant native energy is
$e_{0}\simeq-0.31$ given by $5664$ sequences$,$ but the number of sequences
with the lowest energy ($430$) is not as small a fraction as for $M=24.$ Thus,
it appears that the fraction of sequences among all sequences that gives the
lowest possible native energy is small, this fraction becoming smaller as the
protein size increases. This suggests that the most probable native energy
distribution becomes narrower with the size $M.$ This observation, which seems
to support the emergence of self-averaging for $M\rightarrow\infty$, needs to
be checked further.

For the weakly perturbed model, the same distribution $W_{\text{N}}(e_{0}),$
see Fig. (\ref{F12}), exhibits a clear band structure; the number of bands
seems to be clearly controlled by the number of possible energies in the
corresponding standard model. However, the band structure is "smoothed out"
for the strongly perturbed model because the latter does not allow as many
native state energies as the weakly perturbed model. Because of this
difference in the allowed native state energies, the maximum $W_{\text{N}%
}(e_{0})$ for the weakly perturbed model is much smaller than the maximum
$W_{\text{N}}(e_{0})$ for the strongly perturbed \ model.

We have found that in the majority of cases that we have investigated, the
following sequence containing a repetition of PHHP and which we denote by
$\chi_{0}$%
\begin{equation}
\chi_{0}:(\text{PHHP})_{n} \label{Chi0}%
\end{equation}
gives rise to the lowest energy or very close to it. Because of this, we
mostly present results based on this particular sequence $\chi_{0}$ in this
work, though we have considered other sequences also.

\section{Energetics and Native Conformations\label{Sect_Energetics_NativeConf}%
}

Let us fix $M=24$ and consider unrestricted conformations. The sequence is
fixed to $\chi_{0},$\ i.e. to
\[
\text{PHHPPHHPPHHPPHHPPHHPPHHP}%
\]
for the reason explained in the preceding section. For the standard model,
there are $30$ native states, all of the same energy density $e_{0}=-0.4167,$
as discussed in the following. One of the native states is the following conformation:

$\qquad\qquad\qquad\qquad$%
\begin{equation}%
\begin{array}
[c]{cccc}%
1\text{P} & 2\text{H} & 3\text{H} & 4\text{P}\\
8\text{P} & 7\text{H} & 6\text{H} & 5\text{P}\\
9\text{P} & 10\text{H} & 11\text{H} & 12\text{P}\\
16\text{P} & 15\text{H} & 14\text{H} & 13\text{P}\\
17\text{P} & 18\text{H} & 19\text{H} & 20\text{P}\\
24\text{P} & 23\text{H} & 22\text{H} & 21\text{P}%
\end{array}
, \label{Native0}%
\end{equation}
and can be represented by the string
\begin{equation}
\text{{\normalsize RRRDLLLDRRRDLLLDRRRDLLL},} \label{Native_String0}%
\end{equation}
which is read from the left and refers to the sequential steps from the first
residue along the right (R), left(L), up (U), and down (D) directions. The
first step is always to the right direction, and the first bend is always in
the D direction. This is done to cut down the number of conformations to be
counted. All conformations in which the first bend is in the U direction is
topologically identical to one of the conformations that we generate.
Similarly, conformations that start not in the R direction are also
topologically not distinct. Despite these restrictions, we still duplicate
some conformations if the two ends of the protein are treated identically.
This happens when the last step of the protein is in the L direction and the
bend before the last step is in the U direction. We will explicitly
demonstrate this below. However, this does not affect us as we deal the two
ends as different.

We also report the nine other native states that are given by the strings%

\begin{equation}%
\begin{array}
[c]{c}%
\text{{\normalsize RRRDLLLDRRRDLLLDRRRDLLD,}}\\
\text{{\normalsize RRRDLLLDRRRDLLLDRDDRUUR,}}\\
\text{{\normalsize RRRDLLLDRRRDLDRDLLLURUL,}}\\
\text{{\normalsize RRRDLLLDRDLDRDDRUURULUR,\ }}\\
\text{{\normalsize RRRDLDRDLDRDLLLURULURUL,\ }}\\
\text{{\normalsize RRDLURDRURRULLULDLULDLL,\ }}\\
\text{{\normalsize RRDLURDRURRULLULDLULDLU, }}\\
\text{{\normalsize RDLDRRRULURURRRULLLURRR,\ }}\\
\text{{\normalsize RDLDRRRULURURRRULLLURRU.\ }}%
\end{array}
\label{Native_String1}%
\end{equation}
We notice that the third and the eighth strings above are related by
\begin{equation}
\text{L}\Leftrightarrow\text{R,U}\Leftrightarrow\text{D, and the reversal of
the strings;} \label{Conf_Trans}%
\end{equation}
an example is given below for clarity. Thus, there are only $9$ distinct
native states if the two ends are treated identically. Of course, the above
symmetry transformation does not affect our calculation since we make a
distinction between the N-terminus and the C-terminus.

For the weakly perturbed model B$_{1}$, there are two native states of energy
density $e_{0}=-0.3717$. The native state string
\begin{equation}
\text{{\normalsize RRRDLDRDLDRDLLLURULURUL}} \label{Native_String2}%
\end{equation}
represents the following native state%

\begin{equation}%
\begin{array}
[c]{cccc}%
1\text{P} & 2\text{H} & 3\text{H} & 4\text{P}\\
24\text{P} & 23\text{H} & 6\text{H} & 5\text{P}\\
21\text{P} & 22\text{H} & 7\text{H} & 8\text{P}\\
20\text{P} & 19\text{H} & 10\text{H} & 9\text{P}\\
17\text{P} & 18\text{H} & 11\text{H} & 12\text{P}\\
16\text{P} & 15\text{H} & 14\text{H} & 13\text{P}%
\end{array}
. \label{Native2}%
\end{equation}
{\tiny \ }The other native state string
\begin{equation}
\text{{\normalsize RDLDRDLDRRRULURULURULLL}} \label{Native_String3}%
\end{equation}
represents the native state

$\qquad\qquad\qquad$%
\begin{equation}%
\begin{array}
[c]{cccc}%
24\text{P} & 23\text{H} & 22\text{H} & 21\text{P}\\
1\text{P} & 2\text{H} & 19\text{H} & 20\text{P}\\
4\text{P} & 3\text{H} & 18\text{H} & 17\text{P}\\
5\text{P} & 6\text{H} & 15\text{H} & 16\text{P}\\
8\text{P} & 7\text{H} & 14\text{H} & 13\text{P}\\
9\text{P} & 10\text{H} & 11\text{H} & 12\text{P}%
\end{array}
. \label{Native3}%
\end{equation}
This native state is topologically identical to the previous native state, and
is described by the string obtained by the symmetry transformation
(\ref{Conf_Trans}), as noted above if the two ends are identical. It is clear
that the weak perturbation alone has drastically reduced the native state
multiplicity from $30$ to $2$. This shows the importance of even the weak perturbation.

The strongly perturbed model, surprisingly, has three native states given in
(\ref{Native0}),(\ref{Native2}), and (\ref{Native3}); the last two are related
to each other by the above transformation. This suggests that the relationship
between a given native state and the energetics is quite complex. The set
$\mathbf{N}$ for the first two native conformations are $(10,15,5,0,10,4,0)$,
and $(18,12,9,7,10,4,0)$; the third native conformation has the same
$\mathbf{N}$ as the second one above, which should not come as a surprise. The
energy density of each of the three native conformations is $(-32/72).$ If we
use $e_{\text{hp}}=-2/3=e_{\text{h}},$ then only the last two conformations
survive as the native conformations; the first one is no longer a native
conformation. Now, the native energy density is $(-48/72),$ and $\mathbf{N}$
is $(18,12,9,7,10,4,0),$ the same as for the previous set of energetics. This
is a clear demonstration of the fact that the same native state can occur in
various different models. Therefore, one cannot determine effectively the
energetics of a protein by only studying the native states. For this, one must
also investigate many of the non-native conformations.

\section{Small System Thermodynamics}

\subsection{Microcanonical Entropy}

\subsubsection{Equilibrium\label{Sect_Small_System_Equilibrium}}

The dimensionless ME entropy corresponding to configurations with a given
energy $E$ is given by the Boltzmann relation (\ref{B_S}); as above, we have
set the Boltzmann constant equal to 1. This entropy is relevant if the energy
of the protein is held fixed. Keeping $E$ constant is not the same as keeping
each term $e_{i}N_{i}$ in the sum in (\ref{Energy0}) constant; the latter can
change as long as the sum in (\ref{Energy0}) remains constant. We define the
\emph{equilibrium} to mean that the protein explores all possible
conformations included in $W(E)$ with \emph{equal probability} given in
(\ref{ME_probability})$.$

\begin{figure}
[ptb]
\begin{center}
\includegraphics[
trim=0.668902in 3.167974in 1.504010in 1.173295in,
height=2.5218in,
width=3.4938in
]%
{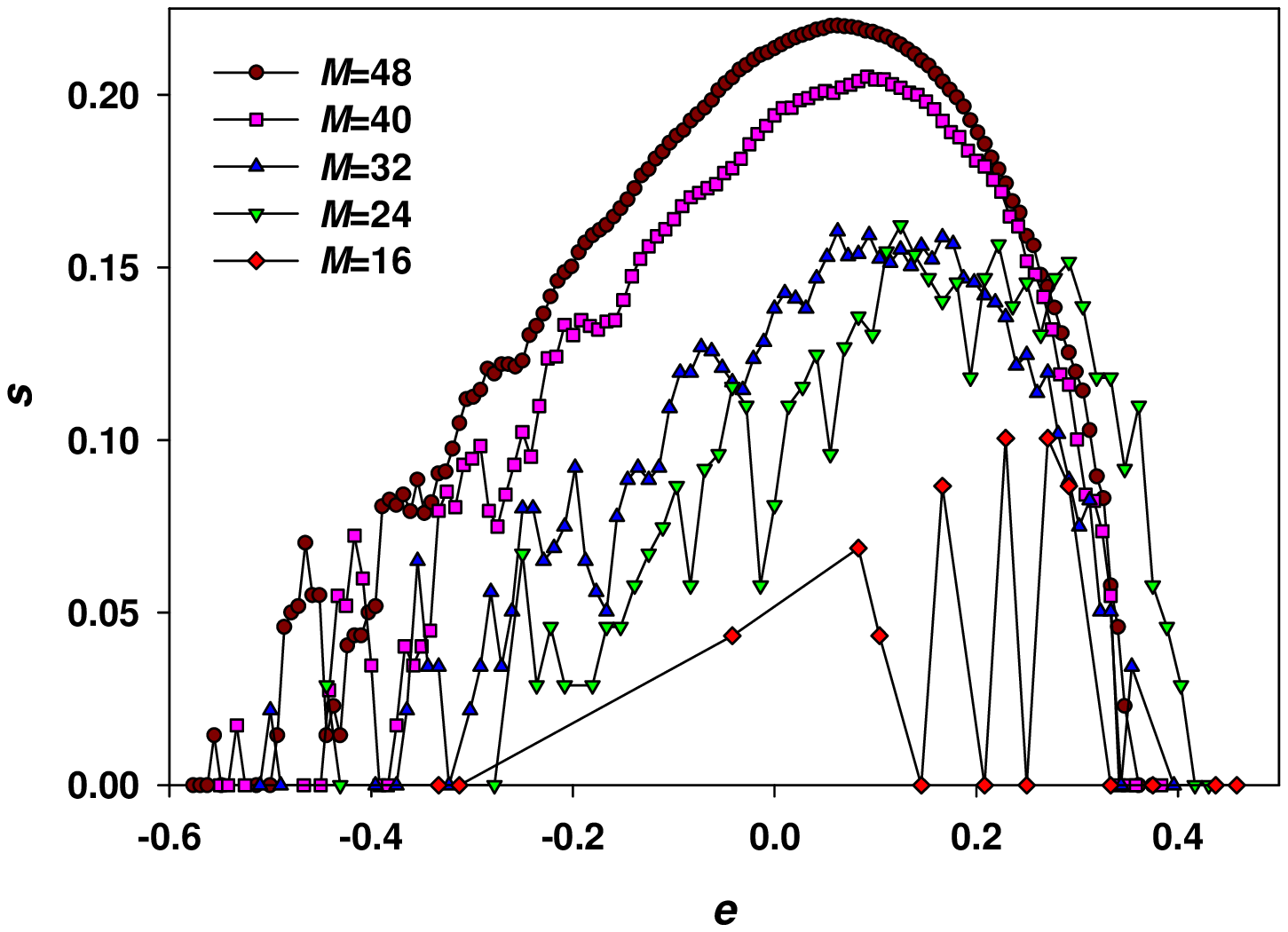}%
\caption{The behavior of $s(e)$ for the strongly perturbed model C$_{1}$ of a
compact protein as a function of the protein size $M$. We observe that the
both the lowest and the highest energy densities decrease with the protein
size.}%
\label{F18}%
\end{center}
\end{figure}

Let us recall the arbitrary positive energy $\epsilon$ (we can take this to be
the magnitude $\left\vert e_{\text{HH}}\right\vert $ for concreteness) that we
have used to introduce the adimensional energy $E,$ which is really
$E/\epsilon$ \cite{note}$.$ For ~a small protein ($M<\infty$), each element in
the set $\mathbf{N}$ is \emph{finite}. Thus, the adimensional energy is also
finite, with the closest spacing $\Delta_{\text{min}}E$\ between two
successive values of $E$ at least $\left\vert e_{\text{min}}\right\vert $
(which is really $\left\vert e_{\text{min}}\right\vert /\epsilon$)$,$ where
$e_{\text{min}}$ is the element with the smallest magnitude in the set
$\mathbf{e.}$ Therefore, for small proteins, $E$ is a discrete variable. The
corresponding energy density per residue
\[
e\equiv E/M
\]
is also discrete and becomes continuous only when $M\rightarrow\infty.$ Thus,
as long as $M$ is finite, the energy and the entropy density per residue%
\[
s(e)\equiv S(E)/M
\]
remain discrete. In addition, they also depend on $M$ for small proteins
\cite{note01}. To show this most clearly, we reproduce $s(e)$ for the strongly
perturbed model C$_{1}$ in Fig. \ref{F18} for $M=16,24,32,40$, and $48;$ we
restrict the conformations of the protein to be compact. There continues to be
a dependence on $M,$ even though the largest value of $M$ is $\ 48.$ We also
note that the discrete nature of the energy and entropy persists. There is a
clear evidence of many local maxima in the entropy, each maximum surrounded by
many energies of lower entropy forming an energy \emph{band. }These bands are
well separated by gaps in the energy, at least near the low end of the energy
even for $M=48.$ It is surprising to observe the erratic form of the entropy
in that the bands are highly irregular in shape, at least near the low energy
end. The entropy function is becoming somewhat smoother (but still discrete)
near it global maximum because the energy levels are becoming denser in this range.%

%

\begin{figure}
[ptb]
\begin{center}
\includegraphics[
trim=0.899811in 3.241814in 1.675569in 3.265990in,
height=2.6273in,
width=3.4627in
]%
{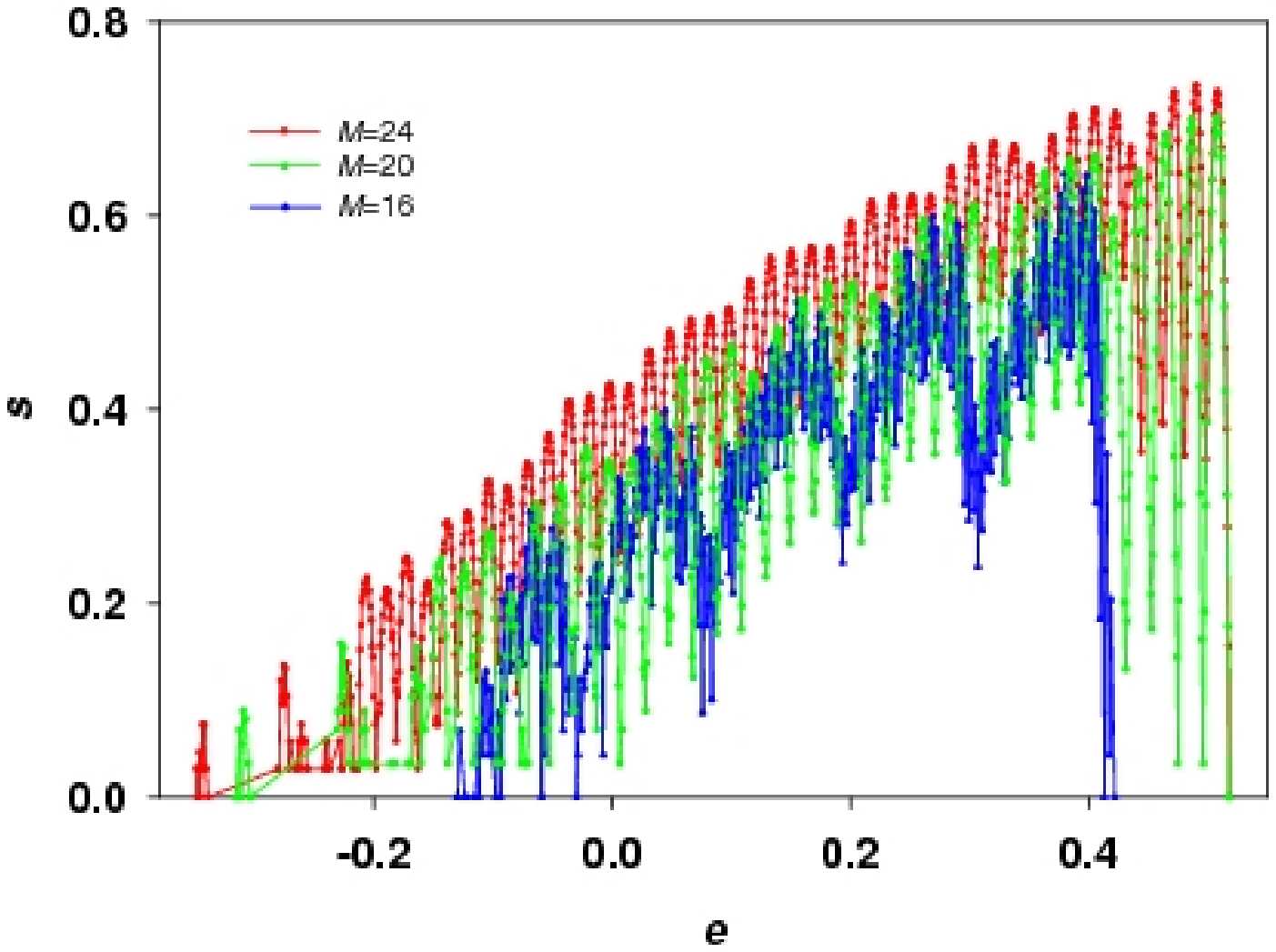}%
\caption{The behavior of $s(e)$ for the weakly perturbed model B$_{1}$ on an
infinite lattice as a function of the protein size $M$. We observe that $s(e)$
for the larger size contains that for the smaller size inside it. }%
\label{F19}%
\end{center}
\end{figure}

In Fig. \ref{F19}, we show the ME entropy $s(e)$ when the protein
conformations are unrestricted. We are considering a weakly perturbed model
B$_{1}.$ We again see a dependence of $s(e)$ on $M$, as before. Similarly, the
allowed energy densities continue to depend on $M.$ This dependence is not so
weak to be negligible, especially near the low energy range, the range more
appropriate and influential in studying protein folding. This is a clear
indication that one cannot treat the densities such as $s$, and $e$ to be
independent of $M$. This point does not seem to be appreciated in the
literature; see for example, \cite{Clementi1}. We notice that $s(e)$ remains
discrete even for $M=24$, close to the largest protein we have investigated in
the case when the conformations are unrestricted.

There are some common features in both figures \ref{F18} and \ref{F19}. The
first feature is the presence of gaps in bands of $s(e)$ at lower energies:
there is a clear energy gap between the two lowest bands for $M$ ranging from
$24$ to $48$ for compact conformations and for $M$ ranging from $20$ to $24$
for unrestricted conformations. The gap decreases with $M$ in both cases. This
is consistent with the claim in Sect. \ref{Sect_Absc_Energy_Gap} of no energy
gap in the model. Another feature we notice is that $s(e)$ is usually higher
for larger $M$ over a wide range of energies. There is a certain pattern in
the undulations present in $s(e):$ they seem to form a band structure with
several peaks within each band; the number of peaks in a band keeps increasing
with $M.$ The presence of these bands will be explained below.

The native state of the protein is, by definition, the lowest energy state at
absolute zero. Depending on the interactions in the protein and the sequence
$\chi$ of H and P residues in it, the native state may or may not be unique.
In the latter case, the multiplicity of the lowest energy state will indicate
that the protein functionality is not simply determined by the native state.
(We will call this multiplicity the \emph{degeneracy} of the native state.)
The way out of this dilemma is to have the energetics tuned in such a way that
the native state becomes unique. At present, our understanding of protein
functionality is not so complete to answer this question unambiguously.
Therefore, we will allow the occurrence of degenerate native states and study
the effect of energetics on this degeneracy to learn how the energetics should
be tuned to give a unique native state. It may be that there exist high energy
barriers between these native states so that it is impossible for the protein
to jump from one native state to another in a finite amount of time. However,
it should be recognized that for a small protein ($M<\infty$), no energy
barrier of any kind except due to excluded volume interactions (which occur
when a site is occupied twice, but do not exist in our lattice model as only
configuration satisfying excluded volume constraints are allowed) can be
infinitely large; hence, the time required to transform from one native state
to another will remain finite, though it may be large in some cases. Thus,
this idea of a large barrier to explain the robustness of a protein may not be
so reliable or relevant.

\subsubsection{Non-equilibrium}

Away from equilibrium, the protein will not explore all the conformations in
$W(E)$ with equal probability. In this case, the entropy of the
non-equilibrium state is given by the Gibbsian relation (\ref{G_S}) in which
$p(\Gamma),$ where $\Gamma$\ is one of the conformations in $W(E)$,\ is
independent of the temperature. This non-equilibrium microcanonical Gibbsian
entropy will eventually achieve its \emph{maximum} under the constraint
\begin{equation}
\sum_{\Gamma}p(\Gamma)\equiv1, \label{Sum_p}%
\end{equation}
as the protein equilibrates.\ This is easily seen by the using the Lagrange
multiplier trick to maximize the combination%
\[
\sum_{\Gamma}p(\Gamma)(-\ln p(\Gamma)+\lambda),
\]
where $\lambda$\ is the Lagrange multiplier. The resulting distribution is
given by%
\[
p(\Gamma)=\exp(\lambda-1).
\]
The use of (\ref{Sum_p}) determines the Lagrange multiplier%
\begin{equation}
\exp(\lambda-1)=1/W(E). \label{Equilibrium_p_ME}%
\end{equation}
Thus, the resulting equilibrium distribution is given by the Boltzmann
relation (\ref{B_S}). This is the conventional law of increase of entropy in
thermodynamics as the system moves towards equilibrium.

This formulation of the second law is obviously applicable to small systems
such as our protein in our approach based on the Conjecture \ref{Conj_3}, and
also justifies our Conjecture \ref{Conj_1}.

\subsection{Behavior of the Compact and Unrestricted Conformations}

The behavior of $S(E)$ is different compact and unrestricted conformations. We
first consider the standard model. For unrestricted conformations, the maximum
energy corresponds to non-compact conformations of which there are many; the
actual value depends on the value of $M.$ Thus, $S(E)$ does not vanish at the
upper end of the energy. Here, the entropy continues to increase as the energy
increases. This can be easily seen in Fig. \ref{F2}. On the other hand, the
situation is drastically different for compact conformations. Here, there are
not that many configurations of the highest energy. Thus, the entropy first
rises and then drops as the energy increases. This remains true for any of the
three models, and we refer the reader to Fig. \ref{F18} where we have shown
the results for compact conformations in the model C$_{1}.$

Let us consider unrestricted conformations of the protein. The standard model
entropy will be perturbed drastically even with weak perturbation of energies.
This is because the number of conformations that contribute to
$W(N_{\text{HH,max}})$ at the highest energy $E_{1}\equiv-N_{\text{HH,max}}$
in the standard model will redistribute themselves in a band due to weak
energy perturbation. The spread of the band will now give zero or very small
entropy at the highest energy in the two perturbed models. This causes a
drastic change in the form of the entropy distribution: each energy level of
the standard model turns into a band; see the bands of the perturbed models in
Fig. \ref{F2}. We see that there are exactly 11 bands, equal in number to the
11 energy levels in the standard model A. The energy gap between the bands at
the low end of the energy spectrum in the weakly perturbed model is also a
manifestation of the energy gap in the standard model. This gap is easy to
notice in Fig. \ref{F19} where we have also shown the entropy at low energies
for the model B$_{1}.$ This gap seems to be almost filled up in the strongly
perturbed model C; see Fig. \ref{F2}.

As $M$ increases, the energy spectrum in $e$ becomes dense so that $e$ and,
therefore, $s(e)$, become continuous.

\subsection{Canonical Partition Function}

A protein in Nature is not a closed system as discussed above. Therefore, the
ME is not the most suitable ensemble to investigate. As the protein interacts
with its surrounding at a given temperature $T$, we need to consider the CE in
which the temperature of the system and its surrounding is held fixed. This
description is more realistic and can be characterized by the canonical
partition function given by%
\begin{equation}
Z(T)\equiv\sum_{E}W(E)\exp(-\beta E), \label{PF}%
\end{equation}
where $\beta\equiv1/T$ is the inverse temperature in the units of the
Boltzmann constant. The reader should be warned that we are using the
partition function formalism, which is believed to give the correct
thermodynamics of large systems, for the current case of a small protein. The
thermodynamics of a small system is far from a complete understanding in that
it is not known if the small system thermodynamics is the same as that
predicted by the use of the above partition function (\ref{PF}). We will not
be concerned with this issue here and adopt the most prevalent view in the
field and use the above small-system partition function formalism to study the
thermodynamics of the small system. A credible justification of this adoption
will be provided at the end of the next section.

It is convenient to rewrite the partition function as a sum over $\mathbf{N}$
as follows:%
\[
Z(T)\equiv\sum_{\mathbf{N}}W(\mathbf{N})\exp[-\beta E(\mathbf{N})].
\]
>From this, we can calculate the thermodynamic averages $\overline{N_{i}}$ as
follows:%
\begin{equation}
\overline{N_{i}}\equiv\frac{\sum_{\mathbf{N}}N_{i}W(\mathbf{N})\exp[-\beta
E(\mathbf{N})]}{Z(T)}=-\left(  \frac{\partial}{\partial\beta e_{i}}\ln
Z(T)\right)  , \label{N_Ave}%
\end{equation}
where the derivative is taken at fixed $\beta\mathbf{e}_{i}^{\prime}$, where
$\mathbf{e}_{i}^{\prime}$ represents the set of all the remaining energies in
the set $\mathbf{e}$ except $e_{i}$, and may be a null set. If we introduce
the fluctuation $\Delta N_{i}\equiv N_{i}-\overline{N_{i}},$ then
\begin{equation}
\overline{\left(  \Delta N_{i}\right)  ^{2}}=\left[  -\frac{\partial}%
{\partial\beta e_{i}}\right]  ^{2}\ln Z(T)=-\left(  \frac{\partial
\overline{N_{i}}}{\partial\beta e_{i}}\right)  \geq0. \label{N_Fluc}%
\end{equation}
It follows, therefore, that
\begin{equation}
\left(  \frac{\partial\overline{N_{i}}}{\partial e_{i}}\right)  \leq0.
\label{N_e_Derivative}%
\end{equation}
As said above, the derivative is taken at fixed $\beta\mathbf{e}_{i}^{\prime}$.

\subsection{Canonical Averages, Fluctuations, and Entropy}

\subsubsection{Equilibrium}

We define the system to be in \emph{equilibrium}, when the canonical
probability distribution for $\Gamma$ is given by%
\begin{equation}
p(\Gamma)\equiv e^{-\beta E(\Gamma)}/Z(T), \label{CE_Prob}%
\end{equation}
where the partition function is given in (\ref{PF}), which can also be written
as a sum over $\Gamma:$
\begin{equation}
Z(T)\equiv%
{\textstyle\sum\limits_{\Gamma}}
e^{-\beta E(\Gamma)}. \label{PF1}%
\end{equation}
One can also introduce the probability for the system to have a given energy
$E:$%
\begin{equation}
p(E)=W(E)e^{-\beta E(\Gamma)}/Z(T). \label{PrE}%
\end{equation}
It is clear that
\begin{equation}%
{\textstyle\sum\limits_{\Gamma}}
p(\Gamma)\equiv%
{\textstyle\sum\limits_{E}}
p(E)\equiv1. \label{Sum_p_T}%
\end{equation}
\qquad

The canonical probability distribution $p(\Gamma)$ can be used to directly
evaluate the thermodynamic average (to be denoted by an overbar in the
following) of any thermodynamically extensive quantity \cite{note3}
$O(\Gamma)$ using
\begin{equation}
\overline{O}\equiv%
{\textstyle\sum\limits_{\Gamma}}
O(\Gamma)p(\Gamma). \label{Conf_Ave}%
\end{equation}
Similarly, we can use $p(E)$\ to directly evaluate the thermodynamic average
(again to be denoted by an overbar in the following) of any thermodynamically
extensive quantity $O(E)$ using%
\begin{equation}
\overline{O}\equiv%
{\textstyle\sum\limits_{E}}
O(E)p(E). \label{Energy_Ave}%
\end{equation}
Both averages are functions of the temperature $T.$\ Two of the examples of
such averages are $\overline{\mathbf{N}}(T),$ and $\overline{E}\equiv
\mathbf{e}\cdot\overline{\mathbf{N}}(T)$; see (\ref{N_Ave}). It is easy to see
that
\[
\overline{E}=-\left(  \frac{\partial}{\partial\beta}\ln Z(T)\right)  ,
\]
and
\begin{equation}
\overline{\left(  \Delta E\right)  ^{2}}=\left[  -\frac{\partial}%
{\partial\beta}\right]  ^{2}\ln Z(T)=-\left(  \frac{\partial\overline{E}%
}{\partial\beta}\right)  \geq0, \label{E_Fluc}%
\end{equation}
where $\Delta E\equiv E-\overline{E}$ is the energy fluctuation. Thus,
$\overline{E}$ is a monotonic increasing function of $T.$\ 

Let $E_{0}$ and $E_{1}$ denote the minimum and maximum allowed energies in the
model, and $\widetilde{E}$ the energy at which $S(E)$ has its maximum. At
absolute zero ($T=0$), it is easy to see that $\overline{E}(0)=E_{0}.$ At
infinite temperatures,
\[
\overline{E}(\infty)=\frac{1}{W}\sum W(E)E,
\]
and can be very different from $\widetilde{E}$ due to the finite size$.$
(Their equality occurs only for a macroscopic system.) Consider $M=48,$ Model
C$_{1},$ and all its conformations in the compact form. There are $1,194,244$
distinct conformations, and the exact calculation provides%
\[
\overline{e}(\infty)=0.0521,\text{ and }\widetilde{e}=0.0625,
\]
where the energy density per residue $\overline{e}(\infty)\equiv\overline
{E}(\infty)/M$ and $\widetilde{e}\equiv\widetilde{E}$ $/M.$ The energy density
per residue $e_{0}\equiv E_{0}/M=-0.5764,$ and $e_{1}\equiv E_{1}$
$/M=0.3750.$ The number of conformations of energy $\widetilde{E}$ is
$38,707,$ so that the entropy density per residue is $s(\widetilde
{e})=0.2201.$ The two energies $\overline{e}(\infty)$ and $\widetilde{e}$ are
very different. One can also obtain $\overline{e}(\infty)>$ $\widetilde{e}.$
Nevertheless, $\overline{E}$ monotonically increases with $T$ from
$\overline{E}(0)$ to $\overline{E}(\infty).$ This does not guarantee that each
$\overline{N_{i}}$ also increases monotonically with $T$ (except in the
trivial case of the when the set $\mathbf{N}$ has a single member such as the
standard model). Indeed, some of them may actually decrease with $T$.

It is convenient to introduce various densities associated with average
extensive quantities of interest by diving by $M:$%
\[
\overline{e}\equiv\overline{E}/M,\overline{n_{i}}\equiv\overline{N_{i}}/M.
\]
It is these densities that will approach a limit as $M$ becomes larger and
larger \cite{note01}; see Figs. \ref{F18} and \ref{F19}. For finite $M,$ they
remain functions of $M.$

\subsubsection{Non-equilibrium}

If the system is not in equilibrium, then the canonical probability
distribution is not given by (\ref{CE_Prob}). However, the entropy of the
non-equilibrium state is still given by (\ref{G_S}), where $p(\Gamma)$ is the
non-equilibrium probability distribution; it will also depend on $T.$ This
distribution should be used to calculate configuration averages by using
(\ref{Conf_Ave}). As the system approaches towards equilibrium, $p(\Gamma)$
changes so as to maximize the entropy under two constraints, one of which is
the above constraint (\ref{Sum_p}). The other one is the constraint on the
constancy of the average energy%
\begin{equation}%
{\textstyle\sum\limits_{\Gamma}}
p(\Gamma)E(\Gamma)=\overline{E}=\text{constant.} \label{Sum_E}%
\end{equation}
Again, using two Lagrange multipliers $\lambda$ and $\gamma,$ and maximizing
the combination
\[
\sum_{\Gamma}p(\Gamma)[-\ln p(\Gamma)+\lambda+\gamma E(\Gamma)],
\]
we find that the resulting probability distribution is given by
\[
p(\Gamma)=\exp[\lambda-1+\gamma E(\Gamma)].
\]
This distribution can be used in (\ref{G_S}) to find the corresponding
entropy. Comparing this entropy with the relation (\ref{S_E_F}) below, we
conclude that the two Lagrange multipliers are
\[
\gamma=-\beta,
\]
and
\begin{equation}
\exp(\lambda-1)=1/Z(T); \label{Equilibrium_p_CE}%
\end{equation}
consequently, the equilibrium probability distribution is given by given by
(\ref{CE_Prob}), as expected.

>From now on, we only carry out equilibrium calculations.

\subsection{Justification of Using (\ref{PF}) for Small
Systems\label{Sect_Justification_Small_System}}

The free energy in the canonical ensemble is the Helmholtz free energy
\begin{equation}
F(T)\equiv-T\ln Z(T), \label{Can_FreeEnergy}%
\end{equation}
from which we can also obtain the canonical entropy $S(T)$ by using
(\ref{Can_S}). This entropy satisfies the conventional thermodynamic relation
\begin{equation}
S(T)\equiv\beta\left[  \overline{E}(T)-F(T)\right]  \label{S_E_F}%
\end{equation}
as can be easily verified by using (\ref{Can_FreeEnergy}) in (\ref{Can_S}).
>From this, we find that (at constant extensive quantities such as the
"lattice volume", numbers of residues, etc.)%
\begin{equation}
d\overline{E}=TdS+SdT+dF=TdS, \label{dE_relation}%
\end{equation}
which is the first law of thermodynamics now valid for a small system.

Let us compare the canonical entropy in (\ref{Can_S}) with the $S(T)$ given by
the Gibbsian relation (\ref{G_S}). We find that
\[
S(T)=%
{\textstyle\sum\limits_{\Gamma}}
\left[  \beta E(\Gamma)+\ln Z(T)\right]  p(\Gamma)=\beta\left[  \overline
{E}(T)-F(T)\right]  ,
\]
and is identical with the canonical entropy above in (\ref{Can_S}). The two
ways of calculating the canonical entropy give the same result even for a
small system. In other words, the Gibbsian relation (\ref{G_S}) is also valid
for a small system. This is a justification of adopting the partition function
formalism for small systems, as discussed in the previous section.

\section{Small System Microcanonical and Canonical Entropies}

\subsection{$\overline{S}(\overline{E})\geq S(\overline{E}%
)\label{Sect_Both_S_E_Inequality}$}

It should be stressed that one must always use the probability of a
conformation (usually called a microstate in statistical mechanics)
$p(\Gamma)$ in the Gibbsian relation (\ref{G_S}). In other words, one cannot
group these microstates and use the probabilities of the groups. We will
demonstrate this by an example. let us group the microstates of a given energy
together and use the probability $p(E)$\ to construct the combination%
\begin{equation}
\Sigma\equiv-\sum_{E}p(E)\ln p(E), \label{Energy_S}%
\end{equation}
which looks similar to the combination in the Gibbsian relation (\ref{G_S}).
It is easily seen that%
\begin{equation}
\Sigma=S(T)-\overline{S}(T), \label{Energy_S1}%
\end{equation}
where%
\begin{equation}
\overline{S}(T)=%
{\textstyle\sum\limits_{E}}
S(E)p(E) \label{Ave_S}%
\end{equation}
is the thermodynamic average entropy, so that $\Sigma$ does not give $S(T)$.
Moreover, since $\Sigma$ is, in general, not zero, $S(T)$ in (\ref{Can_S}) or
(\ref{G_S}) is not the same as the thermodynamic average entropy $\overline
{S}(T)$ in (\ref{Ave_S}). Thus, the concept of microstates (or conformations
in the context of proteins) is crucial in using the Gibbsian relation
(\ref{G_S}) to obtain the canonical entropy.

An important consequence of (\ref{Energy_S}) is the following. Since $0\leq
p(E)\leq1,$ it is evident that $\Sigma\geq0.$ Hence,
\begin{equation}
S(T)\geq\overline{S}(T). \label{Both_S}%
\end{equation}
>From (\ref{Ave_S}), we conclude that$\ \overline{S}(T)\geq0,$ since it is an
average of a non-negative quantity $S(E).$ Thus,%
\[
S(T)\geq0.
\]
This then proves that the free energy $F(T)$ is a monotonically decreasing
function of $T$ even for a small system.

In the thermodynamic limit ($M\rightarrow\infty$), $\Sigma$ will approach zero
from above, as the sum in (\ref{Energy_S}) is replaced by a single term
corresponding to $E=\overline{E}(T),$ for which $p($ $\overline{E})=1.$ Thus,
$S(T)\ $approaches $\overline{S}(T)$ from above$.$

Both $S$ and $\overline{E}$\ are \emph{continuous function} (except possibly
at a phase transition, which is not relevant here as we are dealing with a
finite protein) of the continuous variable $T.$ We now wish to express the
canonical entropy $S(T)$ as a function of the average energy $\overline{E}$.
To do so, we recognize that the derivative $\partial\overline{E}/\partial T$
is non-negative; see (\ref{E_Fluc}). Thus, it can be \emph{inverted} to
express $T$ as a function $T(\overline{e}),$ where $\overline{e}=\overline
{E}/M$. This allows us to express $S(T)$ as an explicit function $\overline
{S}(\overline{E})\equiv S\left[  T(\overline{e})\right]  $\ of $\overline{E}.$
($\overline{S}(\overline{E})$ should not be confused with $\overline{S}(T)$ in
(\ref{Ave_S}), as the two have different arguments.) The entropy $\overline
{S}(\overline{E})$ can be thought of as the \emph{canonical equivalence} of
the microcanonical entropy $S(E).$ However, they are two \emph{different}
quantities for small proteins. In the first place, $S(E)$ is a discrete
function since $E\ $is discrete, while $\overline{S}(\overline{E})$ is a
continuous function of the continuous variable $\overline{E}$. In the second
place,$\ $%
\begin{equation}
\overline{S}(\overline{E})\geq S(\overline{E}), \label{Both_S_E}%
\end{equation}
the equality holding as $M\rightarrow\infty$ \cite{Guj0412548}. This
inequality should not be confused with the above inequality (\ref{Both_S}). To
demonstrate (\ref{Both_S_E}), let us assume that $E=\overline{E}$ is one of
the energies in the sum in the PF (\ref{PF}). We then rewrite
\[
\overline{S}(\overline{E})\equiv S(T)=\ln Z(T)+\overline{E}/T,
\]
and evaluate $\exp[\overline{S}(\overline{E})]$: \
\begin{equation}
\exp[\overline{S}(\overline{E})]=W(\overline{E})+%
{\textstyle\sum\limits_{E\neq\overline{E}}}
W(E)e^{-\beta(E-\overline{E})}. \label{ContW}%
\end{equation}
The sum above is non-negative; hence, $\exp[\overline{S}(\overline{E})]\geq
W(\overline{E}),$ which proves (\ref{Both_S_E}) above.\ The difference between
$\overline{S}(\overline{E})=S(T)$ and $S(\overline{E})$ is due to the last
term in (\ref{ContW}), which is expected to vanish as $M\rightarrow\infty.$

In case, $\overline{E}$ is not one of the energies in the sum, we can use a
suitable interpolation to define $\overline{W}(\overline{E}),$ without
affecting the conclusion. We give a simple interpolation scheme to show this.
Let $\overline{E}$\ lie between two allowed energies $E_{1}$ (should not be
confused with $E_{1}$ introduced earlier as the highest allowed energy in the
model) and $E_{2}>E_{1}$\ in the microcanonical energy spectrum, and introduce
$\delta E=E_{2}-E_{1}>0.$\ Let $\overline{E}=E_{1}+x\delta E,$\ $E_{2}=$
$\overline{E}+(1-x)\delta E,S(E_{1})=S(\overline{E})-xS^{\prime}\delta
E,S(E_{2})=S(\overline{E})+(1-x)S^{\prime}\delta E,$ where $S^{\prime}%
\equiv\left[  S(E_{2})-S(E_{1})\right]  /\delta E.$ The two terms in
$\exp[\overline{S}(\overline{E})]$ in (\ref{ContW}) containing $E_{1}$ and
$E_{2}$ are%
\begin{align*}
&  W(E_{1})e^{-\beta(E_{1}-\overline{E})}+W(E_{2})e^{-\beta(E_{2}-\overline
{E})}\\
&  =W(\overline{E})\left[  e^{x\alpha}+e^{-(1-x)\alpha}\right]  ,
\end{align*}
where $\alpha=\beta\left(  1-TS^{\prime}\right)  \delta E,$ as can be easily
seen. Assuming $\alpha>0,$ we can write
\begin{equation}
e^{x\alpha}=1+\gamma,~~\gamma>0. \label{e_x_alpha}%
\end{equation}
It is then obvious that we can express $\exp[\overline{S}(\overline{E})]$ as%
\begin{align*}
\exp[\overline{S}(\overline{E})]  &  =W(\overline{E})+W(\overline{E})\left[
\gamma+e^{-(1-x)\alpha}\right] \\
&  +%
{\textstyle\sum\limits_{E\neq E_{1},E_{2}}}
W(E)e^{-\beta(E-\overline{E})}\\
&  \geq W(\overline{E}),
\end{align*}
which proves (\ref{Both_S_E}) for this case also. For $\alpha<0,$ we use
$e^{-(1-x)\alpha}$ on the left side of (\ref{e_x_alpha}), and proceed the same
way with a similar conclusion. The same conclusion also remains valid for
$\alpha=0$. Thus, we have succeeded in establishing (\ref{Both_S_E}) in all cases.

The above proof does not depend on the discrete nature of the energies in ME;
thus, it is also valid for continuum models though more care is needed. We
show in Fig. \ref{F2} the entropies per residue
\[
s(e)\equiv(1/M)S(E)
\]
by symbols, and%
\[
\overline{s}(\overline{e})\equiv(1/M)\overline{S}(\overline{E})
\]
by curves, for the\ three models for the case $M=24$ as a function of the
discrete variable $e\equiv E/M$ or $\overline{e}$ from our exact
enumeration$.$ The energy densities have been \emph{shifted} by the lowest
energy density $e_{0}\equiv E_{0}/M$ for each model separately so that all the
the curves have the same origin.

\subsection{Concavity of $\overline{S}(\overline{E})\ $and\ Its Absence in
$S(E)$}

\subsubsection{Concavity of $\overline{S}(\overline{E})$ and Thermodynamic
Stability}

\begin{figure}
[ptb]
\begin{center}
\includegraphics[
trim=0.691714in 3.124344in 1.450237in 3.210541in,
height=2.5261in,
width=3.5111in
]%
{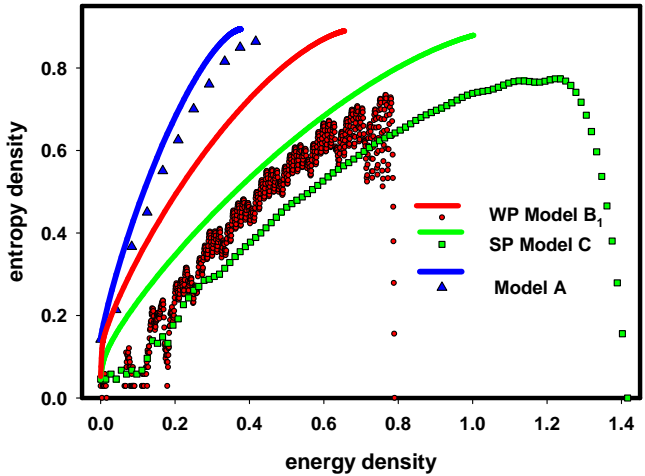}%
\caption{The canonical equivalence of the entropy $\overline{s}$ as a function
of average energy $\overline{e}$ (continuous curves), and the microcanonical
entropy (points) as a function of discreet energy $e$ for a given sequence
$(M=24)$ for the three models. We consider unrestricted conformations here.
The energy density has been \emph{shifted} by the lowest energy density
$e_{0}$ of each model so that the lowest \emph{shifted} energy density is the
same ($=0$) for all models. We notice a clear band band structure in $s$ in
the perturbed models (B$_{1}$ and C$_{1}$). The bands become more pronounced
and their separation also decreases, as $M$ increases (results not shown). We
also see that the native state is almost disjoint from the rest of the bands.
This is merely a reflection of the energy gap in the standard model A at low
energies due to finite size of the protein. }%
\label{F2}%
\end{center}
\end{figure}

We also see a distinct \emph{band structure} in $s(e)$ for the two perturbed
models (B$_{1}$ and C$_{1}$) in Fig. \ref{F2}. The band structure is related
to the nature of the perturbative interactions and has no implication for any
phase transition as we now discuss. From (\ref{E_Fluc}), we see that
\begin{equation}
\left(  \frac{\partial\overline{E}}{\partial T}\right)  \geq0, \label{Cv}%
\end{equation}
which states that the canonical heat capacity is non-negative, and is one of
the requirements of \emph{stability} of the system regardless of the size.
>From the relation (\ref{dE_relation}), it is easily seen that the canonical
entropy function satisfies the conventional thermodynamic relation
\cite{Guj0412548}
\begin{equation}
\partial\overline{S}(\overline{E})/\partial\overline{E}=1/T.
\label{T-relation}%
\end{equation}
>From (\ref{Cv}) and above, we conclude that $\overline{S}(\overline{E})$ is,
therefore, concave%
\[
\partial^{2}\overline{S}(\overline{E})/\partial\overline{E}^{2}<0
\]
\cite{note1} even for a small system; compare with (\ref{Concave_S}) for a
macroscopic system$.$ On the other hand, the microcanonical entropy need
\emph{not }be concave; see Fig. \ref{F2}, where the bands seen in $s(e)$ have
both positive and negative slopes, which is in contradiction with
(\ref{T-relation}) valid for $\overline{s}(\overline{e}).$ The non-concave
$S(E)$ does not violate the finite system thermodynamics. There is ample
evidence that the above convexity is also present in the results presented in
\cite{Sali}. The canonical entropy is the physical entropy for proteins in its
environment and remains concave in Fig. \ref{F2} as required by thermodynamics.%

\subsubsection{Convex Regions in $S(E)$}

To understand the absence of concavity, we first consider the standard model
A. The energy in this model is always negative, so there is no harm in
considering the entropy as a function of the absolute energy $\left\vert
E\right\vert =N_{\text{HH}}.$\ In all cases that we have studied,
$S(N_{\text{HH}})$ is found to be a concave discrete function. The number of
states $W(N_{\text{HH}})$ can be partitioned into $W(N_{\text{HH}}%
,\mathbf{N}^{\prime})$; see (\ref{W_HH}). In the model B, in which the
energies are weakly perturbed, $\mathbf{e}^{\prime}\simeq0;$ therefore, most
of the conformations in\ $W(N_{\text{HH}})$ have energies that are close to
$(-N_{\text{HH}});$ some of them will have energies that are outside the range
($-N_{\text{HH}}-1,-N_{\text{HH}}+1).$ The resulting $S(E)$ associated with
this $N_{\text{HH}}$ is almost concave, as seen in each of the bands in Fig.
\ref{F2}; see the mathematical fits for the two of the bands blown up in Figs.
\ref{F21}, and \ref{F22} where the minibands within each of the bands are also
evident. This then give rise to the lack of concavity or the emergence of
\emph{convexity} in the region where two nearby bands overlap. The number of
bands equals the number of possible values of $N_{\text{HH}}$ in the model A.
These convex portions of $s(e)$ should disappear and $s(e)$ should approach
$\overline{s}(\overline{e})$ from below as $M\rightarrow\infty$ $\ $%
\cite{Guj0412548}. But for small systems, the convex regions persists. The
band structure persists for all sequences that we have checked. The strongly
perturbed energies in the model C provide enough spread for each band to
strongly overlap, especially at the upper end of the energy spectrum, which
reduces the size of convex regions. Even here, we find that the band nature
survives at the upper end of the energies near the maximum; the bands at the
lower end of the energy spectrum continue to persist even for strong
perturbation. This is clear from Fig.(\ref{F2}). Thus, we are confident that
convex regions in $S(E)$ will exist in any realistic model of small proteins.
Their presence, however, does not imply any phase transition, as $\overline
{S}(\overline{E})$ is always concave. This is true even though we note from
Fig. \ref{F2}, that there is a clear gap between the bands at the lowest
energy; see also Fig. \ref{F16} for a clear evidence of such a gap near the
native state where we have shown the entropy density for the model B$_{1}$ for
low energies. The presence of bands alone and not the energy gaps between them
give rise to convexity in $S(E),$ but not in $\overline{S}(\overline{E}).$ One
does not need any energy gap for a convex $S(E)$ as was the case for the
random energy model$.$ The energy gaps between the bands in the present case
are due to the discreteness inherent in small systems. As the bands disappear
in $s(e)$ in the $M\rightarrow\infty$ limit, there will be no energy gap in
this limit$,$ as discussed earlier in Sect. \ref{Sect_Absc_Energy_Gap}.

\subsection{Behavior of $S(E)$ in its bands}

\begin{figure}
[ptb]
\begin{center}
\includegraphics[
trim=1.147154in 5.697459in 2.171282in 1.157795in,
height=2.6775in,
width=3.4065in
]%
{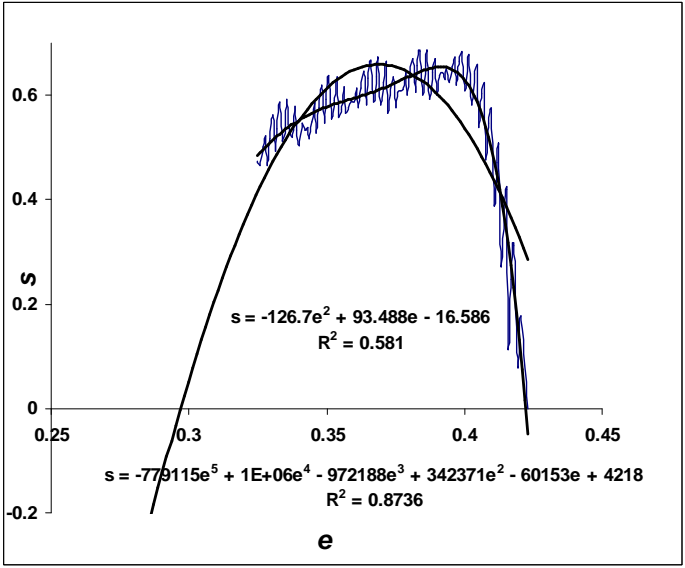}%
\caption{Entropy fit for the last band for the model B$_{1\text{ }}$with
$M=22$ and unrestricted conformations.}%
\label{F21}%
\end{center}
\end{figure}

\begin{figure}
[ptb]
\begin{center}
\includegraphics[
trim=1.316620in 5.603814in 1.186262in 0.886437in,
height=2.5261in,
width=3.4238in
]%
{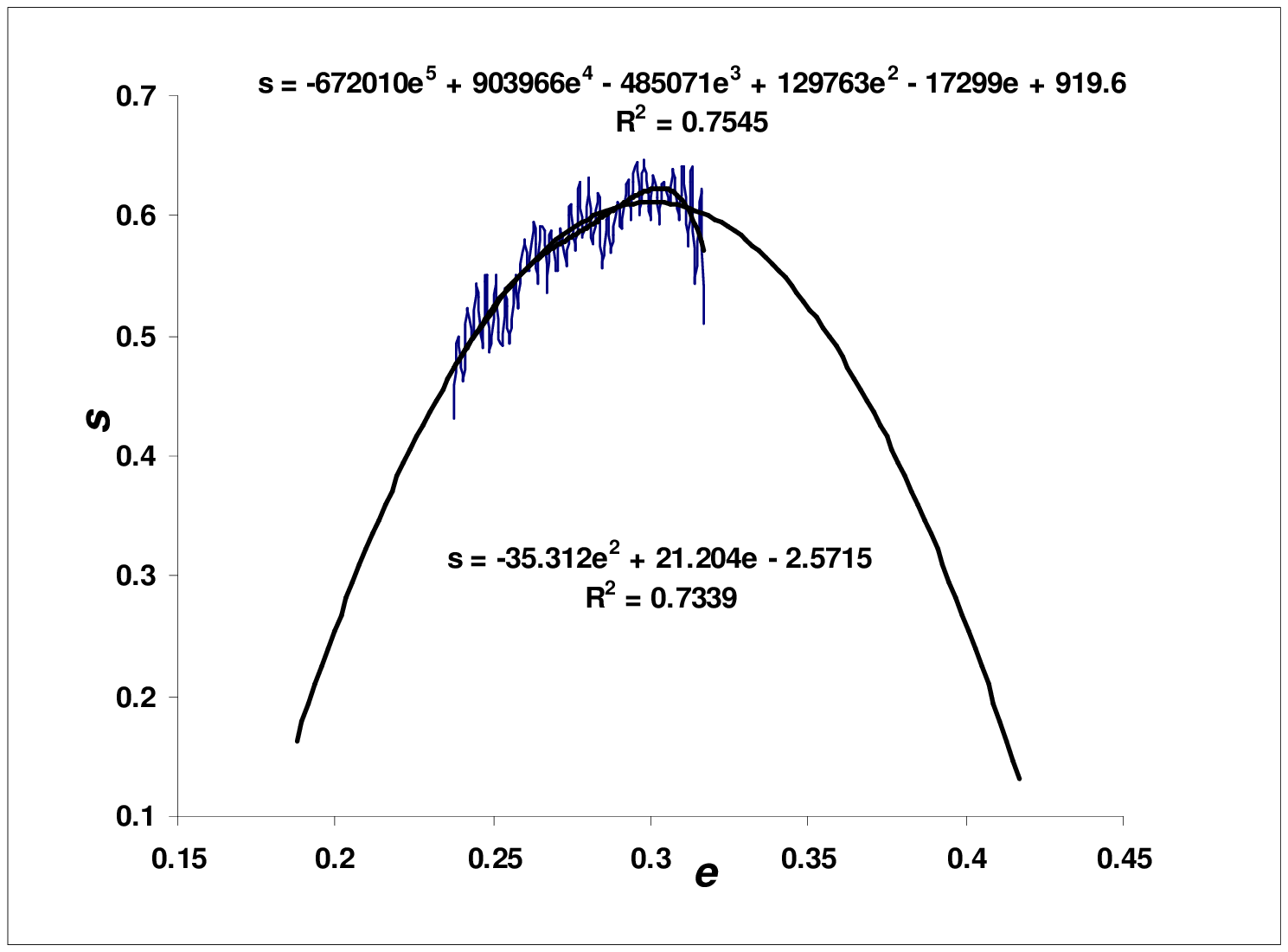}%
\caption{Entropy fit for the next to the last band for the model B$_{1\text{
}}$with $M=22$ and unrestricted conformations.}%
\label{F22}%
\end{center}
\end{figure}

Let us now investigate the behavior of $s(e)$ in these bands by finding some
smooth fits by neglecting its oscillatory pattern. We consider the two top
most bands for the weakly perturbed model B$_{1}$, which are reproduced in
Figs. \ref{F21} and \ref{F22}, respectively, along with the best quadratic and
cubic fits and their R values. It should be noted that the quadratic fit is
equivalent to the Gaussian form (\ref{Gaussian_S}), provided the coefficient
of the quadratic term is negative. Because of the nature of each of these
bands, this is true. If the linear term is positive (negative), then the most
probable energy $\widetilde{E}_{\text{b}}$\ within the band is positive
(negative). From Fig. \ref{F21}, we observe that the Gaussian fit is extremely
poor in comparison with the cubit fit; even the latter fit is not too good. On
the other hand, the result for the next band in Fig. \ref{F22} shows that both
fits are similar in their R-values and that both are poor. This is because of
the oscillating nature of $s(e)$ in the bands.\ It is interesting to note that
the cubic fit is better for the top most band than the next lower one. But
this cubic fit is not a concave function.%

%

%

\begin{figure}
[ptb]
\begin{center}
\includegraphics[
trim=-0.089351in 0.000000in 0.189324in 0.274120in,
height=2.6204in,
width=3.4765in
]%
{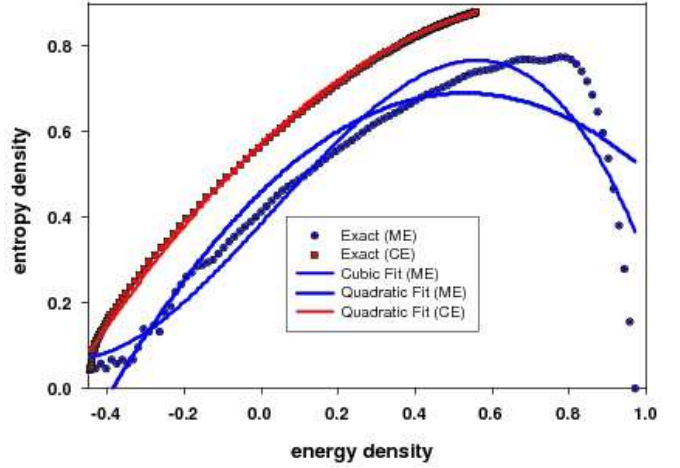}%
\caption{Exact ME and CE entropy density for the SP model C$_{1}$
(unrestricted conformations) along with quadratic and cubic fits. }%
\label{F25}%
\end{center}
\end{figure}

\subsection{Numerical Fits for $S(E)$ and $\overline{S}(\overline
{E})\label{Sect_Gaussian_Fit}$}

In Fig. \ref{F25} we reproduce the ME\ and CE entropy density for the strongly
perturbed C$_{1}$ model (unrestricted conformations); we also show the best
quadratic and cubic fits along their R values. The fits for the ME entropy
are
\begin{align*}
s  &  =0.3852+0.9875e+0.1136e^{2}-1.182e^{3}\ \ (R=0.9594),\\
s  &  =0.4592+0.8717e-0.8221e^{2}\ \ (R=0.9197).
\end{align*}
It is clear that between the two, the cubic fit is the better fit overall.
However, both fits are extremely poor at the low energy end, which is the
relevant range for the folded or the native state. Thus, the quadratic fit,
which as said above is the Gaussian form (\ref{Gaussian_S}), is not suitable
to describe the ME entropy. Moreover, the quadratic fit gives rise to the
vanishing of the entropy at an energy higher than the lowest allowed energy
$e_{0}$, which is most certainly not true of the exact entropy, which is
everywhere non-negative ($e\geq e_{0}$). It is not possible for the entropy to
vanish at the lower end of the energy as $M\rightarrow\infty$, as there is not
an energy gap in our model; see Sect. \ref{Sect_Absc_Energy_Gap}. Hence, to
conclude an ideal glass transition based on the vanishing of the Gaussian fit
of the ME entropy is \emph{misleading} even for small proteins. Even the
prediction of an energy gap is misleading as there are several energy levels
between the energy $E_{\text{F}}$ and $E_{0}.$ The presence of a convex region
in the entropy $s\left(  e\right)  $ in both fits has nothing to do with any
phase transition as the canonical entropy does not show any signature of a
transition, as is clear from the figure.

The fits for the CE entropy are given by%
\begin{align*}
\overline{s}  &  =0.8236+0.9162\overline{e}-2.6895\overline{e}^{2}%
-0.27950\overline{e}^{3}\ (R=0.9843),\\
\overline{s}  &  =0.5711+0.8511\overline{e}-0.5434\overline{e}^{2}%
\ (R=0.9994).
\end{align*}
For the CE case, the quadratic fit is the better one;\ however, both fail in
the low energy range. Thus, these fits also do not do justice to the native
state. It should be noted, however, that both fits yield a positive CE entropy
at all energies $e\geq e_{0}$.

\begin{figure}
[ptb]
\begin{center}
\includegraphics[
trim=-0.090055in 0.045421in 0.201407in 0.200611in,
height=2.5477in,
width=3.2503in
]%
{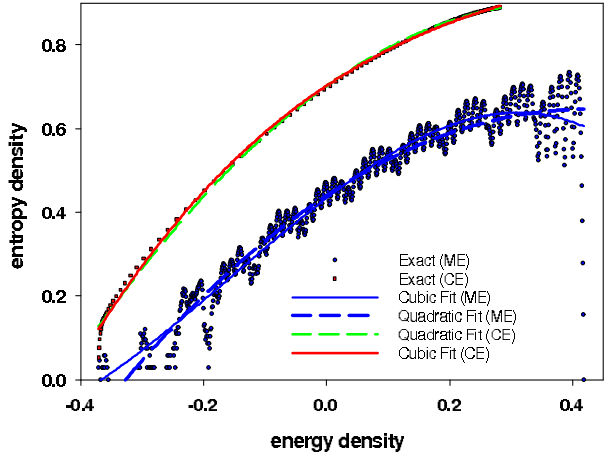}%
\caption{Exact ME and CE entropy density for the WP model B$_{1}$
(unrestricted conformations) along with quadratic and cubic fits. }%
\label{F26}%
\end{center}
\end{figure}

In Fig.(\ref{F26}), we show the entropies and their best quadratic and cubic
fits for the weakly perturbed model B$_{1}$. For the ME case, we have%
\begin{align*}
s  &  =0.4341+1.1203e-0.8863e^{2}-1.9502e^{3}\ (R=0.9663),\\
s  &  =0.4406+0.9705e-1.1454e^{2}\ (R=0.9623).
\end{align*}
Again, both fits are poor at the lower energy range; otherwise, they are very
similar in their R-values. The Gaussian fit again predicts an energy gap, just
as was the case for the strongly perturbed model in Fig.\ref{F25}, and has no
significance for any folding transition. The exact discrete entropy
$s(e)$\ does show an energy gap between the two lower bands, which is expected
to disappear in the limit $M\rightarrow\infty$. The prediction of negative ME
entropy from the Gaussian fit is unphysical as above for the same reason, and
cannot be taken seriously.

For the CE, the fits are:%

\begin{align*}
\overline{s}  &  =0.7029+0.9772\overline{e}-1.2906\overline{e}^{2}%
-0.8344\overline{e}^{3}\ (R=0.9995),\\
\overline{s}  &  =0.7020+1.0448\overline{e}-1.3558\overline{e}^{2}%
\ (R=0.9994).
\end{align*}
The behavior of the two fits are similar to that for the strongly perturbed
case above. Once again, the ME entropy fits give non-negative entropy for all
energies $e\geq e_{0}$.%

\section{Energetics Effects on Densities and Specific Heat}

\subsection{Densities and Energetics}%

\begin{figure}
[ptb]
\begin{center}
\includegraphics[
trim=0.602093in 3.045597in 1.444534in 3.210541in,
height=2.4846in,
width=3.2445in
]%
{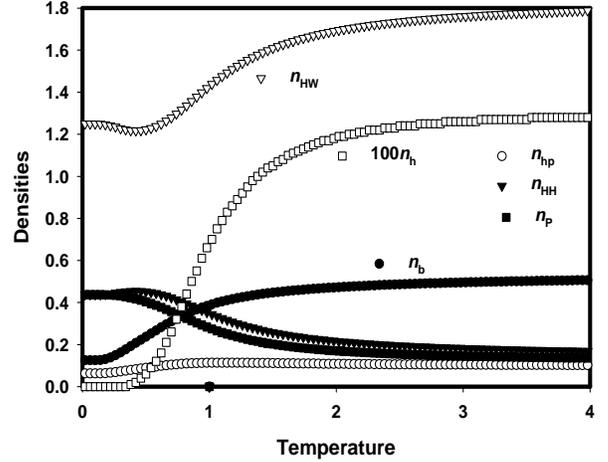}%
\caption{Densities for $M=16$ with only H residues (unrestricted
conformations). The energetics belong to the strongly perturnbed case, with
only $e_{\text{b}}=1,$ but all other elements in $\mathbf{e}^{\prime}$ are
zero.}%
\label{F10}%
\end{center}
\end{figure}

To understand the effect of bending only due to semiflexibility, we consider
an unrestricted protein with $M=16$ that belongs to the strongly perturbed
case; see Fig. \ref{F10}. The only non-zero energies are $e_{\text{b}}=1,$ and
$e_{\text{HH}}=-1.$ All other energies in $\mathbf{e}^{\prime}$ are zero.
Thus, we are considering the model C$_{2}.$ Furthermore, all residues are H;
there is no P residue. This means that $n_{\text{PP}}=n_{\text{PW}}=0$ at all
temperatures. There is a unique native state in which the protein bends around
in a double strand
\[
\text{{\normalsize RRRRRRRDLLLLLLL}}%
\]
with $N_{\text{b}}=2,$ and $N_{\text{HH}}=7$ so that it has the energy $E=-5$.
Other quantities of interest are: $N_{\text{p}}=7,N_{\text{hp}}=1,N_{\text{h}%
}=0,N_{\text{HW}}=20,$ and $N_{\text{PH}}=0.$ The fact that the entire protein
is exposed to the water is understandable, as there is no interaction with
water in this case. This is the state of the protein at $T=0.$ As $T$ is
raised, the various densities behave as shown in Fig. \ref{F10}. It is not
surprising that $n_{\text{HH}}$ mostly decreases monotonically due to the
penetration of water inside the protein.

What one notices from the figure is that around $T\simeq0.5,$ there is not
only a sudden rise in the helix density, a sudden drop in the parallel bond
pair and HH-contact densities, but also a minimum in the HW-contact density.
This minimum is due to the bending penalty as we now discuss. As said above,
the native state corresponds to a double strand $(N_{\text{HH}}=7,$ and
$N_{\text{HW}}=20).$ This state does not have the maximum HH-contact, which
happens in a compact state ($N_{\text{HH}}=9$). However, the compact state
corresponds to at least $4$ additional bends ($N_{\text{b}}=6$), so its energy
is at least $E=-3,$ and is higher relative to the native state. At higher
temperatures, the compact state, which has higher entropy, becomes more
stable. This heuristically justifies the dip in $n_{\text{HW}}.$ While it is
not noticeable in the figure, $n_{\text{HH}}$ has a maximum ($=0.4539$) at
$T=0.42,$ exactly where the dip is in $n_{\text{HW}}$ ($=1.2172$ at $T=0.42$).

\begin{figure}
[ptb]
\begin{center}
\includegraphics[
trim=0.701491in 3.036020in 1.539859in 3.217990in,
height=2.4855in,
width=3.3347in
]%
{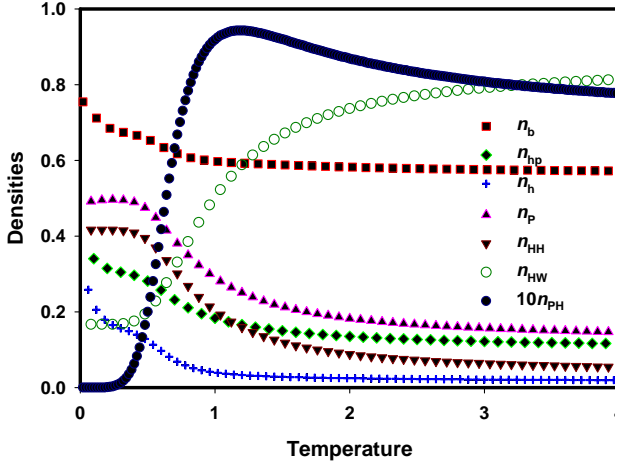}%
\caption{Densities for $M=24$ protein (unrestricted conformations). The
energetics belong to the weakly perturnbed case. The sequence is prseset to
the repeatation of PHHP.}%
\label{F11}%
\end{center}
\end{figure}

To understand the effects of the energetics better, we now give the results
for a $M=24$ protein with a fixed sequence $\chi_{0}$. We consider the weakly
perturbed model B$_{1}.$ As said above, there are two native states related by
the symmetry transformation (\ref{Conf_Trans}). In the native state, we have
$\mathbf{N}=(18,12,9,7,10,4,0),$ and $E=-446/50.$ The results for the
densities as a function of $T$ are presented in Fig. \ref{F11}. We observe
that the rate of $n_{\text{PH}}$ rise is maximum around $T=0.58;$ in the
neighborhood of this temperature, almost all densities in Fig. \ref{F11} have
some unusual behavior. For example, $n_{\text{HH}}$ has a rapid drop around
this temperature. Other densities seem to have a plateau around this
temperature. As a matter of fact, all densities have an inflection point
around this temperature. \ \ \ \ \

\subsection{Shifted Energy, Excitations, and
Energetics\label{Sect_Excitations_NativeState}}

An interesting property of the three models should be noted from the above
Fig.(\ref{F2}). The three entropies $\overline{s}(\overline{e})$ are drawn in
such a way that the upper end of each of them corresponds to $T=4.0.$ What we
see is that the corresponding \emph{shifted} energies in the three models
satisfy the following inequality:%
\begin{equation}
\overline{E}_{\text{C}}(4.0)-\overline{E}_{\text{C}}(0)>\overline{E}%
_{\text{B}}(4.0)-\overline{E}_{\text{B}}(0)>\overline{E}_{\text{A}%
}(4.0)-\overline{E}_{\text{A}}(0). \label{EnergyShift_Relation}%
\end{equation}
Thus, at high temperatures, the excess energy above the native state of a
given model is highest for the strongly perturbed model and lowest for the
unperturbed model. This should not be taken as to mean that the heat capacity
of the strongly perturbed model is the highest. We will return to this issue later.

\begin{figure}
[ptb]
\begin{center}
\includegraphics[
trim=0.752820in 3.074329in 1.507269in 3.178616in,
height=2.4855in,
width=3.2243in
]%
{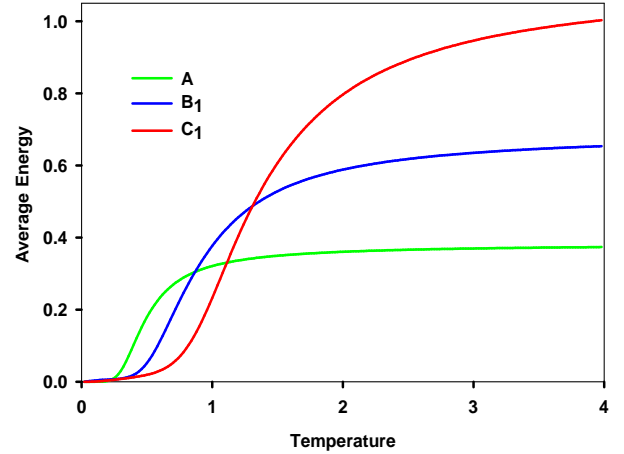}%
\caption{Shifted average energies for the three models for $M=24$
(unrestricted conformations). }%
\label{F17}%
\end{center}
\end{figure}

We see from (\ref{EnergyShift_Relation}) that the excess energy of the
strongly perturbed model is the highest at $T=4.$ In Fig. \ref{F17}, we report
the exact excess energies $\overline{E}(T)-\overline{E}(0)$ for the three
models, A, B$_{1},$ and C$_{1}$ $(M=24)$ on an infinite lattice. We see that
the behavior changes at low temperatures, where the inequality of
(\ref{EnergyShift_Relation}) is completely reversed. In other words, there are
more excitations in the unperturbed model than in the perturbed models. This
means that the net effect of the perturbations is to make the native state
more robust to perturbations: The perturbations stabilize the native state to
higher temperatures.%

\subsection{Energy Fluctuations or Specific Heat\label{Sect_SpHeat}}

\begin{figure}
[ptb]
\begin{center}
\includegraphics[
trim=0.624906in 3.042405in 1.446978in 3.212669in,
height=2.4846in,
width=3.4307in
]%
{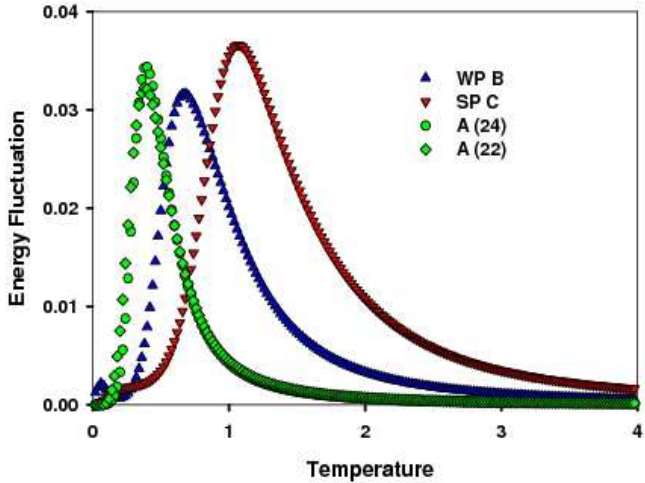}%
\caption{Energy fluctuations $\overline{(\Delta e)^{2}}$ for the three models
for $M=24$ (unrestricted conformations)$.$ For comparision,we also show the
fluctuation in the standard model for $M=22.$ We clearly see that the
fluctuations become stronger as $M$ increases, but the postion of the peak
does not shift much. }%
\label{F15}%
\end{center}
\end{figure}

We report the energy fluctuations in Fig. \ref{F15} for the three models
(unrestricted $M=24$ protein). The fluctuation is related to the specific heat
in the model; see (\ref{E_Fluc}). The peaks in these fluctuations suggest
strong fluctuations due to cooperativity in the models and are located at the
inflection points in the average energies. As is known, these peaks usually
provide a clue to an impending thermodynamically sharp transition in the
thermodynamic limit. To understand such a claim better, we also report in the
same figure the energy fluctuation for the unrestricted protein ($M=22)$ in
model A. The peak of this fluctuation is somewhat lower in height than the
corresponding peak for $M=24,$ thus suggesting that the peak height has
increased with the protein size $M$. Standard statistical mechanical arguments
require the energy fluctuations in the energy density to decrease with the
size $M$ for macroscopic systems as follows:%
\[
\overline{(\Delta e)^{2}}\propto1/M.
\]
Thus, the fluctuations behave differently for small systems. The increase,
however, is not very much, suggesting that the peaks may not diverge as will
be the case for a continuous folding transition. We expect the folding
transition to be a discontinuous one in the thermodynamic limit. However, more
work is needed to settle this point.

The locations of the peak for the weakly perturbed model is at higher
temperatures than the temperatures around $T=0.58,$ where the densities show
unusual behavior. This is most probably due to the finite size effects, and
should not be surprising.%

\section{Conformational Space and Distance\label{Sect_Distance}}

\subsection{Distance Matrix}

\subsubsection{Conformations or Microstates and Configurational Space $%
\mathbb{C}
$}

For monomeric systems, in which each monomer is treated as a particle, the
energy landscape is easy to characterize. One labels the monomers
$\alpha(=1,2,...,M)$ so that each monomer has a unique index $\alpha$. Then
one considers their positions $\mathbf{r}^{(\alpha)}.$ The \emph{ordered} set
\[
\mathbf{R\equiv}\left\{  \mathbf{r}^{(1)},\mathbf{r}^{(2)},\mathbf{r}%
^{(3)},...,\mathbf{r}^{(M)}\right\}
\]
specifies a point in the $3M$-dimensional configuration space $%
\mathbb{C}
$, and the energy $E$ associated with this configuration then defines the
energy landscape in a $(3M+1)$-dimensional hyperspace. As only the ordered set
$\mathbf{R}$\ is used, permutation of particles positions is not allowed.
Thus, each point in the configuration space represents a \emph{distinct}
microstate of the system. The ordered nature of the set $\mathbf{R}$ also
takes into account for the connectivity of the protein: a residue occupying
the lattice site $\mathbf{r}^{(k)}$ is connected to its neighboring residues
located at lattice sites $\mathbf{r}^{(k-1)}$ (for $k>1$) and $\mathbf{r}%
^{(k+1)}$ (for $k<M$). To see this most easily, we proceed as follows. We take
the C-terminus of the protein to be the starting point of the sequence. We
index the starting point as the first residue, which is used to root the
protein. Each successive residue in the sequence is, hereafter, given an index
increasing by one, until the last residue is given the index $M$. The location
of a site on the lattice is also given by a doublet $\mathbf{r}=(x,y)$ with
the location of the root given by the doublet $(0,0).$ Because of the choice
of the lattice spacing ($a=1$), the coordinates $x,y$ are integers. The
conformation of the protein is uniquely given by the ordered sequence of the
doublets $\mathbf{R\equiv}\left\{  \mathbf{r}^{(\alpha)}\right\}  ,$ where the
residue $\alpha(=1,2,...,M)$ is located at the lattice site $\mathbf{r}%
^{(\alpha)}.$ We also require the first bond of the protein to be in a fixed
direction. Each ordered sequence $\mathbf{R}$\ specifies a protein
conformation, a microstate, $\mathbf{\Gamma}$\ uniquely. There are altogether
$W$ distinct conformations or microstates. In the following, we will also use
state to simply refer to a microstate or a conformation.

\subsubsection{Distance between Conformations}

The \emph{distance} between two conformations $\mathbf{R}$ and $\mathbf{R}%
^{\prime}\mathbf{\equiv}\left\{  \mathbf{r}^{\prime(\alpha)}\right\}  $ is
defined here to be the Euclidean distance%
\[
d(\mathbf{R},\mathbf{R}^{\prime})=\sqrt{\sum\limits_{\alpha=1}^{M}%
[\mathbf{r}^{(\alpha)}-\mathbf{r}^{\prime(\alpha)}]^{2}.}%
\]
The distance provides useful information not only about the topology of the
energy landscape but may also be relevant for the dynamical description of the
folding process (even though we are not presently interested in the dynamics)
by introducing the concept of a \emph{neighborhood} of a point in the
conformation space $%
\mathbb{C}
$: two conformations are \emph{neighbors} or are \emph{connected} in $%
\mathbb{C}
$ if their separation is less than or equal to some chosen distance.

\subsubsection{Distance or Neighborhood Matrix}

The distance $d(\mathbf{R},\mathbf{R}^{\prime})$ can be used as an element to
define a $W\times W$ \emph{distance} or \emph{neighborhood} matrix
$\mathcal{D}\mathbf{,}$ whose diagonal elements are the only elements that are
$0\mathbf{.}$ All other elements are non-zero. Thus, $\mathcal{D}$ is not
going to be a sparse matrix. The distance between a compact conformation and a
completely extended conformation will be among the largest. The shortest
distances will usually be between two conformations that differ in a few
elements. For example, assume that only the elements $\mathbf{r}^{(M)}$ and
$\mathbf{r}^{\prime(M)}$ differ. The two elements can only differ in one of
its components, and that too by only one lattice spacing. Thus, the distance
between these two conformations will be $1.$ It is also possible that two
conformations differ in only one interior element at the position $k\neq M.$
The vectors $\mathbf{r}^{(k)}$ and $\mathbf{r}^{\prime(k)}$ must differ in
each of their components by $1.$ Hence, the distance between these
conformations will be $\sqrt{2}.$

\subsubsection{Native (0) and Stretched (S) States}

As $W$ is usually a large number, it is not possible to study the entire
matrix $\mathcal{D}$. Therefore, we will consider the distance of each
conformation from two selective conformations, viz. the native state (to be
denoted by 0 in the following) and the completely \emph{stretched state }(to
be denoted by S in the following); the latter is the conformation in which the
protein is given by the string of only {\normalsize R }steps
\[
\text{{\normalsize RRR...}}{\normalsize .\ }%
\]
so that the conformation is completely in the horizontal direction. If the
native state is not unique, we pick the first one of the generated native
states. The stretched conformation is unique in that it does not depend on the
energetics. On the other hand, the native conformation depends strongly on the
energetics and, therefore, is not unique as far as different energetics are
concerned. This feature makes the stretched conformation a desirable reference
state. This state can be used to compare proteins with different energetics.
We denote the two distances by $d_{\text{0}}(\mathbf{R})$ (from the native
conformation) and $d_{\text{S}}(\mathbf{R})$ (from the stretched
conformation), respectively. In most cases of interest, there is a unique
native state for a given energetics. It is the standard model which invariably
gives rise to degenerate native states.

Let the set $\mathbf{R}_{l}\mathbf{\equiv}\left\{  \mathbf{r}_{l}^{(\alpha
)}\right\}  $ denote the two reference conformations $\left(  l=\text{0,S}%
\right)  $, and%
\[
d_{l}(\mathbf{R})=\sqrt{\sum\limits_{\alpha=1}^{M}[\mathbf{r}^{(\alpha
)}-\mathbf{r}_{l}^{(\alpha)}]^{2}}%
\]
the distance of some conformation $\Gamma$ specified by the set $\mathbf{R}$
from $\mathbf{R}_{l}$.\ This distance of a conformation of energy
$E(\mathbf{R})$ gives information about how close that conformation is to the
native state. Thus, we can classify each conformation by its distance $d_{l}$
and energy $E$ and present them in a two-dimensional plot as in Figs.
\ref{F4},\ref{F5},\ref{F6},\ref{F7},\ref{F8},\ref{F27} and \ref{F9}. In all
these plots, we have shifted the energy so that the native state energy is at
$0,$ so that we can compare the configuration space $%
\mathbb{C}
$ of proteins with different energetics$.$ Moreover, we only consider one of
the native states if there are several native states to save computational
time. In this sense, our results are not complete in such cases. Therefore, we
also present the result for a weakly interacting $M=16$ protein of a sequence
for which there exists only one unique native state so that we can compare
this complete case with the incomplete case. We will find that there is no
dramatic difference.

\subsubsection{Reduction of $%
\mathbb{C}
$ to a 2-dimensional plane $%
\mathbb{C}
_{2l}$}

The use of the two reference states will provide us with two distinct but
partial perspectives of the configuration space $%
\mathbb{C}
$ by projecting it on a lower dimensional space$.$ Let us consider the
perspective of $%
\mathbb{C}
$ while looking at it from the native state. The projected plane is denoted by
$%
\mathbb{C}
_{2\text{0}}.$Imagine the energy distribution of conformations that are a
distance $d_{\text{0}}$ from the native state. All these states are on a
hypersphere of radius $d_{\text{0}}$ and have various energies$.$ Let us
further coalesce all of the conformations of a given energy $E$\ that lie on
this hypersphere to a single point. We will use $W(d_{\text{0}},e)$ to
represent the number of these conformations associated with the single point
in $%
\mathbb{C}
_{2\text{0}}$. Such a transformation allows us to transform $%
\mathbb{C}
$ to a two-dimensional surface $%
\mathbb{C}
_{2\text{0}}$ on which a point is represented by ($d_{\text{0}},e$)$.$ On such
a plane, a constant energy line represents the \emph{equipotential}
conformations at various distances from the native state. All these
conformations are at the same height (from the native state) in the energy
landscape. A constant $d_{\text{0}}$ line represents all conformation with
various energies that lie on a hypersphere centered at the native state. A
similar reduction from $%
\mathbb{C}
$ to the two-dimensional surface ($d_{\text{S}},e$) provides another
perspective of the energy landscape. We will use $W(d_{\text{S}},e)$ to
represent the number of conformations associated with the single point in the
above coalescing on the ($d_{\text{S}},e$) plane. The projected plane is
denoted by $%
\mathbb{C}
_{2\text{S}}.$

It is obvious that
\begin{equation}
W(e)\equiv\sum_{d_{l}}W(d_{l},e),\text{ }l=\text{0,S,} \label{W_sum}%
\end{equation}
so that the two perspectives only differ in the way $W(e)$ is partitioned into
$W(d_{l},e)$\ by the distance $d_{l}$. The total number of microstates $W(e)$
remains the same in the two perspectives. In addition, the allowed energies
also do not change in the two representations of $%
\mathbb{C}
.$

\subsection{Standard Model\label{Sect_Distance_Standard}}

\begin{figure}
[ptb]
\begin{center}
\includegraphics[
trim=0.974429in 2.929605in 1.075457in 3.519144in,
height=2.5425in,
width=3.6867in
]%
{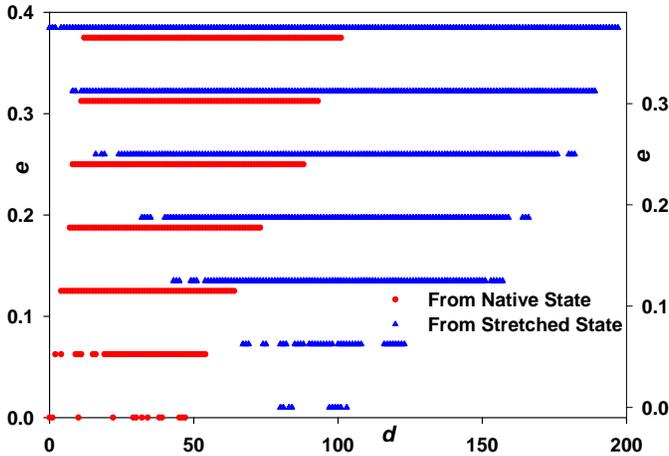}%
\caption{$E$ vs. $d_{\text{S}}$ distribution for for $M=16$ Model A protein
(unrestricted conformations). }%
\label{F4}%
\end{center}
\end{figure}

\begin{figure}
[ptb]
\begin{center}
\includegraphics[
trim=0.848959in 2.900873in 1.868199in 2.363477in,
height=2.3624in,
width=2.3952in
]%
{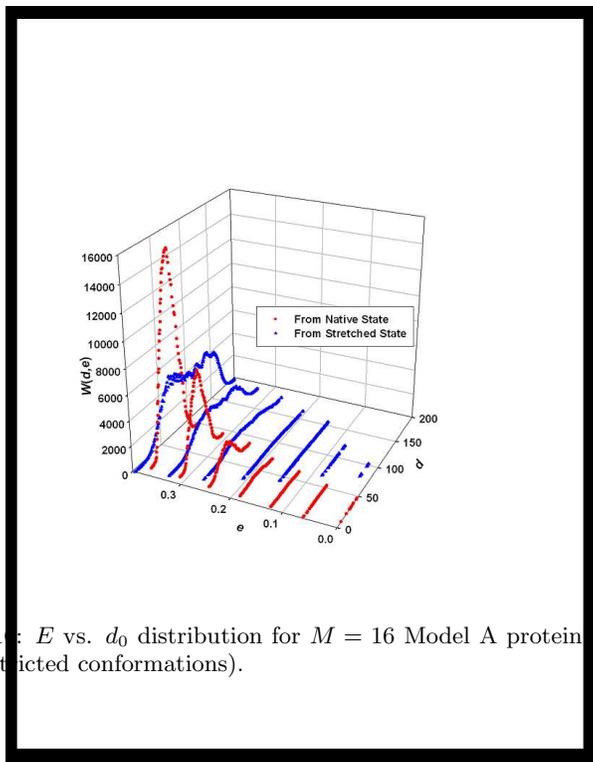}%
\caption{$E$ vs. $d_{\text{0}}$ distribution for $M=16$ Model A protein
(unrestricted conformations). }%
\label{F5}%
\end{center}
\end{figure}

The first two figures, Figs. \ref{F4} and Fig. \ref{F5} are for the standard
model. Fig. \ref{F4} shows the energy density distribution vs. $d_{\text{0}}$
(red circles:\ $%
\mathbb{C}
_{2\text{0}}$) or $d_{\text{S}}$ (blue triangles: $%
\mathbb{C}
_{2\text{S}}$)$,$ respectively; they are two possible perspectives of $%
\mathbb{C}
$. The two conformations at $d=0$ in Fig. \ref{F4}\ represents the native
conformation (red circle at $e=0$) and the extended state (blue triangle at
$e=3/8$) that are used as the origin of the distance for the two perspectives,
respectively. We observe that both the maximum and the minimum $d_{\text{0}}$
increase with $e,$ the former more so than the latter. However, while the
maximum $d_{\text{S}}$ increases with $e,$ the minimum $d_{\text{S}}$ decrease
with $e.$ We observe that the maximum $d_{\text{0}},$ to be denoted by
$d_{\text{0,max}},$ is about 120, while the maximum $d_{\text{S}},$ to be
denoted by $d_{\text{S,max}},$ is about 200. As said above, the number of
conformations $W(e)$ for a given energy, and the allowed energies ($7$ in
total) are the same for both distributions. The left axis shows $e$ for the
red circles and the left axis for the blue triangles. The left axis has been
shifted by $0.02$ so that the two colors do not overlap. In Fig. \ref{F5}, we
show the $3$-d plot $d-e-W(d,e)$ as the projected energy landscape built on $%
\mathbb{C}
_{2\text{0}}$ (red circles) and $%
\mathbb{C}
_{2\text{S}}$ (blue triangles). The energies for blue triangles has been
shifted by 0.02 so that the two symbols will not overlap. We observe that for
a given $e$, $W(d,e)$ has a single peak in $%
\mathbb{C}
_{2\text{0}}$ (red circles), while it has several peaks in $%
\mathbb{C}
_{2\text{S}}$ (blue triangles)$.$ Moreover, the peaks in $%
\mathbb{C}
_{2\text{0}}$ rise and move away from the native state as we approach higher energies.

Because of the sum rule (\ref{W_sum}) and the fact that the allowed $d$-range
of $d_{\text{S}}$ is much larger than of $d_{\text{0}},$ it is not surprising
that $W(d_{\text{0}},e)$ is much higher near its peak than $W(d_{\text{S}},e)$
near its peaks. It is clear that many high energy conformations are far from
the extended conformation of the same high energy, but most of these high
energy conformations are closer in distance from the native conformation. This
suggests a very open landscape for the standard model with the native state in
the middle, and which continues to narrow down with decreasing energy. We also
note that $%
\mathbb{C}
_{2\text{S}}$ is more symmetric than $%
\mathbb{C}
_{2\text{0}}$. We also observe that the native state is around $d_{\text{S}%
}\simeq90$ from the stretched state ($e=3/8$); see blue triangles. It follows
from the figures that there are several other states of energy $e=7/16$ that
are much closer to the native state. Indeed, there are high energy states as
close as about $d_{\text{0}}=10.$%

%

\qquad\ \qquad

We comment on some interesting features that is apparent in the figures.
Consider Fig. \ref{F4}. The best way to understand this figure is to imagine
drawing a (hyper)circle of radius $d$ with its center at the chosen native
state. Now draw a (hyper)cylinder on this circle along the energy direction.
Then the microstates that lie on this cylinder are the microstates (red
circles) that appear on the vertical line drawn at the distance $d$ (from the
native state) in Fig. \ref{F4}. All of these microstates are on the cylinder
of radius $d,$\ but the distances between them may be much different from
$d$.\ In fact, some of them may be closer than $d,$\ while others may be
farther apart.

The conformation closest to the native state in Fig. \ref{F4} is not at
$e=1/16,$ but another native state. Thus, there is no energy barrier between
these two microstates (at the same energy). However, there are other
microstates at the lowest energy that are widely separated in the \emph{radial
direction} of $d_{0}.$ The same is true of states at $e=1/16$. (At higher
energies, the microstates are almost dense in $d_{0},$ so that can be treated
as \emph{connected} in that they lie on neighboring cylinders.) Consider the
microstates at $e=1/16.$ Between various separations (in the direction of
$d_{0}$) in these states exist many higher energy states at $e=2/16$. This is
true in other figures also. Thus, this feature appears to be generic. But this
is true only of the lowest lying microstates. The microstates at higher
energies are connected in the sense note above. Thus, the energy barriers in
the radial direction exist only for low-lying states. There are no barriers in
the radial direction for highly excited states. This does not imply that there
are no barriers in other transverse directions in the configurations space $%
\mathbb{C}
.$ The implication of this for the possible dynamics can be easily appreciated
if we recognize that only local moves are possible in a suitably chosen short
duration $\tau$. During this time $\tau$, the protein can only change its
conformation to a new conformation that is nearby in distance. Thus, in the
process of folding, the protein will more efficiently move to the native state
from $e=2/16$ than from $e=1/16,$ if the former is closer to the native state
than the latter. We will not pursue this point further here as we are only
considering equilibrium properties in this work. We hope to return to this
issue in a future contribution.

\subsection{Weakly Perturbed Model}

\begin{figure}
[ptb]
\begin{center}
\includegraphics[
trim=1.724805in 2.981749in 1.882864in 3.454231in,
height=3.0554in,
width=3.2958in
]%
{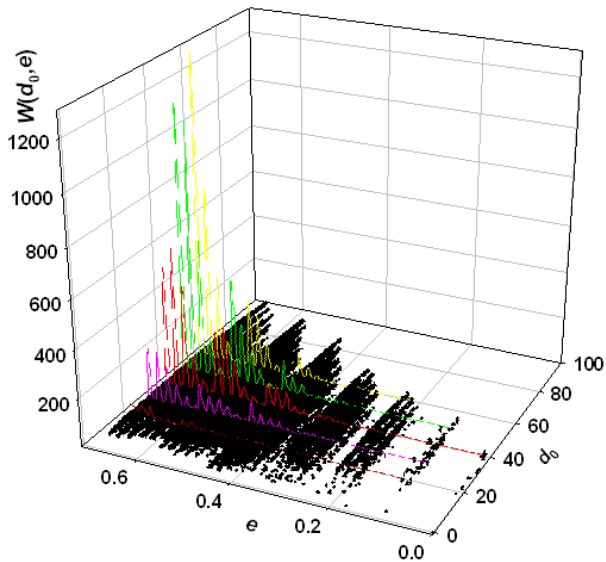}%
\caption{$E$ vs. $d_{\text{0}}$ distribution for for $M=16$ Model B$_{2}$
protein (unrestricted conformations). }%
\label{F6}%
\end{center}
\end{figure}

\begin{figure}
[ptb]
\begin{center}
\includegraphics[
trim=1.071383in 3.011544in 1.252255in 3.399960in,
height=2.354in,
width=3.2309in
]%
{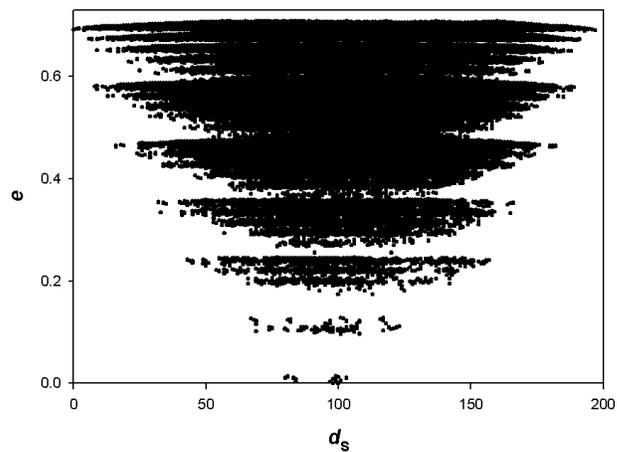}%
\caption{$E$ vs. $d_{\text{S}}$ distribution for $M=16$ Model B$_{2}$ protein
(unrestricted conformations) for the sequence $\chi_{0}$. }%
\label{F7}%
\end{center}
\end{figure}

\begin{figure}
[ptb]
\begin{center}
\includegraphics[
trim=-0.289205in 0.000000in 0.000000in 0.000000in,
natheight=4.776400in,
natwidth=6.000100in,
height=2.5114in,
width=3.2984in
]%
{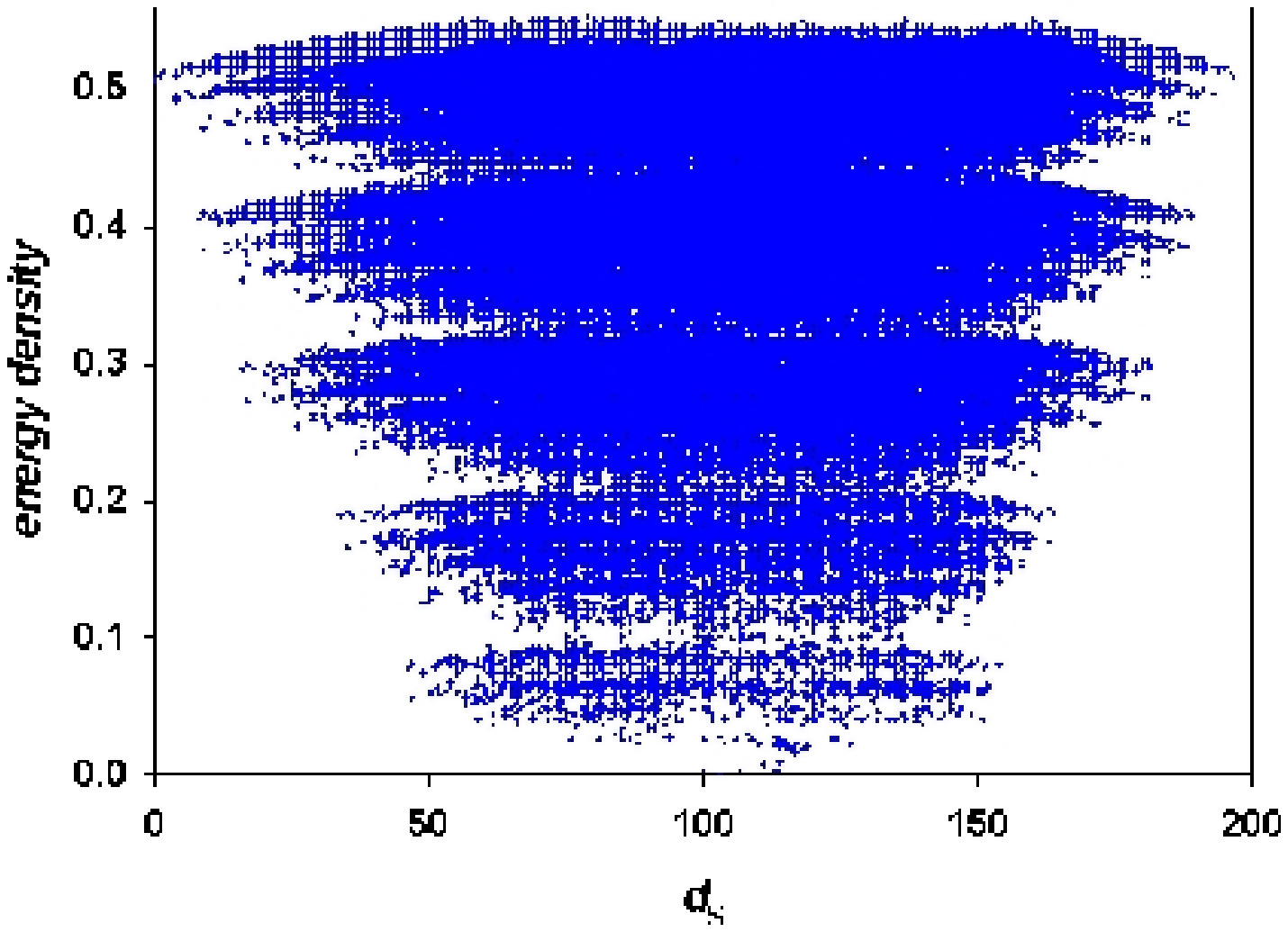}%
\caption{$E$ vs. $d_{\text{S}}$ distribution for $M=16$ Model B$_{2}$ protein
(unrestricted conformations) for the sequence $\chi:$ PPPPHHPPHHHHHHHPP. }%
\label{F27}%
\end{center}
\end{figure}

The energy density $e$ in the standard model changes by a non-zero but
appreciable amount $\Delta e=1/16.$ This can be made smaller by introducing
other energies in the model. For the model B$_{2}$, the results are shown in
Figs. (\ref{F6}, and \ref{F7}) for the sequence $\chi_{0}$. In Fig. \ref{F6},
we show $%
\mathbb{C}
_{2\text{0}}$ along with the distribution $W(d_{\text{0}},e)$ for some
selected distances $d_{\text{0}}=20,30,40,50,$ and $60.$\ In Fig. \ref{F7}, we
show $%
\mathbb{C}
_{2\text{S}}.$ The discrete band structure of Figs. (\ref{F4}, and \ref{F5})
still persists even to the higher energies, except that $\Delta e$ is smaller,
and the energy spectrum begins to look more continuous at higher energies. At
energies close to the native state, the spectrum is still very much discrete.
Otherwise, the features of the model A have not disappeared. For example, the
symmetry in $%
\mathbb{C}
_{2\text{S}}$ is still present; see Figs. (\ref{F5}, and \ref{F7}). In Fig.
\ref{F27}, we show the result for a weakly interacting Model B$_{2}$ $M=16$
unrestricted protein for the following sequence:%
\[
\chi:\text{ PPPPHHPPHHHHHHHPP.}%
\]
In this case, there is only one unique native state, which is given by the
string%
\[
\text{{\normalsize RRRDDDLUULDLULD}}%
\]
starting with the first residue. The energy of the native state has
$\mathbf{N}=(9,6,4,2,4,4,4),$ and $E_{0}=-117/56.$ However, a comparison with
Fig. \ref{F7} shows that the distributions of states in $%
\mathbb{C}
_{2\text{S}}$ for the two cases are almost the same, except at low energies.
Thus, we believe that our incomplete results are not different from the
complete results at intermediate and higher energies.

The distribution $W(d_{\text{0}},e)$ in Fig. \ref{F6} shows that it has an
oscillatory pattern and that the highest peak in it has a maximum around
$d_{\text{0}}=60,$ and $e=0.7.$%

%

%

\subsection{Strongly Perturbed Model}

\begin{figure}
[ptb]
\begin{center}
\includegraphics[
trim=0.637941in 3.096677in 1.309287in 3.270133in,
height=1.3782in,
width=2.4385in
]%
{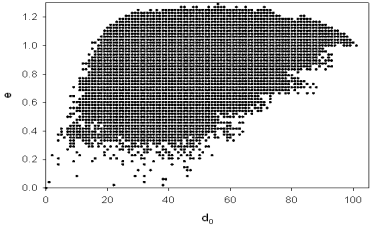}%
\caption{$E$ vs. $d_{\text{0}}$ distribution for $M=16$ Model C$_{1}$ protein
(unrestricted conformations). }%
\label{F8}%
\end{center}
\end{figure}

The projected conformation spaces $%
\mathbb{C}
_{2\text{0}}$ and $%
\mathbb{C}
_{2\text{S}}$\ for model C$_{1}$ are shown in Figs. \ref{F8} and \ref{F9},
respectively. We again see the symmetry present in the distribution of states
in $%
\mathbb{C}
_{2\text{S}}.$ The energetics is such that there is a strong mixing of levels
to the point that the clear cut band pattern is completely absent at high
energies; their discrete nature is still present near the bottom. The
energetics change the native state so that its distance from the extended
state are different in the three models.%

%

\begin{figure}
[ptb]
\begin{center}
\includegraphics[
trim=-0.145202in 0.000000in 0.000000in 0.000000in,
natheight=4.776400in,
natwidth=6.000100in,
height=2.6524in,
width=3.4091in
]%
{E_dS_C16_1.wmf}%
\caption{$E$ vs. $d_{\text{S}}$ distribution for $M=16$ Model C$_{1}$ protein
(unrestricted conformations). }%
\label{F9}%
\end{center}
\end{figure}

\subsection{Small System Energy Landscape and Convexity of $S(E)$%
\label{Sect_Distance_Convexity}}

The distribution of points in the $%
\mathbb{C}
_{2}$ plane allows us to draw certain conclusions about the form of the energy
landscape. Assume that the energy landscape is a single inverted cone of a
fixed (hyper-solid)angle. In that case, all conformations of a given energy
$E$ will have the same radial distance from the native state in $%
\mathbb{C}
$, and the energy-distance distribution in $%
\mathbb{C}
_{2\text{0}}$\ will be represented by points that lie along a single straight
line at a fixed angle in the $e-d_{0}$ plane: For each energy, all states are
collapsed into a single point on this line. This is most obviously not the
case here in any of the $%
\mathbb{C}
_{2\text{0}}$ for the three models shown here. Consider the standard model in
Fig.(\ref{F4}). We see that there are a few allowed energies at a given
distance $d_{\text{0}}$ from the native state. If we draw a hypercylinder of
radius $d_{\text{0}},$\ then this cylinder will cut the landscape at these
energies. These energies are at different angles so they lie on different
cones making different angles at its apex located at the native state. The
number of points the hypercylinder cuts the landscape is given by the sum
\[
W(d_{0})\equiv\sum_{E}W(d_{0},E),
\]
where $W(d_{0},E)$ is the number of conformations of energy $E$ that are at
the radial distance $d_{0}$ from the native state.

Because of conformational changes during folding, the folding is believed to
be governed by the multiplicity $W(E),$ which in turn governs the energy
landscape \cite{Guj0412548}:\ each point on the hypersurface represents a
conformation. The lack of concavity discovered here has a profound effect on
the shape of the landscape. It no longer narrows down as $E$ decreases. It
will be interesting to pursue this point further. This is beyond the scope of
the present work, but we hope to consider it elsewhere. It is evident, and as
discussed above, several different $\mathbf{N}$ will usually mix together for
a given $E$, except in the model (A) [in which $E=-N_{\text{HH}}$]$.$ There
will be a certain landscape topology for the standard model, which will change
with $\mathbf{e}^{\prime}$. From (\ref{W_HH}), it is evident that the
landscape will become narrower for $\mathbf{e}^{\prime}\neq0.$ At the same
time, the total "surface area" $W$ of the landscape will not change (even
though the allowed energies change) with $\mathbf{e}^{\prime}.$ It is possible
that it is this narrowing at constant $W$ that makes the approach to native
state more directional with the consequence that it would be fast. This issue
needs to be probed carefully.

Since it is CE that is relevant for a real protein in its environment, it is
the canonical multiplicity%
\[
W_{\text{CE}}(\overline{E})\equiv\exp[\overline{S}(\overline{E})]
\]
that is relevant for folding. As shown above in (\ref{T-relation}), it
continuously increases with $\overline{E},$ until we reach at infinite
temperatures. Thus, the narrowing of the landscape with non-zero
$\mathbf{e}^{\prime}$ may not be as relevant for protein folding as the
observation that $W_{\text{CE}}(\overline{E})>W(\overline{E}).$\ From
(\ref{ContW}), we observe that $W_{\text{CE}}(\overline{E})$ gets contribution
from \emph{all} conformations, not just the conformations $W$ associated with
$\overline{E}.$ In particular, it also includes the contribution from the
native state(s) though its probability is going to be small unless we are at
very low temperatures$.$ Thus, it is misleading to think that a small protein
at a given $T$ only probes average conformations $W(\overline{E})$ when in
equilibrium. As $T$ is reduced, the protein continues to probe all
conformations although the probability for conformations of lower energies
increases. It would be interesting to pursue the consequence(s) of this observation.

\section{Free Energy Landscape and $\partial S(E)/\partial E$}

\subsection{Free Energy Landscape}

Let us consider the implications of the non-concavity of $S(E)$ on the
\emph{free energy functional}
\[
F(E,T)\equiv E-TS(E),
\]
which should not be confused with $F(T)$ introduced earlier in
(\ref{Can_FreeEnergy}) and (\ref{S_E_F}). The later represents the free energy
of the equilibrium state of the system. It is a monotonic function of $T$, and
because $\overline{E}$ is monotonic in $T,$ it is also a monotonic function of
$\overline{E}.$ On the other hand, the functional $F(E,T)$ is defined at any
$T$ as a function of $E$. Thus, it is also defined for energies different from
the equilibrium energy at $T$. For a macroscopic system, it is well known that
one must minimize globally $F(E,T)$ with respect to $E$ at fixed $\ T$ to
obtain the equilibrium free energy $F(T)=F(\overline{E},T)$ evaluated at the
minimum. For continuous functions, this minimization is equivalent to
(\ref{T-relation}) for $S(E):$%
\begin{equation}
\partial S(E)/\partial E=1/T. \label{T-relation_ME}%
\end{equation}%
\begin{figure}
[ptb]
\begin{center}
\includegraphics[
trim=0.667272in 3.167974in 1.419277in 3.256299in,
height=2.348in,
width=3.3615in
]%
{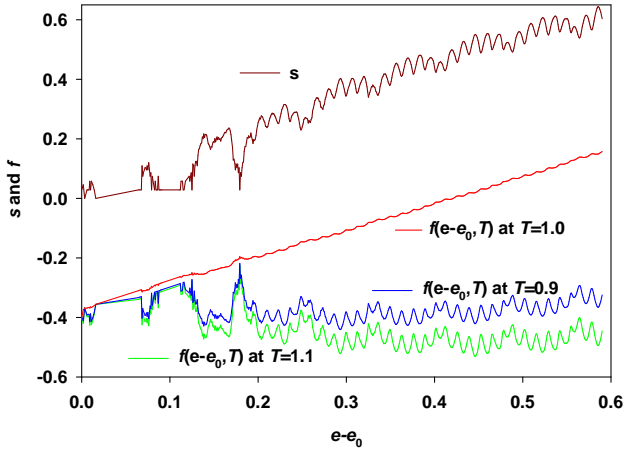}%
\caption{The free energy functional $f(e,T)\equiv F(E,T)/M$ at three different
temperatures~$T=0.9,1.0,$ and $1.1$ for the model B$_{1}.$ We also show the
entropy $s(e)$ for low energies. The free energy functional describe the free
energy landscape at a given temperature $T.$ The functions are discrete and
the curves are drawn through their points only as a guide for the eye and to
clearly show the undulations in them.}%
\label{F16}%
\end{center}
\end{figure}
It is this relation (\ref{T-relation_ME}) that was used for the Gaussian
entropy (\ref{Gaussian_S}) to obtain the Gaussian energy relation
(\ref{Gaussian_E}) earlier. The above discussion makes it clear that the
derivation given there was valid for a macroscopic system, and not for a small system.

At a given temperature, the free energy functional $F(E,T)$ describes, what is
customarily called the free energy landscape at that temperature with the
energy $E$ playing the role of a reaction coordinate of the landscape. We show
in Fig. \ref{F16} this landscape at three different temperatures $T=0.9,1.0,$
and $1.1$ for the weakly perturbed model B$_{1}$ $($unrestricted protein with
$M=24).$ We have also shown the entropy density at low energies, which is a
blow up of the entropy shown in Fig. \ref{F2}.

\subsection{Lack of Physical Significance of Global Minimum of $F(E,T)$}

The global minima of the three landscapes occur at $e=-0.0783,-0.3717,$ and
$0.0742,$ respectively. ($e_{0}=-0.3717.$) The depths of the minima are,
respectively, $f=-0.4410,0,-0.4006,$ and $-0.5309.$ That the energy of the
global minima and their depths as a function of temperature have no
thermodynamic significance is obvious when we recognize that these energies
and free energies are not monotonic in $T,$ whereas proper thermodynamics
requires them to be monotonic even for small systems. It is interesting to
compare these energies and the depth of the free energy minima with the
exactly computed average energy density $\overline{e}(T)$ and the free energy
$f(T).$\ The computed average energies at these temperatures are
$-0.0479,0.0054,$ and $0.0493,$ while the free energy densities are
$-0.6294,-0.697,$ and $-0.7694.$

What is striking is the tremendous error in the computed values and those
obtained by the application of macroscopic thermodynamic principle to small
proteins. Neither the location nor their depth are close to the exact computed
values. This is a sobering realization of the effects of the finite size of
the protein on thermodynamics.%

\begin{figure}
[ptb]
\begin{center}
\includegraphics[
trim=-0.187779in 0.000000in 0.348369in 0.719793in,
height=2.6013in,
width=3.4601in
]%
{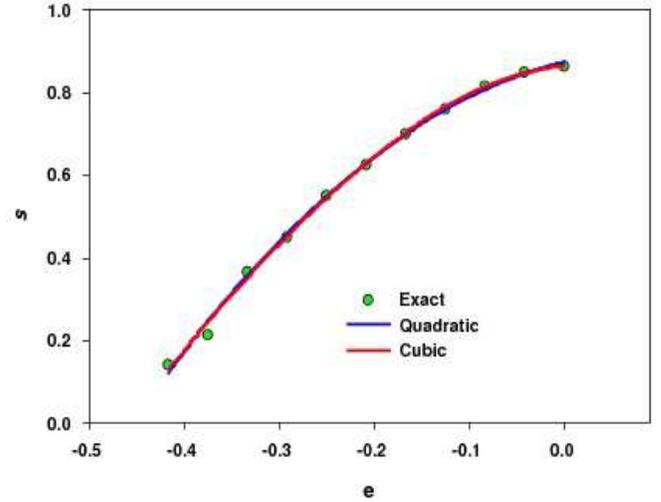}%
\caption{A quadratic and cubic fit for the ME entropy for $M=24$ Model A.}%
\label{F_23}%
\end{center}
\end{figure}
%

\begin{figure}
[ptb]
\begin{center}
\includegraphics[
trim=-0.135954in -0.048297in 0.337166in 0.336670in,
height=2.597in,
width=3.4575in
]%
{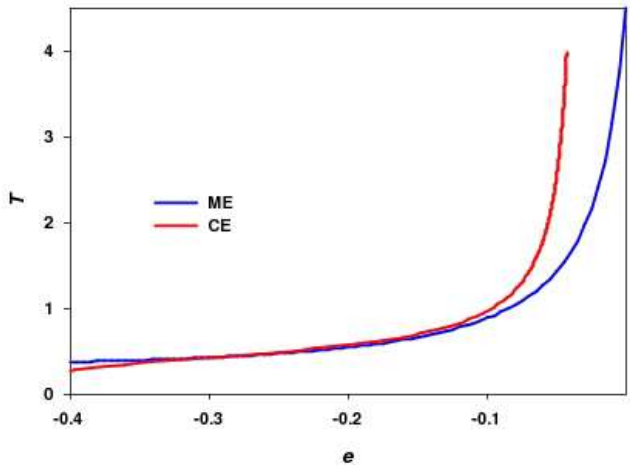}%
\caption{Calculation of $T$ using (\ref{T-relation_ME}) for the cubic fit of
the ME\ entropy in Fig.(\ref{F_23}) and by using the CE\ entropy.}%
\label{F_24}%
\end{center}
\end{figure}

\subsection{Error in Using $\partial S(E)/\partial E=1/T$%
\label{Sect_T_Relation}}

We consider the unperturbed model A for an unrestricted protein ($M=24$),
whose ME entropy density as a function of the energy density is reproduced
again in Fig. \ref{F_23}. As it is a discrete function, we cannot calculate
its derivative to see if (\ref{T-relation_ME}) is valid for small systems.
However, it is possible to find a continuous fit for $s(e).$ We have shown a
quadratic and a cubic fit in Fig. \ref{F_23} along with the $R$-values. They
are respectively%

\begin{align*}
s  &  =0.8737+0.5484e-3.0151e^{2};~\ \ R=0.9987,\ \\
s  &  =0.8650+0.2145e-5.1161e^{2}-3.361e^{3};~\ R=0.9990,\
\end{align*}
and can be used to calculate the inverse derivative $\partial E/\partial S,$
which is plotted in Fig. \ref{F_24} as the blue curve, along with the inverse
derivative $\partial\overline{E}/\partial\overline{S}$ as the red curve$.$
Here, we have used the cubic fit for the calculation of $\partial S/\partial
E$. The difference shows the error caused by using (\ref{T-relation_ME}) to
calculate the inverse temperature. The correct temperature is given by the red
curve. We find that the correct temperature from CE is lower than the
incorrect temperature from ME at lower energies, with their nature reversed at
higher temperatures. In other cases, it is also possible to observe the
opposite relation for the two ways of computing the temperature.

\section{Discussion and Conclusions}

We have considered a lattice model of a small protein as a semiflexible
copolymer in its solvent environment. The copolymer is random due to possible
forms of its residue sequence, but this randomness is considered frozen
(quenched). The model presented here is an extension of the original model of
semiflexible homopolymer due to Flory; this extended model has been
investigated recently by us. However, the model requires a very important
modification because of the heteropolymer nature of the protein. Here, we have
restricted our investigation to an incompressible copolymer representation of
the protein. Another novelty is to restrict the analysis to a single protein
size of a finite size $M$. Our previous investigation has involved either an
infinitely long polymer or an infinite number of finite polymers. Thus,
studying small system effects on the statistical mechanics of the protein has
been a central feature of this investigation.

Our aim is to study exactly the statistical mechanics of the general model.
For this, we take the approach of exact enumeration in which we count exactly
the number of conformations of the protein by anchoring one of its ends, the
C-terminus, at the origin of the lattice. We consider a square lattice and use
its lattice symmetry to generate only those conformations whose first step
from the origin is in the horizontal direction. This reduces the number of
conformations by $4$. We also allow the first bend only in the downward
direction, but not in the upward direction to further reduce the number of
conformations that we generate. We consider two different kinds of
conformations for enumeration. We either consider only compact conformations
or consider unrestricted (compact and non-compact) conformations, and generate
all conformations under the above two restrictions due to lattice symmetry.
For compact conformations, we have considered $M\leq64,$ and for unrestricted
conformations, we have considered $M\leq26$ so that the enumeration can be
done in a reasonable amount of time. As real protein interactions are not
well-understood, we have considered three different model energetics to study
the effects of energetics on protein thermodynamics. One of the models (Model
A) is the standard model, while the other two are obtained by weak
perturbation (Model B), and strong perturbation (Model C).

Using plausible arguments under some very mild assumptions, we show that these
models have no energy gap for $M\rightarrow\infty$, even though there appears
to be some gap in the case of small proteins. Indeed, an energy gap is not the
only way a discontinuous folding transition can occur. The latter is known to
occur even in the absence of an energy gap such as the Flory model of
semiflexible homopolymer as shown recently by us. However, the presence of a
gap endows the microcanonical ensemble (Boltzmann) entropy $S(E)$ with
non-concavity. For a macroscopic system, such a non-concave entropy implies a
discontinuous folding transition. Thus, it is the non-concavity that drives
the discontinuous folding transition and not the energy gap. However, we
demonstrate that it is the canonical ensemble equivalent entropy function
$\overline{S}(\overline{E})$ that shares the concavity requirement for small
or macroscopic systems; the canonical entropy $S(E)$ is not required by
thermodynamics to be concave. Moreover, we prove that $\overline{S}%
(\overline{E})\geq$ $S(\overline{E}).$Our exact enumeration confirm these
facts. We show that a Gaussian fit is not very good for exact entropies that
we calculate, especially at low energies, the energies most relevant for the
folding transition. The Gaussian fit invariably gives rise to negative
entropies that are then avoided by advocating an energy gap. This is despite
the fact that the exact enumeration never leads to a negative entropy. Thus,
the usefulness of the random energy model for small proteins is highly questionable.

It is plausible that infinite random copolymers are self-averaging. As a
consequence, all thermodynamic densities are the same for almost all
sequences. However, we find that small proteins are far from being
self-averaging. Therefore, as is commonly believed, the protein sequence is
extremely relevant for its proper or desired functioning. In other words, we
cannot overlook the importance protein sequences have in determining the
native state. Also, as expected, various densities such as the entropy and
energy densities retain a strong dependence on $M$; this dependence should not
be neglected. While a small protein is not supposed to show a sharp folding
transition, a signature of a rounded folding transition appears in the peak in
the specific heat.

We introduce a notion of a distance between conformations and show how the
multi-dimensional configuration space $%
\mathbb{C}
$ can be mapped onto a two-dimensional configuration space $%
\mathbb{C}
_{2l}$,$l=$0,S. These two-dimensional projections provide a glimpse of the
form of $%
\mathbb{C}
,$ and from which we obtain some limited perspective of the energy landscape.
We also calculate the free energy landscape by using the energy density as the
coordinate. These free energy landscape appear very flat with undulations that
are not very high.

We have also shown that applying thermodynamic relations that are valid for
macroscopic systems to small system microcanonical entropy will cause errors
in estimating thermodynamic properties, and should be avoided.

\begin{acknowledgments}
Acknowledgement is made to the National Science Foundation for support (Brad
Lambeth) of this research through the University of Akron REU Site for Polymer
Science (DMR-0352746). Evan Askanazi participated in this project while he was
a high school student, and Brad Lambeth completed the project and obtained
most of the results. The code for the computation was initially created by
Andrea Corsi, and Evan Askanazi checked its various parts. The code was
finally completed by Brad Lambeth.
\end{acknowledgments}


\begin{thebibliography}{99}
\bibitem {note03}According to our definition, a single protein system is small
if $M$ is \emph{finite}, even if the volume $V$ is infinitely large. In this
case, $c$ will converge to zero. If $M$ also diverges with $V$ (the limiting
concentration $c\geq0$), we will refer to this system as macroscopic. A system
with $c=0$ can be either small or macroscopic depending on whether
$N_{\text{R}}$ remains finite or not as $V$ diverges. A bulk system will
always refer to the case of many proteins, which we do not consider here. For
a single protein of finite $M$, it is clear that if one of the ends of the
protein is rooted at the center of the volume $V$, then one does not need to
take the limit $V\rightarrow\infty$, as long as the interactions are
short-ranged. Then the rooted protein will never feel the effects of the
boundary of $V$ as long as $V$ is sufficiently large.
\bibitem {note01}The number of conformations $W$ depends on the size $M$ of
the protein, the residence sequence $\chi,$ the topology of the lattice, etc.
In the rest of the paper, we will not show this dependence explicitly for the
sake of notational economy, but should not be forgotten. As we will discuss
later, by rooting or anchoring the protein, as we do in this work, we make $W$
independent of $\chi.$
\bibitem {note}Since $W$ is a number, it can only be a function of a
dimensionless quantity. Indeed, the arguments of any dimensionless function
must be dimensionless. For $E$, this requires introducing an arbitrary energy
scale $\epsilon$ so that $E/\epsilon$ becomes a pure number. Then, we can
introduce $W$ as a function of this dimensionless energy $E/\epsilon.$
However, following the tradition, in the rest of this work, we will continue
to express various quantities as functions of variables that customarily have
dimensions such as $E$ or $T$ (in the units of the Boltzmann constant),
keeping in mind$\ $that they have been made dimensionless by dividing by
$\epsilon.$
\bibitem {note0}In the thermodynamic limit ($M\rightarrow\infty$) of an
infinitely large macroscopic system, $E$ does not exist mathematically as it
will be usually infinitely large in magnitude, and one must consider its
density $e$. For a small system under investigation here, one can study either
$E$ or $e,$ as both remain bounded.
\bibitem {note3}An extensive quantity is a thermodynamic quantity that
increases asymptotically linearly with the size (in our case $M$) of the
system. Thus, the energy $E$ is extensive, but $W(E)$ is not.
\bibitem {note1}The graph of a concave function (discrete or continuous) is
one in which the line connecting its values at any two points must never lie
above the function at intermediate points.
\bibitem {note02}It should be noted that the Gaussian distribution in
(\ref{Gaussian_W(E)}) cannot be truly a realistic distribution as it violates
the constraint $W(E)\geq1.$ However, this violation is usually interpreted to
imply a folding transition in proteins because of the resulting non-concavity
of the entropy, as discussed later in this section. It should also be noted
that (\ref{Gaussian_W(E)}) cannot be valid for small energies, where
$W(E)\sim1,$ since $W(E)$ is supposed to be an integer, while $W(E)$ in
(\ref{Gaussian_W(E)}) is a continuous variable. This is a serious matter, as
the distribution function at small energies are crucial in determining the
behavior of the folded state.
\bibitem {note2}A better choice for the fit would have been $W\sim
aM^{\gamma-1}e^{b(M-2)},$ which is known to be an appropriate asymptotic
behavior for large $M$. We have not done that. Our simple fit gives the
$R$-value shown in Fig. (\ref{F3}).
\end{thebibliography}
\end{document}